\documentclass[11pt]{article}

\usepackage{amsmath, amsfonts, amssymb, amsthm,  graphicx, mathtools, enumerate,multirow}

\usepackage[affil-it]{authblk}

\usepackage[numbers, square]{natbib}
\usepackage[CJKbookmarks=true,
            bookmarksnumbered=true,
			bookmarksopen=true,
			colorlinks=true,
			citecolor=red,
			linkcolor=blue,
			anchorcolor=red,
			urlcolor=blue]{hyperref}
\usepackage[usenames]{color}

\usepackage[letterpaper, left=1.2truein, right=1.2truein, top = 1.2truein, bottom = 1.2truein]{geometry}
\usepackage[ruled, vlined, lined, commentsnumbered]{algorithm2e}

\usepackage{prettyref,soul}

\newtheorem*{hp}{Hypothesis A}
\newtheorem{lemma}{Lemma}[section]
\newtheorem{proposition}{Proposition}[section]
\newtheorem{thm}{Theorem}[section]

\theoremstyle{definition}
\newtheorem{remark}{Remark}[section]

\newrefformat{eq}{(\ref{#1})}
\newrefformat{chap}{Chapter~\ref{#1}}
\newrefformat{sec}{Section~\ref{#1}}
\newrefformat{algo}{Algorithm~\ref{#1}}
\newrefformat{fig}{Fig.~\ref{#1}}
\newrefformat{tab}{Table~\ref{#1}}
\newrefformat{rmk}{Remark~\ref{#1}}
\newrefformat{clm}{Claim~\ref{#1}}
\newrefformat{def}{Definition~\ref{#1}}
\newrefformat{cor}{Corollary~\ref{#1}}
\newrefformat{lmm}{Lemma~\ref{#1}}
\newrefformat{lemma}{Lemma~\ref{#1}}
\newrefformat{prop}{Proposition~\ref{#1}}
\newrefformat{app}{Appendix~\ref{#1}}
\newrefformat{ex}{Example~\ref{#1}}
\newrefformat{exer}{Exercise~\ref{#1}}
\newrefformat{soln}{Solution~\ref{#1}}
\newrefformat{cond}{Condition~\ref{#1}}

\def\Cov{\text{Cov}} 

\def\text#1{\mbox{\rm #1}}

\DeclarePairedDelimiter{\ceil}{\lceil}{\rceil}

\newcommand{\reals}{{\mathbb{R}}}
\newcommand{\integers}{{\mathbb{Z}}}
\newcommand{\naturals}{{\mathbb{N}}}

\newcommand{\argmin}{\mathop{\rm argmin}}

\newcommand{\lnorm}[2]{\|{#1} \|_{{#2}}}
\newcommand{\indc}[1]{{\mathbf{1}_{\left\{{#1}\right\}}}}
\newcommand{\norm}[1]{\|{#1} \|}
\newcommand{\hf}{{1/2}}
\newcommand{\wh}{\widehat}
\newcommand{\wt}{\widetilde}
\newcommand{\wl}{\overline}
\newcommand{\Norm}[1]{\|{#1} \|}
\newcommand{\Fnorm}[1]{\lnorm{#1}{\rm F}}
\newcommand{\fnorm}[1]{\|#1\|_{\rm F}}
\newcommand{\opnorm}[1]{\|#1\|_{\rm op}}
\newcommand{\nnorm}[1]{\|#1\|_{\rm *}}
\newcommand{\Prob}{\mathbb{P}}
\newcommand{\Expect}{\mathbb{E}}

\newcommand{\Tr}{\mathop{\sf Tr}}
\newcommand{\diag}{\mathop{\text{diag}}}
\newcommand{\supp}{{\rm supp}}
\newcommand{\iprod}[2]{ \langle #1, #2 \rangle}
\newcommand{\pth}[1]{\left( #1 \right)}

\newcommand{\sth}[1]{\left\{ #1 \right\}}
\newcommand{\calL}{\mathcal{L}}
\newcommand{\calF}{\mathcal{F}}
\newcommand{\bbP}{\mathbb{P}}
\newcommand{\bbQ}{\mathbb{Q}}
\newcommand{\TV}{{\sf TV}}
\newcommand{\floor}[1]{{\left\lfloor {#1} \right \rfloor}}
\newcommand{\diff}{\mathrm{d}}

\newcommand{\magenta}{\color{magenta}}
\newcommand{\nb}[1]{{\sf\magenta[#1]}}

\title{Sparse CCA: Adaptive Estimation and Computational Barriers
\thanks{The research
of C.~Gao and H.~H. Zhou is supported in part by NSF Career Award
DMS-0645676 and NSF FRG Grant DMS-0854975.
The research of Z.~Ma is supported in part by NSF Career Award DMS-1352060.}
}
\author[1]{Chao Gao}
\author[2]{Zongming Ma}
\author[1]{Harrison H.~Zhou}
\affil[1]{
Yale University
}
\affil[2]{
University of Pennsylvania
}
\date{}

\begin{document}
\maketitle

\begin{abstract}
Canonical correlation analysis is a classical technique for exploring the relationship between two sets of variables. 
It has important applications in analyzing high dimensional datasets originated from genomics, imaging and other fields.
This paper considers
adaptive minimax and computationally tractable estimation of leading sparse canonical coefficient vectors in high dimensions. 
First, we establish separate
minimax estimation rates for
canonical coefficient vectors of each set of random variables under 
no structural assumption on marginal covariance matrices.
Second, we propose a computationally feasible estimator to attain the optimal rates adaptively under an additional sample size condition.
Finally, we show that a sample size condition of this kind is needed for any randomized polynomial-time estimator to be consistent, assuming hardness of certain instances of the Planted Clique detection problem.  
The result is faithful to the Gaussian models used in the paper.
As a byproduct, we obtain the first computational lower bounds for sparse PCA under the Gaussian single spiked covariance model.
\smallskip

\end{abstract}



\section{Introduction}

Canonical correlation analysis (CCA) \citep{hotelling36} is a classical and important tool in multivariate statistics \citep{anderson58,mkb79}. 
For two random vectors $X\in\mathbb{R}^p$ and $Y\in\mathbb{R}^m$, at the population level,
CCA finds successive vectors ${u}_j\in \mathbb{R}^p$ and ${v}_j\in \mathbb{R}^m$ (called \emph{canonical coefficient vectors}) that solve
\begin{equation}
\label{eq:CCAcov}
\begin{aligned}
\max_{a,b} ~~& a' {\Sigma}_{xy}b,\\
\text{subject to} ~~& a' {\Sigma}_x a=b' {\Sigma}_y b=1, \,\,
a'\Sigma_x {u}_l = b' \Sigma_y {v}_l = 0, \,\, \forall 0\leq l\leq j-1,
\end{aligned}
\end{equation}
where $\Sigma_x=\Cov(X), \Sigma_y=\Cov(Y), \Sigma_{xy}=\Cov(X,Y)$, ${u}_0 = 0$, and ${v}_0 = 0$.
Since our primary interest lies in the covariance structure among $X$ and $Y$, we assume that their means are zeros from here on.
Then the linear combinations $(u_j'X, v_j'Y)$ are the $j$-th pair of \emph{canonical variates}.
This technique has been widely used in various scientific fields to explore the relationship between two sets of variables. 
In practice, one does not have knowledge about the population covariance, and $\Sigma_x$, $\Sigma_y$, and $\Sigma_{xy}$ are replaced by their sample versions $\wh{\Sigma}_x$, $\wh{\Sigma}_y$, and $\wh{\Sigma}_{xy}$ in (\ref{eq:CCAcov}).


Recently, there have been growing interests in applying CCA to analyzing high-dimensional datasets, where the dimensions $p$ and $m$ could be much larger than the sample size $n$. 
It has been well understood that classical CCA breaks down in this regime
\cite{Johnstone07,Bao14,gao14}.
Motivated by genomics, neuroimaging and other applications, 
people have become interested in seeking sparse leading canonical coefficient vectors.
Various estimation procedures imposing sparsity on canonical coefficient vectors have been developed in the literature, which are usually termed \emph{sparse CCA}.
See, for example, \cite{wiesel08, witten09, parkhomenko09, hardoon2011sparse,
le2009sparse, waaijenborg2009sparse,avants2010dementia}.

The theoretical aspect of sparse CCA has also been investigated in the literature. 
A useful model for studying sparse CCA is the \emph{canonical pair model} proposed in \cite{chen13}.
In particular, suppose there are $r$ pairs of canonical coefficient vectors (and canonical variates) among the two sets of variables, then the model reparameterizes the cross-covariance matrix as
\begin{equation}
\Sigma_{xy}=\Sigma_x U\Lambda V'\Sigma_y,
\quad\text{where}\quad U'\Sigma_xU=V'\Sigma_yV=I_r.
\label{eq:CCA}
\end{equation}
Here $U=[u_1,...,u_r]$ and $V=[v_1,...,v_r]$ collect the canonical coefficient vectors and $\Lambda = \diag(\lambda_1,\dots,\lambda_r)$ with $1 > \lambda_1 \geq \cdots \geq \lambda_r > 0$ are the ordered canonical correlations. 
Let $S_u=\supp(U)$ and $S_v=\supp(V)$ be the indices of nonzero rows of $U$ and $V$. 
One way to impose sparsity on the canonical coefficient vectors is to require the sizes of $S_u$ and $S_v$ to be small, namely $|S_u|\leq s_u$ and $|S_v|\leq s_v$ for some $s_u\leq p$ and $s_v\leq m$.
Under this model, \citet{gao14} showed that the minimax rate for estimating $U$ and $V$ under the joint loss function $\fnorm{\wh{U}\wh{V}'-UV'}^2$ is
\begin{equation}
\frac{1}{n\lambda_r^2}\Big(r(s_u+s_v)+s_u\log\frac{ep}{s_u}+s_v\log\frac{em}{s_v}\Big). \label{eq:minimax-joint}
\end{equation}
However, to achieve the rate, \citet{gao14} used a computationally infeasible and nonadaptive procedure, which requires exhaustive search of all possible subsets with the given cardinality and the knowledge of $s_u$ and $s_v$. 
Moreover, it is unclear from \eqref{eq:minimax-joint} whether the estimation error of $U$ depends on the sparsity and the ambient dimension of $V$ and vice versa.

The goal of the present paper is to study three fundamental questions in sparse CCA:
(1) What are the minimax rates for estimating the canonical coefficient vectors on the two sets of variables separately?
(2) Is there a computationally efficient and sparsity-adaptive method that achieves the optimal rates?
(3) What is the price one has to pay to achieve the optimal rates in a computationally efficient way?

\subsection{Main contributions}

We now introduce the main contributions of the present paper from three different viewpoints as suggested by the three questions we have raised.

\paragraph{Separate minimax rates}

The joint loss $\fnorm{\wh{U}\wh{V}'-UV'}^2$ studied by \cite{gao14} characterizes the joint estimation error of both 
$U$ and $V$.
In this paper, we provide a finer analysis by studying individual estimation errors of $U$ and $V$ under a natural loss function that can be interpreted as prediction error of canonical variates. 
The exact definition of the loss functions is given in Section \ref{sec:problem}.
Separate minimax rates are obtained for $U$ and $V$.
In particular, we show that the minimax rate of convergence in estimating $U$ depends only on $n,r,\lambda_r,p$ and $s_u$, but not on either $m$ or $s_v$. 
Consequently, if $U$ is sparser than $V$, then convergence rate for estimating $U$ can be faster than that for estimating $V$. 
Such a difference is not reflected by the joint loss, since its minimax rate (\ref{eq:minimax-joint}) is determined by the slower of the rates of estimating $U$ and $V$. 

\paragraph{Adaptive estimation}

As pointed out in \cite{chen13} and \cite{gao14}, sparse CCA is a more difficult problem than the well-studied sparse PCA. 
A naive application of sparse PCA algorithm to sparse CCA can lead to inconsistent results \cite{chen13}. 
The additional difficulty in sparse CCA mainly comes from the presence of the nuisance parameters $\Sigma_x$ and $\Sigma_y$, which cannot be estimated consistently in a high-dimensional regime in general.
Therefore, our goal is to design an estimator that is adaptive to both the nuisance parameters and the sparsity levels.
Under the canonical pair model, we propose a computationally efficient algorithm. The algorithm has two stages.
In the first stage, 
we propose a convex program for sparse CCA based on a tight convex relaxation of a combinatorial program in \cite{gao14} by considering the smallest convex set containing all matrices of the form $AB'$ with both $A$ and $B$ being rank-$r$ orthogonal matrices.
The convex program can be efficiently solved by the Alternating Direction Method with Multipliers (ADMM) \citep{douglas1956, Boyd11}. 
Based on the output of the first stage, we formulate a sparse linear regression problem in the second stage to improve estimation accuracy, and the final estimator $\wh{U}$ and $\wh{V}$ can be obtained via a group-Lasso algorithm \citep{yuan06}. 
Under the sample size  condition that 
\begin{equation}
n\geq C{s_us_v\log(p+m)}/{\lambda_r^2}
\label{eq:asscomp}
\end{equation}
for some sufficiently large constant $C>0$, 
we show $\wh{U}$ and $\wh{V}$ recover the true canonical coefficient matrices $U$ and $V$ within optimal error rates adaptively with high probability.

\paragraph{Computational lower bound}

We require the sample size condition (\ref{eq:asscomp}) for the adaptive procedure to achieve optimal rates of convergence.
Assuming hardness of certain instances of the Planted Clique detection problem, we provide a computational lower bound to show that a condition of this kind is unavoidable for any computationally feasible estimation procedure to achieve consistency.
Up to an asymptotically equivalent discretization which is necessary for computational complexity to be well-defined, our computational lower bound is established directly for the Gaussian canonical pair model used throughout the paper.

An analogous sample size condition has been imposed in the sparse PCA literature \cite{johnstone09, ma13, cai13a, Vu13}, namely $n \geq C{s^2\log p}/{\lambda^2}$ where $s$ is the sparsity of the leading eigenvector and $\lambda$ the gap between the leading eigenvalue and the rest of the spectrum. 
Berthet and Rigollet \citep{berthet2013complexity} showed that if there existed a polynomial-time algorithm for a generalized sparse PCA detection problem while such a condition is violated, the algorithm could be made (in randomized polynomial-time) into a detection method for the Planted Clique problem in a regime where it is believed to be computationally intractable.
However, both the null and the alternative hypotheses in the sparse PCA detection problem were generalized in \cite{berthet2013complexity} to include all multivariate distributions whose quadratic forms satisfy certain uniform tail probability bounds and so the distributions need not be Gaussian or having a spiked covariance structure \cite{johnstone09}. 
The same remark also applies to the subsequent work on sparse PCA estimation \cite{wang2014statistical}.
Hence, the computational lower bound in sparse PCA was only established for such enlarged parameter spaces.
As a byproduct of our analysis, we establish the desired computational lower bound for sparse PCA in the Gaussian single spiked covariance model. 

\subsection{Organization} 
After an introduction to notation, the rest of the paper is organized as follows. In Section \ref{sec:problem}, we formulate the sparse CCA problem by defining its parameter space and loss function. 
Section \ref{sec:minimax} presents separate minimax rates for estimating $U$ and $V$.
Section \ref{sec:adaptive} proposes a two-stage adaptive estimator that is shown to be minimax rate optimal under an additional sample size condition. 
Section \ref{sec:cb} shows a condition of this kind is necessary for all randomized polynomial-time estimator to achieve consistency by establishing new computational lower bounds for sparse PCA and sparse CCA. 
Section \ref{sec:proof} presents proofs of theoretical results in \prettyref{sec:adaptive}.
Implementation details of the adaptive procedure, numerical studies, additional proofs and technical details are deferred to the supplement \cite{supp}.

\subsection{Notation}

For any $t\in \mathbb{Z}_+$, $[t]$ denotes the set $\{1, 2, ..., t\}$. 
For any set $S$, $|S|$ denotes its cardinality and $S^c$ its complement.
For any event $E$, $\indc{E}$ denotes its indicator function.
For any $a,b\in \reals$, $\ceil{a}$ denotes the smallest integer no smaller than $a$, $\floor{a}$ the largest integer no larger than $a$,
$a\vee b = \max(a,b)$ and $a\wedge b = \min(a,b)$. For a vector $u$, $||u||=\sqrt{\sum_i u_i^2}$, $||u||_0=\sum_i\indc{u_i\neq 0}$, and $||u||_1=\sum_i|u_i|$.
For any matrix $A=(a_{ij})\in \reals^{p\times k}$, $A_{i\cdot}$ denotes its $i$-th row and 
$\supp(A)=\{i\in[p]: \|A_{i\cdot}\|>0\}$, the index set of nonzero rows, is called its support.
For any subset $J\subset [p]\times [k]$,  $A_{J}=(a_{ij}\indc{(i,j)\in J})\in \reals^{p\times k}$ is obtained by keeping all entries in $J$ and replacing all entries in $J^c$ with zeros.
We write $A_{J_1J_2}$ for $A_{J_1\times J_2}$ and $A_{(J_1J_2)^c}$ for $A_{(J_1\times J_2)^c}$. 
Note that $A_{J_1*}=A_{J_1\times [k]}\in \reals^{p\times k}$ while $A_{J_1\cdot}$ stands for the corresponding nonzero submatrix of size $|J_1|\times k$.
In addition, $P_A\in \reals^{p\times p}$ stands for the projection matrix onto the column space of $A$, 
$O(p,k)$
denotes the set of all $p\times k$ orthogonal matrices and $O(k)=O(k,k)$.
Furthermore, $\sigma_i(A)$ stands for the $i$-th largest singular value of $A$ and $\sigma_{\max}(A)=\sigma_1(A)$, $\sigma_{\min}(A)=\sigma_{p\wedge k}(A)$. The Frobenius norm and the operator norm of $A$ are $\fnorm{A}=\sqrt{\sum_{i,j}a_{ij}^2}$ and $\opnorm{A}=\sigma_1(A)$, respectively. The $l_1$ norm and the nuclear norm are  $||A||_1=\sum_{ij}|a_{ij}|$ and $\nnorm{A}=\sum_i\sigma_i(A)$, respectively.
If $A$ is a square matrix, its trace is $\Tr(A) = \sum_{i}a_{ii}$. For two square matrices $A$ and $B$, we write $A\preceq B$ if $B-A$ is positive semidefinite.
For any positive semi-definite matrix $A$, $A^{1/2}$ denotes its principal square root that is positive semi-definite and satisfies $A^{1/2} A^{1/2} = A$. 
The trace inner product of two matrices $A,B\in\mathbb{R}^{p \times k}$ is $\iprod{A}{B}=\Tr(A'B)$. 
Given a random element $X$, $\mathcal{L}(X)$ denotes its probability distribution.
The symbol $C$ and its variants $C_1,C'$, etc. are generic constants and may vary from line to line, unless otherwise specified. The symbols $\mathbb{P}$ and $\mathbb{E}$ stand for generic probability and expectation when the distribution is clear from the context.


\section{Problem Formulation} 
\label{sec:problem}

%

\subsection{Parameter space}
\label{sec:para-space}
Consider a canonical pair model where the observed pairs of measurement vectors $(X_i', Y_i')'$, $i = 1,\dots, n$ are i.i.d.~from a multivariate Gaussian distribution $N_{p+m}(0, \Sigma)$ where
\begin{align*}
\Sigma=\begin{bmatrix}
\Sigma_x & \Sigma_{xy} \\
\Sigma_{yx} & \Sigma_y
\end{bmatrix},
\end{align*}
with the cross-covariance matrix $\Sigma_{xy}$ satisfying \eqref{eq:CCA}.
We are interested in the situation where the leading canonical coefficient vectors are sparse.
One way to quantify the level of sparsity is to bound how many nonzero rows there are in the $U$ and $V$ matrices. 
This notion of sparsity has been used previously in both sparse PCA \cite{cai13a,Vu13} and sparse CCA \cite{gao14} problems when one seeks multiple sparse vectors simultaneously. 

Recall that for any matrix $A$, $\supp(A)$ collects the indices of nonzero rows in $A$.
Adopting the above notion of sparsity, we define 
$\mathcal{F}(s_u,s_v,p,m,r,\lambda;M)$ to be the collection of all covariance matrices $\Sigma$ with the structure (\ref{eq:CCA}) satisfying
\begin{equation}
\begin{aligned}
1.&~~ \mbox{$U\in\mathbb{R}^{p\times r}$ and $V\in\mathbb{R}^{m\times r}$ with $|\supp(U)|\leq s_u$ and $|\supp(V)|\leq s_v$;}
\label{eq:para-space1}\\
2.&~~ \mbox{$\sigma_{\min}(\Sigma_x)\wedge\sigma_{\min}(\Sigma_y)\geq M^{-1}$ and $\sigma_{\max}(\Sigma_x)\vee\sigma_{\max}(\Sigma_y)\leq M$;}\\
3.&~~ \mbox{$\lambda_r\geq\lambda$ and $\lambda_1\leq 1-M^{-1}$.}
\end{aligned}
\end{equation}
The probability space we consider is
\begin{equation}
\begin{aligned}
	\label{eq:para-space}
&\mathcal{P}(n,s_u,s_v,p,m,r,\lambda;M)=\big\{\mathcal{L}((X_1',Y_1')',...,(X_n',Y_n')'): \\
& \quad\quad
\quad\quad
(X_i',Y_i')'\stackrel{iid}{\sim} N_{p+m}(0,\Sigma)
\quad\text{with }\Sigma\in \mathcal{F}(s_u,s_v,p,m,r,\lambda;M)\big\},
\end{aligned}
\end{equation}
where $n$ is the sample size. We shall allow $s_u,s_v,p,m,r,\lambda$ to vary with $n$, while $M > 1$ is restricted to be an absolute constant.


\subsection{Prediction loss}
\label{sec:loss}
From now on, 
the presentation of definitions and results will focus on $U$ only since those for $V$ can be obtained via symmetry.
Given an estimator $\wh{U} = [\wh{u}_1,\dots, \wh{u}_r]$ of the leading canonical coefficient vectors for $X$, a natural way of assessing its quality is to see how well it predicts the values of the canonical variables $U'X^\star \in \mathbb{R}^r$ for a new observation $X^\star$ which is independent of and identically distributed as the training sample used to obtain $\wh{U}$.
This leads us to consider the following loss function 
\begin{equation}
L(\wh{U}, U) = 
\inf_{W\in O(r)}\mathbb{E}^\star\|W'\wh{U}'X^\star-U'X^\star\|^2,
\label{eq:recerror}
\end{equation}
where $\Expect^\star$ means taking expectation only over $X^\star$ and so $L(\wh{U},U)$ is still a random quantity due to the randomness of $\wh{U}$. Since $L(\wh{U},U)$ is the expected squared error for predicting the canonical variables $U'X^{\star}$ via $\wh{U}'X^{\star}$, we refer to it as prediction loss from now on.
It is worth noting that the introduction of an $r\times r$ orthogonal matrix $W$ is unavoidable. 
To see this, we can simply consider the case where $\lambda_1 = \cdots = \lambda_r = \lambda$ in \eqref{eq:CCA}, then we can replace the pair $(U, V)$ in \eqref{eq:CCA} by $(UW, VW)$ for any $W\in O(r)$.
In other words, the canonical  coefficient vectors are only determined up to a joint orthogonal transform.
If we work out the $\Expect^\star$ part in the definition \eqref{eq:recerror}, then the loss function can be equivalently defined as 
\begin{equation}
L(\wh{U},U)=
\inf_{W\in O(r)}\Tr[(\wh{U}W-U)'\Sigma_x(\wh{U}W-U)].
\label{eq:Loss}
\end{equation}
By symmetry, we can define $L(\wh{V},V)$ by simply replacing $U$, $\wh{U}$, $X^\star$ and $\Sigma_x$ in \eqref{eq:recerror} and \eqref{eq:Loss} with $V$, $\wh{V}$, $Y^\star$ and $\Sigma_y$.

A related loss function is  $\fnorm{P_{\wh{U}}-P_U}^2$ measuring the difference between two subspaces. By Proposition \ref{prop:loss} in the supplementary material \cite{supp}, the prediction loss $L(\wh{U},U)$ is a stronger loss function. That is, $\fnorm{P_{\wh{U}}-P_U}^2\leq CL(\wh{U},U)$ for some constant $C>0$ only depending on $M$. 
Actually, $L(\wh{U},U)$ is strictly stronger. 
To see this, let $\Sigma_x=I_p$, $U\in O(p,r)$ and $\wh{U}=2U$. Then, $\fnorm{P_{\wh{U}}-P_U}^2=0$, while $L(\wh{U},U)=\inf_{W\in O(r)}\fnorm{\Sigma_x^{1/2}(\wh{U}W-U)}^2=\inf_{W\in O(r)}(5r-\Tr(W))=r>0$.
In this paper, we will focus on the stronger loss $L(\wh{U},U)$, and provide brief remarks on results for $\fnorm{P_{\wh{U}}-P_U}^2$.

\section{Minimax Rates} 
\label{sec:minimax}

We first provide a minimax upper bound using a combinatorial optimization procedure, and then show that the resulting rate is optimal by further providing a matching minimax lower bound.

Let $(X_i', Y_i')'\in \mathbb{R}^{p+m}, i=1,\dots,n$, be i.i.d.~observations following $N_{p+m}(0,\Sigma)$ for some $\Sigma\in \mathcal{F}(s_u,s_v,p,m,r,\lambda;M)$.
For notational convenience, we assume the sample size is divisible by three, i.e., $n = 3n_0$ for some $n_0\in \mathbb{N}$.

\paragraph{Procedure}
To obtain minimax upper bound, we propose a two-stage combinatorial optimization procedure. 
We split the data into three equal size batches 
$\mathcal{D}_0=\{(X_i',Y_i')'\}_{i=1}^{n_0}$, $\mathcal{D}_1=\{(X_i',Y_i')'\}_{i=n_0+1}^{2n_0}$ and $\mathcal{D}_2=\{(X_i',Y_i')'\}_{i=2n_0+1}^{n}$,
and denote the sample covariance matrices computed on each batch by $\wh\Sigma_x^{(j)}, \wh\Sigma_y^{(j)}$ and $\wh\Sigma_{xy}^{(j)}$ for $j\in \{0,1,2\}$.

In the first stage, we find $(\wh{U}^{(0)}, \wh{V}^{(0)})$ which solves the following program:
\begin{equation}
\label{eq:optexp}
\begin{aligned}
\max_{L\in \mathbb{R}^{p\times r},R\in \mathbb{R}^{m\times r}} &~~ \Tr(L'\wh{\Sigma}^{(0)}_{xy}R), \\
\text{subject to } &~~ L'\wh{\Sigma}^{(0)}_xL=R'\wh{\Sigma}^{(0)}_yR=I_r, \text{ and }
\\
& ~~
|\text{supp}(L)|\leq s_u, |\text{supp}(R)|\leq s_v.
\end{aligned}
\end{equation}
In the second stage, we further refine the estimator for $U$ by finding $\wh{U}^{(1)}$ solving 
\begin{equation}
\label{eq:refineexp}
\begin{aligned}
\min_{L\in \mathbb{R}^{p\times r}} & ~~ \Tr(L'\wh{\Sigma}^{(1)}_xL)-2\Tr(L'\wh{\Sigma}_{xy}^{(1)}\wh{V}^{(0)})\\
\text{subject to} &~~ |\text{supp}(L)|\leq s_u.
\end{aligned}
\end{equation}
The final estimator is a normalized version of $\wh{U}^{(1)}$, defined as
\begin{align}
	\label{eq:mmxexp}
\wh{U}=\wh{U}^{(1)}((\wh{U}^{(1)})'\wh{\Sigma}^{(2)}_x\wh{U}^{(1)})^{-1/2}.
\end{align}

The purpose of sample splitting employed in the above procedure is to facilitate the proof.

\paragraph{Theory and discussion}


The program \eqref{eq:optexp} was first proposed in \cite{gao14} as a sparsity constrained version of the classical CCA formulation. 
However, the resulting estimator will have a convergence rate that involves the sparsity level $s_v$ and the ambient dimension $m$ of the $V$ matrix \cite[Theorem 1]{gao14}, which is sub-optimal. 
The second stage in the procedure is thus proposed to further pursue the optimal estimation rates.
First, if we were given the knowledge of $V$, then the least square solution of regressing $V'Y\in \mathbb{R}^r$ on $X\in \mathbb{R}^p$ is
\begin{equation}
\begin{aligned}
U\Lambda &  = \argmin_{L\in \mathbb{R}^{p\times r}}\mathbb{E}\norm{Y'V - X'L}^2\\
& = \argmin_{L\in \mathbb{R}^{p\times r}}\Tr(L'{\Sigma}_xL)-2\Tr(L'{\Sigma}_{xy}V)+\Tr(V'\Sigma_yV)\\
& = \argmin_{L\in \mathbb{R}^{p\times r}}\Tr(L'{\Sigma}_xL)-2\Tr(L'{\Sigma}_{xy}V),
\end{aligned}
\label{eq:ls}
\end{equation}
where the expectation is with respect to the distribution $(X',Y')'\sim N_{p+m}(0,\Sigma)$. 
The second equality results from taking expectation over each of the three terms in the expansion of the square Euclidean norm, and the last equality holds since $\Tr(V'\Sigma_y V)$ does not involve the argument to be optimized over. 
{In fact, from the canonical pair model, one can easily derive a regression interpretation of CCA, $V'Y=\Lambda U'X+E$, where $E\sim N(0,I_r-\Lambda^2)$. Then, \eqref{eq:refineexp} is a least square formulation of the regression interpretation. 
However, CCA is different from regression because the response $V'Y$ depends on an unknown $V$.}
Comparing \eqref{eq:refineexp} with \eqref{eq:ls}, it is clear that \eqref{eq:refineexp} is a sparsity constrained version of \eqref{eq:ls} where the knowledge of $V$ and the covariance matrix $\Sigma$ are replaced by the initial estimator $\wh{V}^{(0)}$ and sample covariance matrix from an independent sample.
Therefore, $\wh{U}^{(1)}$ can be viewed as an estimator of $U\Lambda$. 
Hence, a final normalization step is taken in \eqref{eq:mmxexp} to transform it to an estimator of $U$.

We now state a bound for the final estimator (\ref{eq:mmxexp}).
\begin{thm}
	\label{thm:uppersep}
Assume
\begin{equation}
\frac{1}{n}\Big(r(s_u+s_v)+s_u\log\frac{ep}{s_u}+s_v\log\frac{em}{s_v}\Big) \leq c
\label{eq:ass}
\end{equation}
for some sufficiently small constant $c>0$.
Then there exist constants $C,C'>0$ only depending on $c$ such that
\begin{eqnarray}
L(\wh{U},U) \leq \frac{C}{n\lambda^2}s_u\Big(r+\log\frac{ep}{s_u}\Big), 
\label{eq:minimaxsep}
\end{eqnarray}
with $\mathbb{P}$-probability at least $1-\exp\left(-C'(s_u+\log(ep/s_u))\right)-\exp\left(-C'(s_v+\log(em/s_v))\right)$ uniformly over $\mathbb{P}\in \mathcal{P}(n,s_u,s_v,p,m,r,\lambda;M)$.
\end{thm}

\begin{remark}\label{remark:rmk}
The paper assumes that $M$ is a constant. However,
it is worth noting that the minimax upper bound
of Theorem \ref{thm:uppersep} 
does not depend on $M$ even if $M$ is allowed to grow with $n$. 
To be specific, assume the eigenvalues of $\Sigma_x$ are bounded in the interval $[M_1,M_2]$. 
The convergence rate of $L(\wh{U},U)$ would still be $\frac{1}{n\lambda^2}s_u\big(r+\log\frac{ep}{s_u}\big)$, because the dependence on $M_1,M_2$ has been implicitly built into the prediction loss. 
On the other hand, a convergence rate for the loss $\fnorm{P_{\wh{U}}-P_U}^2$ would be $\big(\frac{M_2}{M_1}\big)\frac{1}{n\lambda^2}s_u\big(r+\log\frac{ep}{s_u}\big)$, with an extra factor of the condition number of $\Sigma_x$.
\end{remark}

Under assumption \eqref{eq:ass}, \prettyref{thm:uppersep} achieves a convergence rate for the prediction loss in $U$ that does not depend on any parameter related to $V$. 
Note that the probability tail still involves $m$ and $s_v$.
However, it can be shown that $\exp\left(-C'(s_v+\log(em/s_v))\right)\leq m^{-C'/2}$, and so the corresponding term in the tail probability goes to $0$ as long as $m\rightarrow\infty$.
The optimality of this upper bound can be justified by the following minimax lower bound.
\begin{thm} \label{thm:lower}
Assume that $r\leq \frac{s_u\wedge s_v}{2}$. 
Then there exists some constant $C>0$ only depending on $M$  and an absolute constant $c_0>0$, such that
\begin{eqnarray*}
\inf_{\wh{U}}\sup_{\mathbb{P}\in\mathcal{P}}\mathbb{P}\left\{L(\wh{U},U)\geq c_0 \wedge \frac{C}{n\lambda^2}s_u\Big(r+\log\frac{ep}{s_u}\Big) \right\} &\geq& 0.8,
\end{eqnarray*}
where $\mathcal{P}=\mathcal{P}(n,s_u,s_v,p,m,r,\lambda;M)$.
\end{thm}

By Theorem \ref{thm:uppersep} and Theorem \ref{thm:lower}, the rate in (\ref{eq:minimaxsep}), whenever it is upper bounded by a constant, 
is the minimax rate of the problem.

\section{Adaptive and Computationally Efficient Estimation} 
\label{sec:adaptive}

\prettyref{sec:minimax} determines the minimax rates for estimating $U$ under the prediction loss. 
However, there are two drawbacks of the procedure \eqref{eq:optexp} -- \eqref{eq:mmxexp}.
One is that it requires the knowledge of the sparsity levels $s_u$ and $s_v$. It is thus not adaptive.
The other is that in both stages one needs to conduct exhaustive search over all subsets of given sizes in the optimization problems \eqref{eq:optexp} and \eqref{eq:refineexp}, and hence the computation cost is formidable.

In this section, we overcome both drawbacks by proposing a two-stage convex program approach towards sparse CCA. 
The procedure is named CoLaR, standing for Convex program with group-Lasso Refinement. 
It is not only computationally feasible but also achieves the minimax estimation error rates adaptively over a large collection of parameter spaces under an additional sample size condition.
The issues related to this additional sample size condition will be discussed in more detail in the subsequent \prettyref{sec:cb}.

\subsection{Estimation scheme}
\label{sec:adapt-est}

The basic principle underlying the computationally feasible estimation scheme is to seek tight convex relaxations of the combinatorial programs \eqref{eq:optexp} -- \eqref{eq:refineexp}.
In what follows, we introduce convex relaxations for the two stages in order.
As in \prettyref{sec:minimax}, we assume that the data is split into three batches $\mathcal{D}_0,\mathcal{D}_1$ and $\mathcal{D}_2$ of equal sizes and for $j=0,1,2$, let $\wh\Sigma_{x}^{(j)},\wh\Sigma_y^{(j)}$ and $\wh\Sigma_{xy}^{(j)}$ be defined as before.

\paragraph{First stage}
By the definition of trace inner product, the objective function in \eqref{eq:optexp} can be rewritten as
$\Tr(L'\wh{\Sigma}_{xy}R)=\iprod{\wh{\Sigma}_{xy}}{LR'}$.
Since it is linear in $F = LR'$, this suggests treating $LR'$ as a single argument rather than optimizing over $L$ and $R$ separately. 
Next, the support size constraints $|\supp(L)|\leq s_u, |\supp(R)|\leq s_v$ imply that the vector $\ell_0$ norm $\|LR'\|_0 \leq s_u s_v$.
Applying the convex relaxation of $\ell_0$ norm by $\ell_1$ norm and including it as a Lagrangian term, we are led to consider a new objective function
\begin{equation}
\max_{F\in \mathbb{R}^{p\times m}}\iprod{\wh{\Sigma}_{xy}^{(0)}}{F}-\rho||F||_1, \label{eq:newob}
\end{equation}
where $F$ serves as a surrogate for $LR'$, $\|F\|_1 = \sum_{i\in[p], j\in [m]} |F_{ij}|$ denotes the vector $\ell_1$ norm of the matrix argument, and $\rho$ is a penalty parameter controlling sparsity. 
Note that \eqref{eq:newob} is the maximization problem of a concave function, which becomes a convex program if the constraint set is convex.
Under the identity $F= LR'$, the normalization constraint in \eqref{eq:optexp} reduces to 
\begin{align}
	\label{eq:oldcons}
(\wh\Sigma_x^{(0)})^{1/2} F (\wh\Sigma_y^{(0)})^{1/2} \in \mathcal{O}_r 
= \{ AB': A \in O(p,r), B\in O(m,r) \}.
\end{align}
Naturally, we relax it to $(\wh\Sigma_x^{(0)})^{1/2} F (\wh\Sigma_y^{(0)})^{1/2} \in \mathcal{C}_r$ where 
\begin{align}
	\label{eq:newcons}
\mathcal{C}_r = \{G \in \mathbb{R}^{p\times m}: \|G\|_{*}\leq r, \opnorm{G}\leq 1 \} = \mathrm{conv}(\mathcal{O}_r)
\end{align}
is the smallest convex set containing $\mathcal{O}_r$. The relation (\ref{eq:newcons}) is stated in the proof of Theorem 3 of \cite{watson1993matrix}.
Combining \eqref{eq:newob} -- \eqref{eq:newcons}, we use the following convex program for the first stage in our adaptive estimation scheme:
\begin{equation}
\label{eq:optpoly}
\begin{aligned}
\max_{F\in \mathbb{R}^{p\times m}} &~~\iprod{\wh{\Sigma}^{(0)}_{xy}}{F}-\rho||F||_1 \\
\text{subject to} &~~ \nnorm{(\wh{\Sigma}^{(0)}_x)^{1/2}F(\wh{\Sigma}^{(0)}_y)^{1/2}}\leq r, \,\opnorm{(\wh{\Sigma}^{(0)}_x)^{1/2}F(\wh{\Sigma}^{(0)}_y)^{1/2}}\leq 1.
\end{aligned}
\end{equation}

Implementation of \eqref{eq:optpoly} is discussed in
Section \ref{sec:admm} in the supplement \cite{supp}.

\begin{remark}
A related but different convex relaxation was proposed in \citep{Vu13} for the sparse PCA problem, where the set of all rank $r$ projection matrices (which are symmetric) is relaxed to its convex hull -- the Fantope $\left\{P: \Tr(P)=r, 0\preceq P\preceq I_p\right\}$.
Such an idea is not directly applicable in the current setting due to the asymmetric nature of the matrices included in the set $\mathcal{O}_r$ in \eqref{eq:oldcons}.
\end{remark}
\begin{remark}
The risk of the solution to \eqref{eq:optpoly} for estimating $UV'$ is sub-optimal compared to the optimal rates determined in \cite{gao14}.
See \prettyref{thm:adajoint} below.
Nonetheless, it leads to a reasonable estimator for the subspaces spanned by the first $r$ left and right canonical  coefficient vectors under a sample size condition, which is sufficient for achieving the optimal estimation rates for $U$ and $V$ in a further refinement stage to be introduced below. 
Although it is possible that some other penalty function rather than the $\ell_1$ penalty in \eqref{eq:optpoly} could also achieve this goal, $\ell_1$ is appealing due to its simplicity.
\end{remark}

\paragraph{Second stage} 
Now we turn to the convex relaxation to \eqref{eq:refineexp} in the second stage. 
By the discussion following \prettyref{thm:uppersep}, if we view the rows of $L$ as groups, then \eqref{eq:refineexp} becomes a least square problem with a constrained number of active groups.
A well-known convex relaxation for such problems is the group-Lasso \cite{yuan06}, where the number of active groups constraint is relaxed by bounding the sum of $\ell_2$ norms of the coefficient vector of each group.
Let $\wh{A}$ be the solution to \eqref{eq:optpoly} and $\wh{U}^{(0)}$ (resp.~$\wh{V}^{(0)}$) be the matrix consisting of its first $r$ left (resp.~right) singular vectors.
Thus, in the second stage of the adaptive estimation scheme, we propose to solve the following group-Lasso problem:
\begin{equation}
\min_{L\in \mathbb{R}^{p\times m}}
\Tr(L'\wh{\Sigma}^{(1)}_xL)-2\Tr(L'\wh{\Sigma}^{(1)}_{xy}\wh{V}^{(0)})+
\rho_u \sum_{j=1}^p\|L_{j\cdot}\|,
\label{eq:refinepoly}
\end{equation}
where $\sum_{j=1}^p \|L_{j\cdot}\|$ is the group sparsity penalty, defined as the sum of the $\ell_2$ norms of all the row vectors in $L$, and $\rho_u$ is a penalty parameter controlling sparsity. Note that the group sparsity penalty is crucial, since if one uses an $\ell_1$ penalty instead, only a sub-optimal rate can be achieved.
Suppose the solution to \eqref{eq:refinepoly} is $\wh{U}^{(1)}$, then our final estimator in the adaptive estimation scheme is its normalized version
\begin{align}
\label{eq:mmxpoly}
\wh{U}=\wh{U}^{(1)} ((\wh{U}^{(1)})'\wh{\Sigma}^{(2)}_x \wh{U}^{(1)} )^{-1/2}.
\end{align}
As before, sample splitting is only used for technical arguments in the proof. 
Simulation results in Section \ref{sec:num} in the supplement \cite{supp} show that using the whole dataset repeatedly in \eqref{eq:optpoly} -- \eqref{eq:mmxpoly} yields satisfactory performance and
the improvement by the second stage is considerable.

\subsection{Theoretical guarantees}
\label{sec:adapt-thm}

We first state the upper bound for the solution $\wh{A}$ to the convex program \eqref{eq:optpoly}.
\begin{thm}\label{thm:adajoint}
Assume that
\begin{equation}
n \geq C_1 \frac{s_us_v\log(p+m)}{\lambda^2}, \label{eq:ass1}
\end{equation}
for some sufficiently large constant $C_1 > 0$. 
Then there exist positive constants $\gamma_1, \gamma_2$ and $C,C'$ only depending on $M$ and $C_1$, such that when $\rho = \gamma\sqrt{{\log(p+m)}/{n}}$ for $\gamma\in[\gamma_1,\gamma_2]$,
$$\fnorm{\wh{A}-UV'}^2\leq 
C\frac{s_us_v\log(p+m)}{n\lambda^2},
$$
with $\mathbb{P}$-probability at least $1-\exp(-C'(s_u+\log(ep/s_u)))-\exp(-C'(s_v+\log(em/s_v)))$ for any $\mathbb{P}\in\mathcal{P}(n,s_u,s_v,p,m,r,\lambda;M)$.
\end{thm}
Note that the error bound in \prettyref{thm:adajoint} can be much larger than the optimal rate for joint estimation of $UV'$ established in
\cite{gao14}.
Nonetheless, under the sample size condition \eqref{eq:ass1}, it still ensures that $\wh{A}$ is close to $UV'$ in Frobenius norm distance.
This fact, together with the proposed refinement scheme \eqref{eq:refinepoly} -- \eqref{eq:mmxpoly}, guarantees the optimal rates of convergence for the estimator \eqref{eq:mmxpoly} as stated in the following theorem.

\begin{thm}\label{thm:adasep}
Assume
\eqref{eq:ass1} holds
for some sufficiently large $C_1 \geq 0$.  
Then there exist constants $\gamma$ and $\gamma_u$ only depending on $C_1$ and $M$ such that if we set $\rho=\gamma'\sqrt{{[\log(p+m)]}/{n}}$ and $\rho_u=\gamma'_u\sqrt{({r+\log p})/{n}}$ for any $\gamma'\in[\gamma,C_2\gamma]$ and $\gamma_u'\in[\gamma_u,C_2\gamma_u]$ for some absolute constant $C_2>0$, there exist a constants $C,C'>0$ only depending on $C_1,C_2$ and $M$, such that
\begin{eqnarray*}
L(\wh{U},U) &\leq& C\frac{s_u\left(r+\log p\right)}{n\lambda^2},
\end{eqnarray*}
with $\mathbb{P}$-probability at least $1-\exp(-C'(s_u+\log(ep/s_u)))-\exp(-C'(s_v+\log(em/s_v)))-\exp(-C'(r+\log(p\wedge m)))$ uniformly over $\mathbb{P}\in \mathcal{P}(n,s_u,s_v,p,m,r,\lambda;M)$.
\end{thm}

\begin{remark}
The result of Theorem \ref{thm:adasep} assumes a constant $M$. Explicit dependence on the eigenvalues of the marginal covariance can be tracked even when $M$ is diverging. Assuming the eigenvalues of $\Sigma_x$ all lie in the interval $[M_1,M_2]$, then the convergence rate of $L(\wh{U},U)$ would be $\big(\frac{M_2}{M_1}\big)^2\frac{s_u\left(r+\log p\right)}{n\lambda^2}$ and a convergence rate of $\fnorm{P_{\wh{U}}-P_U}^2$ would be $\big(\frac{M_2}{M_1}\big)^3\frac{s_u\left(r+\log p\right)}{n\lambda^2}$. Compared with Remark \ref{remark:rmk}, there is an extra factor $\big(\frac{M_2}{M_1}\big)^2$, which is also present for the Lasso error bounds \citep{Bickel09,negahban2012unified}. Evidence has been given in the literature that such an extra factor can be intrinsic to all polynomial-time algorithms \citep{zhang2014lower}.
\end{remark}

Although both Theorem \ref{thm:adajoint} and Theorem \ref{thm:adasep} assume Gaussian distributions, a scrutiny of the proofs shows that the same results hold if the Gaussian assumption is weakened to subgaussian.
By \prettyref{thm:lower}, the rate in \prettyref{thm:adasep} is optimal.
By \prettyref{thm:adajoint} and \prettyref{thm:adasep}, the choices of the penalty parameters $\rho$ and $\rho_u$ in \eqref{eq:optpoly} and \eqref{eq:refinepoly} do not depend on $s_u$ or $s_v$. 
Therefore, the proposed estimation scheme \eqref{eq:optpoly} -- \eqref{eq:mmxpoly} achieves the optimal rate adaptively over sparsity levels.
A full treatment of adaptation to $M$ is beyond the scope of the current paper, though it seems possible in view of the recent proposals in  \citep{belloni2011square, sun2012scaled,bunea2014}.
A careful examination of the proofs shows that the dependence of $\rho$ and $\rho_u$ on $M$ is through $\opnorm{\Sigma_x}^{1/2}\opnorm{\Sigma_y}^{1/2}$ and $\opnorm{\Sigma_x}$, respectively.
When $p$ and $m$ are bounded from above by a constant multiple of $n$, we can upper bound the operator norms by the sample counterparts to remove the dependence of these penalty parameters on $M$.
We conclude this section with two more remarks.
\begin{remark}
The group sparsity penalty used in the second stage (\ref{eq:refinepoly}) plays an important role in achieving the optimal rate ${s_u(r+\log p)}/{(n\lambda^2)}$. 
Except for the extra $\lambda^{-2}$ term, this convergence rate is a typical one for group Lasso \citep{lounici2011oracle}.
If we simply use an $\ell_1$ penalty, then we will obtain the rate ${rs_u
\log p}/{(n\lambda^2)}$, which is clearly sub-optimal.
\end{remark}
\begin{remark}
Comparing Theorem \ref{thm:uppersep} with Theorem \ref{thm:adasep},
the adaptive estimation scheme achieves the optimal rates of convergence for a smaller collection of parameter spaces of interest due to the more restrictive sample size condition \eqref{eq:ass1}.
We examine the necessity of this condition in more details in \prettyref{sec:cb} below.
\end{remark}

\section{Computational Lower Bounds} 
\label{sec:cb}


In this section, we provide evidence that the sample size condition \eqref{eq:ass1} imposed on the adaptive estimation scheme in Theorems \ref{thm:adajoint} and \ref{thm:adasep} is probably unavoidable for any computationally feasible estimator to be consistent.
To be specific, we show that for a sequence of parameter spaces in \eqref{eq:para-space1} -- \eqref{eq:para-space}, 
if the condition is violated, then any computationally efficient consistent estimator of sparse canonical coefficients leads to a computationally efficient and statistically powerful test for the Planted Clique detection problem in a regime where it is believed to be computationally intractable.

\paragraph{Planted Clique}
Let $N$ be a positive integer and $k\in[N]$. We denote by $\mathcal{G}(N,1/2)$ the Erd\H{o}s-R\'{e}nyi graph on $N$ vertices where each edge is drawn independently with probability $1/2$,  
and
by $\mathcal{G}(N,1/2,k)$ the random graph generated by first sampling from $\mathcal{G}(N,1/2)$ and then selecting $k$ vertices uniformly at random and forming a clique of size $k$ on these vertices. 
For an adjacency matrix $A\in\{0,1\}^{N\times N}$ of an instance from either $\mathcal{G}(N,1/2)$ or $\mathcal{G}(N,1/2,k)$, the \emph{Planted Clique detection problem} of parameter $(N,k)$ refers to testing the following hypotheses
\begin{align}
	\label{eq:pc}
H_0^G: A\sim \mathcal{G}(N,1/2)\quad \mbox{v.s.}\quad
H_1^G: A\sim \mathcal{G}(N,1/2,k).	
\end{align}

It is widely believed that when $k=O(N^{1/2-\delta})$, the problem (\ref{eq:pc}) cannot be solved by any randomized polynomial-time algorithm. 
In the rest of the paper, we formalize the conjectured hardness of Planted Clique problem into the following hypothesis.
\begin{hp}
For any sequence $k=k(N)$ such that $\limsup_{N\to\infty}\frac{\log k}{\log N}<\frac{1}{2}$ and any randomized polynomial-time test $\psi$,
\begin{align*}
\liminf_{N\to\infty}\left(\mathbb{P}_{H_0^G}\psi+\mathbb{P}_{H_1^G}(1-\psi)\right)\geq\frac{2}{3}.	
\end{align*}
\end{hp}
Evidence supporting this hypothesis has been provided in \cite{rossman2010average, feldman2013statistical}.
Computational lower bounds in several statistical problems have been established by assuming the above hypothesis and its close variants, including sparse PCA detection \cite{berthet2013complexity} and estimation \cite{wang2014statistical} in classes defined by a restricted covariance concentration condition, submatrix detection \citep{ma2013computational} and community detection \citep{hajek14}. 

\paragraph{Necessity of the sample size condition \eqref{eq:ass1}}
Under Hypothesis A, the necessity of condition \eqref{eq:ass1} is supported by the following theorem.
\begin{thm}\label{thm:clCCA-cont}
Suppose that Hypothesis A holds and that as $n\to \infty$, $p = m$ satisfying $2n\leq p\leq n^{a}$ for some constant $a>1$,  $s_u =s_v$, $n(\log n)^5\leq cs_u^4$ for some sufficiently small $c>0$, and $\lambda=\frac{s_us_v}{7290n(\log(12n))^2}$. 
If for some $\delta\in (0,1)$, 
\begin{align}
	\label{eq:sample-size}
\liminf_{n\to\infty} \frac{(s_u s_v)^{1-\delta} \log(p+m)}{n\lambda^2} > 0,	
\end{align}
then for any randomized polynomial-time estimator $\wh{u}$,
\begin{equation}
\liminf_{n\to\infty}
{\sup_{\mathbb{P}\in\mathcal{P}(n,s_u,s_v,p,m,1,\lambda;3)}}
\mathbb{P}\Big\{L(\wh{u},u)>\frac{1}{300}\Big\}>\frac{1}{4}.
\label{eq:ceCCA-cont}
\end{equation}
\end{thm}

Comparing \eqref{eq:ass1} with \eqref{eq:sample-size}, we see that subject to a sub-polynomial factor, the condition \eqref{eq:ass1} is necessary to achieve consistent sparse CCA estimation within polynomial time complexity.

\begin{remark}
	\label{rmk:real-computation}
The statement in \prettyref{thm:clCCA-cont} is rigorous only if we assume the computational complexities of basic arithmetic operations on real numbers and sampling from univariate continuous distributions with analytic density functions are all $\Theta(1)$ \cite{Blum12}.
To be rigorous under the probabilistic Turing machine model \cite{Arora09}, we need to introduce appropriate discretization of the problem and be more careful with the complexity of random number generation. 
To convey the key ideas in our computational lower bound construction, we focus on the continuous case throughout this section and defer the formal discretization arguments to \prettyref{sec:discrete} in the supplement \cite{supp}.
\end{remark}
%

In what follows, we divide the reduction argument leading to \prettyref{thm:clCCA-cont} into two parts. 
In the first part, we show Hypothesis A implies the computational hardness of the sparse PCA problem under the Gaussian spiked covariance model.
In the second part, we show computational hardness of sparse PCA implies that of sparse CCA as stated in \prettyref{thm:clCCA-cont}.

\subsection{Hardness of sparse PCA under Gaussian spiked covariance model}
\label{sec:spca-clb}
Gaussian single spiked model \cite{johnstone09} refers to the distribution $N_p(0,\Sigma)$ where 
$\Sigma = \tau \theta\theta' + I_p$.
Here, $\theta$ is the eigenvector of unit length and $\tau > 0$ is the eigenvalue. 
Define the following Gaussian single spiked model parameter space for sparse PCA
\begin{equation}
	\label{eq:gss-model}
\begin{aligned}
\mathcal{Q}(n,s,p,\lambda) = \{\calL(W_1,\dots, W_n): & W_i\stackrel{iid}{\sim} N_p(0,\tau\theta\theta'+I_p),\\
&~~ \|\theta\|_0 \leq s, \tau\in [\lambda, 3\lambda] \}.	
\end{aligned}
\end{equation}
The minimax estimation rate for $\theta$ under the loss $\fnorm{P_{\wh{\theta}}-P_{\theta}}^2$ is 
$\frac{\lambda+1}{n\lambda^2}s\log\frac{ep}{s}$.
See, for instance, \cite{cai13a}.
However, 
to achieve the above minimax rate via computationally efficient methods
such as those proposed in \cite{Birnbaum12, ma13, cai13a},
researchers
have required the sample size to satisfy
$n\geq C\frac{s^2\log p}{\lambda^2}$
for some sufficiently large constant $C>0$. 
Moreover, no computationally efficient estimator is known to achieve consistency when the sample size condition is violated.
As a first step toward the establishment of \prettyref{thm:clCCA-cont}, we show that Hypothesis A implies hardness of sparse PCA under Gaussian spiked covariance model \eqref{eq:gss-model} when $\liminf_{n\to\infty}\frac{s^{2-\delta}\log p}{n\lambda^2} > 0$ for some $\delta > 0$.

We note that previous computational lower bounds for sparse PCA in \cite{berthet2013complexity,wang2014statistical} cannot be used here directly because they are only valid for parameter spaces defined via the restricted covariance concentration (RCC) condition.
As pointed out in \cite{wang2014statistical}, such parameter spaces include (but are not limited to) all subgaussian distributions with sparse leading eigenvectors and the covariance matrices need not be of the spiked form $\Sigma = \tau \theta\theta' + I_p$.
Therefore, the Gaussian single spiked model parameter space defined in \eqref{eq:gss-model} only constitutes a small subset of such RCC parameter spaces.
The goal of the present subsection is to establish the computational lower bound for the Gaussian single spiked model directly.

Suppose we have an estimator $\wh\theta = \wh\theta(W_1,\dots, W_n)$ of the leading sparse eigenvector, we propose the following reduction scheme to transform it into a test for \eqref{eq:pc}.
To this end, we first introduce some additional notation.
Consider integers $k$ and $N$.
Define 
\begin{align}
\label{eq:delta-N}
\delta_N = \frac{k}{N}, \quad
\eta_N = \frac{k}{45 N(\log N)^2}.
\end{align}
For any $\mu\in \mathbb{R}$, let $\phi_\mu$ denote the density function of the $N(\mu,1)$ distribution, and let
\begin{align}
	\label{eq:bar-phi-mu}
\bar\phi_\mu = \frac{1}{2} (\phi_\mu + \phi_{-\mu})
\end{align}
denote the density function of the Gaussian mixture $\frac{1}{2}N(\mu,1)+\frac{1}{2}N(-\mu,1)$.
Next, let $\wt\Phi_0$ be the restriction of the $N(0,1)$ distribution on the interval $[-3\sqrt{\log N}, 3\sqrt{\log N}]$.
For any $|\mu|\leq  3\sqrt{\eta_N\log N}$, 
define two probability distributions $\calF_{\mu,0}$ and $\calF_{\mu,1}$ with densities
\begin{eqnarray}
\label{eq:f0} f_{\mu,0}(x) &=& M_0\left(\phi_0(x)-\delta_N^{-1}[\bar\phi_{\mu}(x)-\phi_0(x)]\right)
\indc{|x|\leq 3\sqrt{\log N}}, \\
\label{eq:f1} f_{\mu,1}(x) &=& M_1\left(\phi_0(x)+\delta_N^{-1}[\bar\phi_{\mu}(x)-\phi_0(x)]\right)
\indc{|x|\leq 3\sqrt{\log N}},
\end{eqnarray}
where the $M_i$'s are normalizing constants such that $\int_\reals f_{\mu, i} =1$ for $i=0,1$. It can be verified that $f_{\mu,i}$ are properly defined probability density function when $|\mu|\leq 3\sqrt{\eta_N\log N}$. For details, see \prettyref{lem:1} in the supplement \cite{supp}.

With the foregoing definition, the proposed reduction scheme can be summarized as \prettyref{algo:reduction1}. Here, the starting point is the adjacency matrix $A$ of the random graph, and the reduction is well defined for all instances of $N\geq 12 n$ and $p\geq 2n$.

\begin{algorithm}[h]
\caption{Reduction from Planted Clique to Sparse PCA (in Gaussian Single Spiked Model)}
\label{algo:reduction1}
\SetAlgoLined
\KwIn{}
1.~Graph adjacency matrix $A\in \{0,1\}^{N\times N}$;\\
2.~Estimator $\wh\theta$ for the leading eigenvector $\theta$. 

\KwOut{A solution to the hypothesis testing problem \eqref{eq:pc}.}

\nl \textbf{Initialization}. 
Generate i.i.d.~random variables $\xi_1,\dots, \xi_{2n}\sim \wt\Phi_0$.
Set
\begin{align}
	\label{eq:red-1}
\mu_i = \eta_N^{1/2}\, \xi_i , \quad i =1,\dots, 2n.
\end{align}

\nl \textbf{Gaussianization}. 
Generate two matrices $B_0, B_1 \in \mathbb{R}^{2n\times 2n}$ where conditioning on the $\mu_i$'s, all the entries are mutually independent satisfying
\begin{align}
	\label{eq:red-2}
\calL((B_0)_{ij}|\mu_i) = \calF_{\mu_i,0}
\quad \mbox{and} \quad
\calL((B_1)_{ij}|\mu_i) = \calF_{\mu_i,1}.
\end{align}
Let $A_0\in\{0,1\}^{2n\times 2n}$ be the lower--left $2n\times 2n$ submatrix of the matrix $A$.
Generate a matrix $W = [W_1',\dots,W_{2n}']'\in \mathbb{R}^{2n\times p}$ where
for each $i\in[2n]$, if $j\in [2n]$, we set
\begin{equation}
\label{eq:red-3}
W_{ij}=(B_0)_{ij}\left(1-(A_0)_{ij}\right)+(B_1)_{ij}(A_0)_{ij}.
\end{equation}
If $2n < j\leq p$, we let $W_{ij}$ be an independent draw from $N(0,1)$.

\nl \textbf{Test Construction}. 
Let $\wh{\theta} = \wh{\theta}(W_{1},\dots, W_{n})$ be the estimator of the leading eigenvector by treating $\{W_{i}\}_{i=1}^{n}$ as data. It is normalized to be a unit vector.
We reject $H_0^G$ if
\begin{align}
\label{eq:red-6}
\wh{\theta}' (\frac{1}{n}\sum_{i=n+1}^{2n} W_{i} W_i') \wh{\theta} \geq 1 + \frac{1}{4} \,k\,\eta_N.
\end{align}
\end{algorithm}


We now explain how the reduction achieves its goal. 
For simplicity, focus on the case where $p = 2n$.
Let ${\epsilon}=({\epsilon}_1,...,{\epsilon}_{2n})\in \{0,1\}^{2n}$ where $\epsilon_i$ is the indicator of whether the $i$-th row of $A_0$ (defined in Step 2 of \prettyref{algo:reduction1}) belongs to the planted clique or not, and
${\gamma}=({\gamma}_1,...,{\gamma}_{2n})$ the indicators of the columns of $A_0$. 
In what follows, we discuss the distributions of $W$ when $A\sim H_0^G$ and $H_1^G$, respectively.

When $A\sim H_0^G$, the $\epsilon_i$'s and $\gamma_j$'s are all zeros. 
In this case, we can verify that the entries of $W$ are mutually independent and for each $(i,j)$ the marginal distribution of $W_{ij}$ is close to the $N(0,1)$ distribution (c.f., Lemma \ref{lem:fixedTV} in the supplement \cite{supp}).
Hence, the rows of $W$ are close to i.i.d.~random vectors from the $N_p(0, I_p)$ distribution. 
Since $\wh{\theta} = \wh{\theta}(W_1,\dots,W_n)$ is independent of $\{W_i\}_{i=n+1}^{2n}$, the LHS of \eqref{eq:red-5} is close in distribution to a $\chi^2_n$ random variable scaled by $n$ which concentrates around its expected value one.
Indeed, it is upper bounded by $1 + O(\sqrt{\log (n)/n})$ with high probability.

If $A\sim H_1^G$, then the $(i,j)$-th entry of $A_0$ is an edge in the planted clique if and only if $\epsilon_i = \gamma_j = 1$.
Moreover, the joint distribution of $\{\epsilon_1,\dots, \epsilon_{2n}, \gamma_1,\dots, \gamma_{2n} \}$ is close to that of $4n$ i.i.d.~Bernoulli random variables $\{ \wt\epsilon_1,\dots, \wt\epsilon_{2n}, \wt\gamma_1,\dots, \wt\gamma_{2n} \}$ with success probability $\delta_N = k/N$.
For simplicity,
suppose that these indicators are indeed i.i.d.~Bernoulli($\delta_N$) variables $\{ \wt\epsilon_1,\dots, \wt\epsilon_{2n}, \wt\gamma_1,\dots, \wt\gamma_{2n} \}$. Then, one can show that conditioning on $\wt\gamma_j = 0$, for any $i\in [2n]$, the conditional distribution of $(W_{ij}|\wt\gamma_j = 0)$, after integrating over the conditional distribution of $\wt\epsilon_i$, $\mu_i$ and $(A_0)_{ij}$, is approximately $N(0,1)$. 
In contrast, conditioning on $\wt\gamma_j = 1$, for any $i\in [2n]$, the conditional distribution of $(W_{ij}|\wt\gamma_j = 1)$
is approximately $N(0, 1 + \eta_N)$.
Therefore, conditioning on $\wt\gamma$ the distribution of the $W_i$'s is close to that of $2n$ i.i.d.~random vectors sampled from 
\begin{equation*}
N_p(0, \tau \theta\theta' + I_p), \quad
\mbox{where $\theta = {\wt\gamma}/{\| \wt\gamma\|}$ and $\tau = \eta_N\|\wt\gamma\|^2$,}
\end{equation*}
i.e., a Gaussian spiked covariance model in \prettyref{eq:gss-model}. 
Here, the leading eigenvector $\theta$ has sparsity level $|\supp(\theta)| = |\supp(\wt\gamma)| =\sum_{j} \wt\gamma_j$, which concentrates around its mean value $n\delta_N \asymp k$ if $N\asymp n$.
Thus, if $\wh{\theta}$ estimates $\theta$ well, 
then the LHS of \eqref{eq:red-5} approximately follows a non-central $\chi_n$ distribution scaled by $n$, 
which should exceed $1 + O(\sqrt{\log(n)/n})$ with high probability under the alternative hypothesis.
Hence, \prettyref{algo:reduction1} is expected to yield a test with small error for the Planted Clique problem \eqref{eq:pc} when $\wh\theta$ is a good estimator.

The materialization of the foregoing discussion leads to the following result
which demonstrates quantitatively that a decent estimator of the leading sparse eigenvector results in a good test (by applying the reduction \eqref{eq:red-1} -- \eqref{eq:red-6}) for the Planted Clique detection problem \eqref{eq:pc}.
\begin{thm}
\label{thm:test-error-pca}
For some sufficiently small constant $c > 0$,
assume $\frac{N(\log N)^5}{k^4}\leq c$, $cN\leq n\leq N/12$ and $p\geq 2n$. 
Then, for any $\wh\theta$ such that
\begin{equation}
\sup_{\mathbb{Q}\in \mathcal{Q}(n,{3k}/{2}, p, k\eta_N/2)} 
\mathbb{Q}\Big\{\fnorm{P_{\wh{\theta}}- P_\theta}^2 > \frac{1}{3} \Big\} \leq \beta,\label{eq:mazongming}
\end{equation}
the test $\psi$ defined by \eqref{eq:red-1} -- \eqref{eq:red-6}
satisfies
\begin{align*}
\mathbb{P}_{H_0^G}\psi+\mathbb{P}_{H_1^G}(1-\psi) < \beta+\frac{4n}{N}+C(n^{-1}+N^{-1}+e^{-C'k}),
\end{align*}
for sufficiently large $n$
with some constants $C,C'>0$.
\end{thm}

If the estimator $\wh\theta$ is uniformly consistent over $\mathcal{Q}(n,3k/2,p,k\eta_N/2)$, then $\beta$ is close to zero. 
Hence the conclusion of \prettyref{thm:test-error-pca} implies that for appropriate growing sequences of $n, N$ and $k$, the testing error for \eqref{eq:pc} can be made smaller than any fixed nonzero probability. 
Further invoking Hypothesis A, we obtain the following computational lower bounds for sparse PCA.
\begin{thm}
	\label{thm:clPCA-cont}
Suppose that Hypothesis A holds and that as $n\rightarrow\infty$, $2n\leq p\leq n^{a}$ for some constant $a>1$, $n(\log n)^5\leq cs^4$ for some sufficiently small $c>0$, and $\lambda=\frac{s^2}{2430n(\log(12n))^2}$.
If for some $\delta\in(0,2)$,
\begin{equation}
\label{eq:pca-cond}
\liminf_{n\rightarrow\infty}\frac{s^{2-\delta}\log p}{n\lambda^2}>0,
\end{equation}
then for any randomized polynomial-time estimator $\wh{\theta}$,
\begin{equation}
\liminf_{n\to\infty}\sup_{\mathbb{Q}\in\mathcal{Q}(n,s,p,\lambda)}
\mathbb{Q}\Big\{\fnorm{P_{\wh{\theta}}-P_\theta}^2> \frac{1}{3}\Big\}>\frac{1}{4}.
\label{eq:clower}
\end{equation}
\end{thm}

In addition to estimation, we can also consider the following sparse PCA detection problem: Let $\mathbb{Q}$ denote the joint distribution of $W_1,\dots, W_n$, and we want to test
\begin{align}
\label{eq:PCA-detection}
H_0: \mathbb{Q}\in \mathcal{Q}(n,s,p,0), \quad
\mbox{v.s.}\quad H_1: \mathbb{Q}\in \mathcal{Q}(n,s,p,\lambda).
\end{align}
Note that $\mathcal{Q}(n,s,p,0)$ contains only one distribution $\mathbb{Q}_0$ where $W_i\stackrel{iid}{\sim} N_p(0,I_p)$.
Given any testing procedure $\phi = \phi(W_1,\dots, W_{n})$, we can obtain a solution to \eqref{eq:pc} by replacing the third step in \prettyref{algo:reduction1} with the direct testing result of $\phi(W_1,\dots, W_{n})$.
Following the lines of the proof of \prettyref{thm:clPCA-cont}, we have the following theorem.
\begin{thm}
\label{thm:clPCA-test}
Under the same condition of \prettyref{thm:clPCA-cont}, for any randomized polynomial-time test $\phi$ for testing \eqref{eq:PCA-detection}, 
\begin{align}
\liminf_{n\to\infty} \Big(\mathbb{Q}_{0}\phi + \sup_{\mathbb{Q}\in \mathcal{Q}(n,s,p,\lambda)}\mathbb{Q}(1-\phi) \Big) \geq \frac{1}{4}.
\end{align}
\end{thm}

\begin{remark}
Theorems \ref{thm:clPCA-cont}--\ref{thm:clPCA-test} are the first computational lower bounds for sparse PCA that are valid in the setting of Gaussian single spiked covariance models \eqref{eq:gss-model}.
\end{remark}

\subsection{Hardness of sparse CCA}
\label{sec:pca2cca}
In the second step, we show that computational hardness of sparse PCA under Gaussian spiked covariance model implies the desired result in \prettyref{thm:clCCA-cont}.
To this end, we propose the following reduction.

\begin{algorithm}[h]
\caption{Reduction from Sparse PCA to Sparse CCA}
\label{algo:reduction2}
\SetAlgoLined
\KwIn{}
1.~Observations $W_1,\dots, W_n\in \reals^p$;\\
2.~Estimator $\wh{u}$ of the first leading canonical correlation coefficient $u$. 

\KwOut{An estimator $\wh\theta$ of the leading eigenvector of $\calL(W_1)$.}

\nl Generate i.i.d.~random vectors $Z_1,\dots, Z_{n}\sim N_p(0,I_p)$.
Set
\begin{align}
\label{eq:red-4}
X_i = \frac{1}{\sqrt{2}}(W_{i} + Z_i), \qquad Y_i = \frac{1}{\sqrt{2}}(W_{i} - Z_i),\qquad i=1,\dots,n.
\end{align}

\nl Compute $\wh{u} = \wh{u}(X_1,Y_1,\dots, X_n, Y_n)$. 
Set
\begin{align}
\label{eq:red-5}
\wh\theta = \wh\theta(W_1,\dots,W_n) = \wh{u} / \|\wh{u}\|.
\end{align}
\end{algorithm}

To see why \prettyref{algo:reduction2} is effective,
one can verify that if $W_i\stackrel{iid}{\sim} N_p(0,\tau\theta\theta'+I_p)$, 
then $(X_i', Y_i')'\stackrel{iid}{\sim} N_{p+m}(0, \Sigma)$ where
\begin{align}
	\label{eq:scca}
\Sigma_x=\Sigma_y=\frac{\tau}{2}\theta\theta'+I_p,\,\, \Sigma_{xy}=\Sigma_x(\lambda uv')\Sigma_y
\end{align}
with	
$u=v=\frac{\theta}{\sqrt{\tau/2+1}},\, \lambda=\frac{\tau/2}{\tau/2+1}$.
This is a special case of the Gaussian canonical pair model \eqref{eq:CCA}.
Thus, the leading eigenvector of $W_i$ aligns with the leading canonical coefficient vectors of $(X_i, Y_i)$. 
Exploiting this connection, we obtain the following theorem.

\begin{thm}
\label{thm:pca-cca-reduction}
Consider $p=m$, $s_u=s_v$ and $\lambda\leq 1$.
Then for any $\wh{u}$ such that
\begin{equation}
\sup_{\mathbb{P}\in \mathcal{P}\left(n, s_u, s_v, p, m, 1, \lambda/3; 3\right)} 
\Prob\Big\{L(\wh{u}, u) > \frac{1}{300} \Big\} \leq \beta, \label{eq:ce}
\end{equation}
the estimator $\wh\theta$ defined by \prettyref{algo:reduction2} satisfies
\begin{align*}
	\sup_{\mathbb{Q}\in \mathcal{Q}(n, s, p, \lambda)} 
	\mathbb{Q}\Big\{\fnorm{P_{\wh{\theta}}- P_\theta}^2 > \frac{1}{3} \Big\} \leq \beta.
\end{align*}
\end{thm}

If we start with an estimator $\wh{u}$ of the leading canonical coefficient vector, then we can construct the reduction from Planted Clique to sparse CCA directly by essentially following the steps in \prettyref{algo:reduction1} while using \prettyref{algo:reduction2} to construct $\wh\theta$ from $\wh{u}$ in the third step.
Finally, the desired \prettyref{thm:clCCA-cont} is a direct consequence of Theorems \ref{thm:clPCA-cont} and \ref{thm:pca-cca-reduction}.



\section{Proofs}
\label{sec:proof}
This section presents proofs of Theorems \ref{thm:adajoint} and \ref{thm:adasep}. The proofs of the other theoretical results are given in the supplement \cite{supp}.

\subsection{Proof of Theorem \ref{thm:adajoint}}
\label{sec:proof-adajoint}

\newcommand{\epsnu}{\epsilon_{n,u}}
\newcommand{\epsnv}{\epsilon_{n,v}}

Before presenting the proof, we state some technical lemmas. 
The proofs of all the lemmas are given in Section \ref{sec:pftech} in the supplement \cite{supp}. 
First, note that the estimator is normalized with respect to $\wh{\Sigma}^{(0)}_x$ and $\wh{\Sigma}^{(0)}_y$, while the truth $U$ and $V$ is normalized with respect to $\Sigma_x$ and $\Sigma_y$. 
To address this issue, we normalize the truth with respect to $\wh\Sigma_x^{(0)}$ and $\wh\Sigma_y^{(0)}$ to obtain
 $\wt{U}=U(U'\wh{\Sigma}_x^{(0)}U)^{-1/2}$ and $\wt{V}=V(V'\wh{\Sigma}_y^{(0)}V)^{-1/2}$. 
Also define $\wt{\Lambda}=(U'\wh{\Sigma}_x^{(0)}U)^{1/2}\Lambda (V'\wh{\Sigma}_y^{(0)}V)^{1/2}$. 
For notational convenience, define 
\begin{align}
\epsnu = \sqrt{\frac{1}{n}\Big(s_u+\log\frac{ep}{s_u}\Big)},\quad
\epsnv = \sqrt{\frac{1}{n}\Big(s_v+\log\frac{em}{s_v}\Big)}.
\end{align}
The following lemma bounds the normalization effect.

\begin{lemma} \label{lem:normalize}
Assume 
$\epsnu^2 + \epsnv^2 \leq c$ for some sufficiently small constant $c\in(0,1)$. Then there exist some constants  $C,C'>0$ only depending on $c$ such that
\begin{align*}
\opnorm{\Sigma_x^{1/2}(\wt{U}-U)} \leq C\epsnu,~
\opnorm{\Sigma_y^{1/2}(\wt{V}-V)} \leq C\epsnv, ~
\opnorm{\wt{\Lambda}-\Lambda} \leq C(\epsnu + \epsnv),
\end{align*}
with probability at least $1-\exp\left(-C'(s_u+\log(ep/s_u))\right)-\exp\left(-C'(s_v+\log(em/s_v))\right)$.
\end{lemma}

Using the definitions of $\wt{U}$ and $\wt{V}$, let us state the following lemma, which asserts that the matrix $\wt{A} = \wt{U}\wt{V}'$ is feasible to the optimization problem (\ref{eq:optpoly}).
\begin{lemma} \label{lem:feasible}
Define $\wt{A}=\wt{U}\wt{V}'$.
When $\wt{A}$ exists, we have
\begin{align*}
\nnorm{(\wh{\Sigma}_x^{(0)})^{1/2}\wt{A}(\wh{\Sigma}_y^{(0)})^{1/2}}=r\quad \mbox{and}\quad
\opnorm{(\wh{\Sigma}_x^{(0)})^{1/2}\wt{A}(\wh{\Sigma}_y^{(0)})^{1/2}}=1.
\end{align*}
\end{lemma}

As was argued in Section \ref{sec:adapt-est}, the set $\mathcal{C}_r$ is the convex hull of $\mathcal{O}_r$. The following curvature lemma shows that the relaxation $\mathcal{C}_r$ preserves the restricted strong convexity of the objective function.
\begin{lemma} \label{lem:identify}
Let $F\in O(p,r)$, $G\in O(m,r)$, $K\in\mathbb{R}^{r\times r}$ and $D=\text{diag}(d_1,...,d_r)$ with $d_1\geq...\geq d_r>0$. If $E$ satisfies $\opnorm{E}\leq 1$ and $\nnorm{E}\leq r$, then
\begin{equation}
\iprod{FKG'}{FG'-E}\geq\frac{{d_r}}{2}\fnorm{FG'-E}^2-\fnorm{K-D}\fnorm{FG'-E}. \label{eq:identify}
\end{equation}
\end{lemma}

Define
\begin{equation}
\wt{\Sigma}_{xy}=\wh{\Sigma}_x^{(0)}U\Lambda V'\wh{\Sigma}_y^{(0)}. \label{eq:wtsigmadef}
\end{equation}
Lemma \ref{lem:rho} is instrumental in determining the proper value of the tuning parameter required in the program (\ref{eq:optpoly}).
\begin{lemma} \label{lem:rho}
Assume $r\sqrt{{[\log (p+m)]}/{n}}\leq c$ for some sufficiently small constant $c\in(0,1)$. Then there exist some constants $C,C'>0$ only depending on $M$ and $c$ such that
$||\wh{\Sigma}_{xy}^{(0)}-\wt{\Sigma}_{xy}||_{\infty}\leq C\sqrt{{[\log (p+m)]}/{n}}$,
with probability at least $1-(p+m)^{-C'}$.
\end{lemma}

We also need a lemma on restricted eigenvalue.
For any p.s.d. matrix $B$, define
$$\phi_{\max}^{B}(k)=\max_{||u||_0\leq k,u\neq 0}\frac{u'Bu}{u'u},\quad \phi_{\min}^{B}(k)=\min_{||u||_0\leq k,u\neq 0}\frac{u'Bu}{u'u}.$$
The following lemma is adapted from Lemma 12 in \cite{gao14}, and its proof is omitted.

\begin{lemma} \label{lem:sparsespec}
Assume $\frac{1}{n}\big((k_u\wedge p)\log(ep/(k_u\wedge p))+(k_v\wedge m)\log(em/(k_v\wedge m))\big)\leq c$ for some sufficiently small constant $c>0$. Then there exist some constants $C,C'>0$ only depending on $M$ and $c$ such that 
for $\delta_u(k_u) = \sqrt{\frac{(k_u\wedge p)\log (ep/(k_u\wedge p))}{n}}$ and $\delta_v(k_v) = \sqrt{\frac{(k_v\wedge m)\log (em/(k_v\wedge m))}{n}}$, we have
\begin{align*}
M^{-1}-C \delta_u(k_u)\leq \phi_{\min}^{\wh{\Sigma}_x^{(j)}}(k_u) & \leq\phi_{\max}^{\wh{\Sigma}_x^{(j)}}(k_u)\leq M+C\delta_u(k_u),\\
M^{-1}-C\delta_v(k_v) \leq \phi_{\min}^{\wh{\Sigma}_y^{(j)}}(k_v)& \leq\phi_{\max}^{\wh{\Sigma}_y^{(j)}}(k_v)\leq M+C\delta_v(k_v),
\end{align*}
with probability at least $1-\exp\big(-C'(k_u\wedge p)\log(ep/(k_u\wedge p))\big)-\exp\big(-C'(k_v\wedge m)\log(em/(k_v\wedge m))\big)$, for $j=0,1,2$.
\end{lemma}

Finally, we need a result on subspace distance. 
Recall that for a matrix $F$, $P_F$ denotes the projection matrix onto its column subspace.
\begin{lemma} \label{lem:proj}
For any matrix $F\in O(d,r)$ and any matrix $G\in\mathbb{R}^{d\times r}$, we have
$$\inf_W\fnorm{F-GW}^2=\frac{1}{2}\fnorm{P_F-P_G}^2.$$
If further $G\in O(d,r)$,  then $\inf_{W\in O(r,r)}\fnorm{F-GW}^2=\frac{1}{2}\fnorm{P_F-P_G}^2.$
\end{lemma}
Proofs of Lemma \ref{lem:normalize}-\ref{lem:proj} are given in Section \ref{sec:pf-pf} of the supplement \cite{supp}.

\begin{proof}[Proof of Theorem \ref{thm:adajoint}]
In the rest of this proof, we denote $\wh{\Sigma}_x^{(0)}$, $\wh{\Sigma}_y^{(0)}$ and $\wh{\Sigma}_{xy}^{(0)}$ by $\wh{\Sigma}_x$, $\wh{\Sigma}_y$ and $\wh{\Sigma}_{xy}$ for notational convenience. 
We also let $\Delta=\wh{A}-\wt{A}$. 
The proof consists of two steps. In the first step, we are going to derive an upper bound for $\fnorm{\wh{\Sigma}_x^{1/2}\Delta\wh{\Sigma}_y^{1/2}}$. In the second step, we derive a generalized cone condition and use it to lower bound $\fnorm{\wh{\Sigma}_x^{1/2}\Delta\wh{\Sigma}_y^{1/2}}$ by a constant multiple of $\fnorm{\Delta}$ and hence the upper bound on $\fnorm{\wh{\Sigma}_x^{1/2}\Delta\wh{\Sigma}_y^{1/2}}$ leads to an upper bound on $\fnorm{\Delta}$.
\smallskip

\noindent\textbf{Step 1. } 
By Lemma \ref{lem:normalize}, $\wt{U}$ and $\wt{V}$ are well-defined with high probability. Thus, $\wt{A}$ is well-defined with high probability, and we have
\begin{equation}
\opnorm{\Sigma_x^{1/2}(\wt{A}-UV')\Sigma_y^{1/2}}\leq C
(\epsnu + \epsnv).
\label{eq:early}
\end{equation}
with probability at least $1-\exp\left(-C'(s_u+\log(ep/s_u))\right)-\exp\left(-C'(s_v+\log(em/s_v))\right)$.
According to Lemma \ref{lem:feasible}, $\wt{A}$ is feasible. Then, by the definition of $\wh{A}$, we have
$$\iprod{\wh{\Sigma}_{xy}}{\wh{A}}-\rho||\wh{A}||_1\geq \iprod{\wh{\Sigma}_{xy}}{\wt{A}}-\rho||\wt{A}||_1.$$
After rearrangement, we have
\begin{equation}
-\iprod{\wt{\Sigma}_{xy}}{\Delta}\leq \rho\big(||\wt{A}||_1-||\wt{A}+\Delta||_1\big)+\iprod{\wh{\Sigma}_{xy}-\wt{\Sigma}_{xy}}{\Delta}, \label{eq:basic}
\end{equation}
where $\wt{\Sigma}_{xy}$ is defined in (\ref{eq:wtsigmadef}). For the first term on the right hand side of (\ref{eq:basic}), we have
\begin{align*}
||\wt{A}||_1-||\wt{A}+\Delta||_1 &= ||\wt{A}_{S_uS_v}||_1-||\wt{A}_{S_uS_v}+\Delta_{S_uS_v}||_1-||\Delta_{(S_uS_v)^c}||_1 \\
&\leq ||\Delta_{S_uS_v}||_1-||\Delta_{(S_uS_v)^c}||_1.
\end{align*}
For the second term on the right hand side of (\ref{eq:basic}), we have $\iprod{\wh{\Sigma}_{xy}-\wt{\Sigma}_{xy}}{\Delta}\leq ||\wh{\Sigma}_{xy}-\wt{\Sigma}_{xy}||_{\infty}||\Delta||_1$. Thus when
\begin{equation}
\rho \geq 2||\wh{\Sigma}_{xy}-\wt{\Sigma}_{xy}||_{\infty}, \label{eq:rhosigma}
\end{equation}
we have
\begin{equation}
-\iprod{\wt{\Sigma}_{xy}}{\Delta}\leq \frac{3\rho}{2}||\Delta_{S_uS_v}||_1 - \frac{\rho}{2}||\Delta_{(S_uS_v)^c}||_1.\label{eq:basic2}
\end{equation}
Using Lemma \ref{lem:identify}, we can lower bound the left hand side of (\ref{eq:basic2}) as
\begin{eqnarray}
\nonumber-\iprod{\wt{\Sigma}_{xy}}{\Delta} &=& \iprod{\wh{\Sigma}_x^{1/2}U\Lambda V'\wh{\Sigma}_y^{1/2}}{\wh{\Sigma}_x^{1/2}(\wt{A}-\wh{A})\wh{\Sigma}_y^{1/2}} \\
\nonumber&=& \iprod{\wh{\Sigma}_x^{1/2}\wt{U}\wt{\Lambda}\wt{V}'\wh{\Sigma}_y^{1/2}}{\wh{\Sigma}_x^{1/2}(\wt{A}-\wh{A})\wh{\Sigma}_y^{1/2}} \\
\label{eq:corrected}&\geq& \frac{1}{2}\lambda_r\fnorm{\wh{\Sigma}_x^{1/2}(\wt{A}-\wh{A})\wh{\Sigma}_y^{1/2}}^2-\delta\fnorm{\wh{\Sigma}_x^{1/2}(\wt{A}-\wh{A})\wh{\Sigma}_y^{1/2}},
\end{eqnarray}
where $\delta=\fnorm{\wt{\Lambda}-\Lambda}$.
Combining (\ref{eq:basic2}) and (\ref{eq:corrected}), we have
\begin{align}
\label{eq:cc0}\lambda_r\fnorm{\wh{\Sigma}_x^{1/2}\Delta\wh{\Sigma}_y^{1/2}}^2 &\leq  3\rho||\Delta_{S_uS_v}||_1-\rho||\Delta_{(S_uS_v)^c}||_1 +2\delta\fnorm{\wh{\Sigma}_x^{1/2}\Delta\wh{\Sigma}_y^{1/2}} \\
\label{eq:cc1}&\leq  3\rho||\Delta_{S_uS_v}||_1 +2\delta\fnorm{\wh{\Sigma}_x^{1/2}\Delta\wh{\Sigma}_y^{1/2}}.
\end{align}
Solving the quadratic equation (\ref{eq:cc1}) by Lemma 2 of \cite{cai13a}, we have
\begin{equation}
\fnorm{\wh{\Sigma}_x^{1/2}\Delta\wh{\Sigma}_y^{1/2}}^2\leq {6\rho||\Delta_{S_uS_v}||_1}/{\lambda_r}+
{4\delta^2}/{\lambda_r^2}. \label{eq:cc9}
\end{equation}
Combining (\ref{eq:cc0}) and (\ref{eq:cc9}), we have
\begin{eqnarray}
\nonumber 0 &\leq& 3\rho||\Delta_{S_uS_v}||_1-\rho||\Delta_{(S_uS_v)^c}||_1 +{\delta^2}/{\lambda_r}+\lambda_r\fnorm{\wh{\Sigma}_x^{1/2}\Delta\wh{\Sigma}_y^{1/2}}^2 \\
\label{eq:cone} &\leq& 9\rho||\Delta_{S_uS_v}||_1-\rho||\Delta_{(S_uS_v)^c}||_1 + {5\delta^2}/{\lambda_r},
\end{eqnarray}
which gives rise to the generalized cone condition that we are going to use in Step 2. Finally, by the bound $||\Delta_{S_uS_v}||_1\leq \sqrt{s_us_v}\rho\fnorm{\Delta_{S_uS_v}}$ and (\ref{eq:cc9}), we have
\begin{equation}
\fnorm{\wh{\Sigma}_x^{1/2}\Delta\wh{\Sigma}_y^{1/2}}^2\leq {6\sqrt{s_us_v}\rho\fnorm{\Delta_{S_uS_v}}}/{\lambda_r}
+{4\delta^2}/{\lambda_r^2},\label{eq:basic3}
\end{equation}
which completes the first step.

\noindent\textbf{Step 2. } 
By (\ref{eq:cone}), we have obtained the following condition
\begin{equation}
||\Delta_{(S_uS_v)^c}||_1\leq 9||\Delta_{S_uS_v}||_1+{5\delta^2}/{(\rho\lambda_r)}. \label{eq:gencone}
\end{equation}
Due to the existence of the extra term $5\delta^2/(\rho\lambda_r)$ on the RHS, we call it a \emph{generalized cone condition}.
In this step, we are going to lower bound $\fnorm{\wh{\Sigma}_x^{1/2}\Delta\wh{\Sigma}_y^{1/2}}$ by $\fnorm{\Delta}$ on the generalized cone.
Motivated by the argument in \cite{Bickel09},
let the index set $J_1=\{(i_k,j_k)\}_{k=1}^t$ in $(S_u\times S_v)^c$ correspond to the entries with the largest absolute values in $\Delta$, and we define the set $\wt{J}=(S_u\times S_v)\cup J_1$.
Now we partition $\wt{J}^c$ into disjoint subsets $J_2,...,J_K$ of size $t$ (with $|J_K|\leq t$), such that $J_k$ is the set of (double) indices corresponding to the entries of $t$ largest absolute values in $\Delta$ outside $\wt{J}\cup \bigcup_{j=2}^{k-1}J_j$. By triangle inequality,
\begin{align*}
& \fnorm{\wh{\Sigma}_x^{1/2}\Delta\wh{\Sigma}_y^{1/2}}
\geq \fnorm{\wh{\Sigma}_x^{1/2}\Delta_{\wt{J}}\wh{\Sigma}_y^{1/2}}-
{\sum_{k=2}^K}\fnorm{\wh{\Sigma}_x^{1/2}\Delta_{J_k}\wh{\Sigma}_y^{1/2}} \\
&\geq \sqrt{\phi_{\min}^{\wh{\Sigma}_x}(s_u+t)\phi_{\min}^{\wh{\Sigma}_y}(s_v+t)}\fnorm{\Delta_{\wt{J}}} -  \sqrt{\phi_{\max}^{\wh{\Sigma}_x}(t)\phi_{\max}^{\wh{\Sigma}_y}(t)}
 {\sum_{k=2}^K}\fnorm{\Delta_{J_k}}.
\end{align*}
By the construction of $J_k$, we have
\begin{eqnarray}
\nonumber  \sum_{k=2}^K\fnorm{\Delta_{J_k}} &\leq&  \sqrt{t}\sum_{k=2}^K||\Delta_{J_k}||_{\infty}\leq t^{-1/2}\sum_{k=2}^K||\Delta_{J_{k-1}}||_1\leq t^{-1/2}||\Delta_{(S_uS_v)^c}||_1 \\
\label{eq:niubility} && \leq t^{-1/2}\left(9||\Delta_{S_uS_v}||_1+\frac{5\delta^2}{\rho\lambda_r}\right)\leq 9\sqrt{\frac{s_us_v}{t}}\fnorm{\Delta_{\wt{J}}}+\frac{5\delta^2}{\rho\lambda_r\sqrt{t}},
\end{eqnarray}
where we have used the generalized cone condition (\ref{eq:gencone}). Hence, we have the lower bound
$$\fnorm{\wh{\Sigma}_x^{1/2}\Delta\wh{\Sigma}_y^{1/2}} \geq \kappa_1 \fnorm{\Delta_{\wt{J}}}-{\kappa_2\delta^2}/({\rho\lambda_r\sqrt{t}}),$$
with
\begin{align}
\label{eq:kappa}\kappa_1 &= \sqrt{\phi_{\min}^{\wh{\Sigma}_x}(s_u+t)\phi_{\min}^{\wh{\Sigma}_y}(s_v+t)}-9\sqrt{\frac{s_us_v}{t}}\sqrt{\phi_{\max}^{\wh{\Sigma}_x}(t)\phi_{\max}^{\wh{\Sigma}_y}(t)}, \\
\nonumber\kappa_2 &= 5\sqrt{\phi_{\max}^{\wh{\Sigma}_x}(t)\phi_{\max}^{\wh{\Sigma}_y}(t)}.
\end{align}
Taking $t=c_1s_us_v$ for some sufficiently large constant $c_1 > 1$, with high probability, $\kappa_1$ can be lower bounded by a positive constant $\kappa_0$ only depending on $M$.
To see this, note that by Lemma \ref{lem:sparsespec}, (\ref{eq:kappa}) can be lower bounded by the difference of
$\sqrt{M^{-1}-C\delta_u(2c_1 s_u s_v)
}\sqrt{M^{-1}-C\delta_v(2c_1 s_u s_v)
}$
and
$9c_1^{-1/2}\sqrt{M+C\delta_u(c_1 s_u s_v)
}\sqrt{M+C \delta_v(c_1 s_u s_v)
}$,
where $\delta_u$ and $\delta_v$ are defined as in \prettyref{lem:sparsespec}.
It is sufficient to show that $\delta_u(2c_1 s_u s_v)$, $\delta_v(2c_1 s_u s_v)$, $\delta_u(c_1 s_u s_v)$ and $\delta_v(c_1 s_u s_v)$
are sufficiently small to get a positive absolute constant $\kappa_0$. For the first term, when $2c_1 s_us_v\leq p$, it is bounded by $\frac{2c_1 s_us_v\log (ep)}{n}$ and is sufficiently small under the assumption (\ref{eq:ass}). When $2c_1 s_us_v>p$, it is bounded by $ \frac{2c_1 s_us_v}{n}$ and is also sufficiently small under (\ref{eq:ass}). The same argument also holds for the other terms. Similarly, $\kappa_2$ can be upper bounded by some constant.

Together with (\ref{eq:basic3}), this brings the inequality
$$\fnorm{\Delta_{\wt{J}}}^2
\leq
C_1(\sqrt{s_us_v}\rho/{\lambda_r})\fnorm{\Delta_{\wt{J}}}+
C_2\big({\delta^2}/{\lambda_r^2}+({\delta^2}/({\rho\lambda_r\sqrt{t}}))^2 \big).$$
Solving this quadratic equation, we have
\begin{equation}
\fnorm{\Delta_{\wt{J}}}^2\leq C\Big(\frac{s_us_v\rho^2}{\lambda_r^2}+\frac{\delta^2}{\lambda_r^2}+\Big(\frac{\delta^2}{\rho\lambda_r\sqrt{t}}\Big)^2\Big). \label{eq:part1}
\end{equation}
By (\ref{eq:niubility}), we have
\begin{equation}
\fnorm{\Delta_{\wt{J}^c}}\leq \sum_{k=2}^K\fnorm{\Delta_{J_k}}\leq 9\sqrt{\frac{s_us_v}{t}}\fnorm{\Delta_{\wt{J}}}+\frac{5\delta^2}{\rho\lambda_r\sqrt{t}}. \label{eq:part2}
\end{equation}
Summing (\ref{eq:part1}) and (\ref{eq:part2}), we obtain a bound for $\fnorm{\Delta}$.
According to Lemma \ref{lem:rho}, we may choose $\rho= \gamma\sqrt{[{\log (p+m)}]/{n}}$ for some large $\gamma$, so that (\ref{eq:rhosigma}) holds with high probability. By Lemma \ref{lem:normalize}, $\delta\leq C\sqrt{{r(s_u+s_v+\log(p+m))}/{n}}\leq C'\rho\sqrt{t}$ with high probability.
Hence,
\begin{equation}
\fnorm{\Delta}\leq 
C {\sqrt{s_u s_v} \rho}/{\lambda_r},
\label{eq:bounddelta}
\end{equation}
with high probability. This completes the second step. 
Finally, the triangle inequality leads to $\fnorm{\wh{A}-UV'}\leq \fnorm{\Delta}+\fnorm{\wt{A}-UV'}$. By (\ref{eq:early}) and (\ref{eq:bounddelta}), the proof is complete.
\end{proof}

\subsection{Proof of Theorem \ref{thm:adasep}}

Define
$
U^*=U\Lambda V'\Sigma_y \wh{V}^{(0)}$ and $\Delta=\wh{U}^{(1)}-U^*$.

\begin{lemma}\label{lem:ep}
Assume $\frac{r+\log p}{n}\leq c$ for some sufficiently small constant $c\in(0,1)$. Then there exist some constants $C,C'>0$ only depending on $M$ and $c$ such that
$\max_{1\leq j\leq p}||[\wh{\Sigma}^{(1)}_{xy}\wh{V}^{(0)}-\wh{\Sigma}^{(1)}_xU^*]_{j\cdot}||\leq C\sqrt{{(r+\log p)}/{n}}$,
with probability at least $1-\exp\big(-C'(r+\log p)\big)$.
\end{lemma}
The proof of Lemma \ref{lem:ep} is given in Section \ref{sec:pf-pf} of the supplement \cite{supp}.

\begin{proof}[Proof of Theorem \ref{thm:adasep}]
In the rest of this proof, we denote $\wh{\Sigma}_x^{(1)}$, $\wh{\Sigma}_y^{(1)}$ and $\wh{\Sigma}_{xy}^{(1)}$ by $\wh{\Sigma}_x$, $\wh{\Sigma}_y$ and $\wh{\Sigma}_{xy}$ for simplicity of notation. Note that they depends on $\mathcal{D}_1$, while the estimator $\wh{V}^{(0)}$ depends on $\mathcal{D}_0$. 
Hence, $\wh{V}^{(0)}$ is independent of the sample covariance matrices occurring in this proof.
The proof consists of three steps. In the first step, we derive a bound for $\Tr(\Delta'\wh{\Sigma}_x\Delta)$. In the second step, we derive a cone condition and use it to obtain a bound for $\fnorm{\Delta}$ by arguing that $\Tr(\Delta'\wh{\Sigma}_x\Delta)$ upper bounds $\fnorm{\Delta}$. In the last step, we derive the desired bound for $L(\wh{U},U)$.
\smallskip

\noindent\textbf{Step 1. } By definition of $\wh{U}^{(1)}$, we have
$\Tr((\wh{U}^{(1)})'\wh{\Sigma}_x\wh{U}^{(1)})-2\Tr((\wh{U}^{(1)})'\wh{\Sigma}_{xy}\wh{V}^{(0)})+\rho_u  {\sum_{j=1}^p||\wh{U}^{(1)}_{j\cdot}|| } \leq \Tr((U^*)'\wh{\Sigma}_xU^*)-2\Tr((U^*)'\wh{\Sigma}_{xy}\wh{V}^{(0)})+\rho_u {\sum_{j=1}^p||U^*_{j\cdot}||}$.
After rearrangement, we have
\begin{equation}
\Tr(\Delta'\wh{\Sigma}_x\Delta)
\leq \rho_u\sum_{j=1}^p\Big[ ||U_{j\cdot}^*||-||U_{j\cdot}^*+\Delta_{j\cdot}||\Big]
+2\Tr\Big[\Delta'(\wh{\Sigma}_{xy}\wh{V}^{(0)}-\wh{\Sigma}_xU^*)\Big]. \label{eq:Basic}
\end{equation}
For the first term on the right hand side of (\ref{eq:Basic}), we have
\begin{align*}
 \sum_{j=1}^p\big(||U_{j\cdot}^*||-||U_{j\cdot}^*+\Delta_{j\cdot}||\big) 
&=  \sum_{j\in S_u}||U_{j\cdot}^*||-\sum_{j\in S_u}||U_{j\cdot}^*+\Delta_{j\cdot}||-\sum_{j\in S_u^c}||\Delta_{j\cdot}|| \\
&\leq  \sum_{j\in S_u}||\Delta_{j\cdot}||-\sum_{j\in S_u^c}||\Delta_{j\cdot}||.
\end{align*}
For the second term on the right hand side of (\ref{eq:Basic}), we have
\begin{eqnarray*}
\Tr\Big(\Delta'(\wh{\Sigma}_{xy}\wh{V}^{(0)}-\wh{\Sigma}_xU^*)\Big) 
\leq \Big(\sum_{j=1}^p||\Delta_{j\cdot}||\Big)\max_{1\leq j\leq p}||[\wh{\Sigma}_{xy}\wh{V}^{(0)}-\wh{\Sigma}_xU^*]_{j\cdot}||,
\end{eqnarray*}
where $[\cdot]_{j\cdot}$ means the $j$-th row of the corresponding matrix.
When
\begin{equation}
\rho_u\geq 4\max_{1\leq j\leq p}||[\wh{\Sigma}_{xy}\wh{V}^{(0)}-\wh{\Sigma}_xU^*]_{j\cdot}||, \label{eq:rhou}
\end{equation}
we have
\begin{equation}
\Tr(\Delta'\wh{\Sigma}_x\Delta)\leq \frac{3\rho_u}{2}
 {\sum_{j\in S_u}}||\Delta_{j\cdot}||-
\displaystyle{\frac{\rho_u}{2}}
 {\sum_{j\in S_u^c}}||\Delta_{j\cdot}||. \label{eq:Basic2}
\end{equation}
Since $\sum_{j\in S_u}||\Delta_{j\cdot}||\leq \sqrt{s_u}\sqrt{\sum_{j\in S_u}||\Delta_{j\cdot}||^2}$, (\ref{eq:Basic2}) can be upper bounded by
\begin{equation}
\Tr(\Delta'\wh{\Sigma}_x\Delta)\leq \frac{3\sqrt{s_u}\rho_u}{2}\sqrt{ {\sum_{j\in S_u}}||\Delta_{j\cdot}||^2}. \label{eq:Basic3}
\end{equation}
This completes the first step.

\noindent\textbf{Step 2. } The inequality (\ref{eq:Basic2}) implies the cone condition
\begin{equation}
 {\sum_{j\in S_u^c}||\Delta_{j\cdot}||}\leq  
3
 {\sum_{j\in S_u}||\Delta_{j\cdot}||}. \label{eq:Cone}
\end{equation}
Let the index set $J_1=\{j_1,...,j_t\}$ in $S_u^c$  correspond to the rows with the largest $\ell_2$ norm in $\Delta$, and we define the extended support $\wt{S}_u=S_u\cup J_1$. Now we partition $\wt{S}_u^c$ into disjoint subsets $J_2,...,J_K$ of size $t$ (with $|J_K|\leq t$), such that $J_k$ is the set of indices corresponding to the $t$ rows with largest $\ell_2$ norm in $\Delta$ outside $\wt{S}_u\cup \bigcup_{j=2}^{k-1}J_j$. Note that $\Tr(\Delta'\wh{\Sigma}_x\Delta)=\fnorm{n^{-1/2}X\Delta}^2$, where $X=[X_1,...,X_n]'\in\mathbb{R}^{n\times p}$ denotes the data matrix. By triangle inequality, we have
\begin{eqnarray*}
\fnorm{n^{-1/2}X\Delta} &\geq& \fnorm{n^{-1/2}X\Delta_{\wt{S}_u*}}-\sum_{k\geq 2}\fnorm{n^{-1/2}X\Delta_{J_k*}} \\
&\geq& \sqrt{\phi_{\min}^{\wh{\Sigma}_x}(s_u+t)}\fnorm{\Delta_{\wt{S}_u*}} - \sqrt{\phi_{\max}^{\wh{\Sigma}_x}(t)}\sum_{k\geq 2}\fnorm{\Delta_{J_k*}},
\end{eqnarray*}
where for a subset $B\subset [p]$, $\Delta_{B*}=(\Delta_{ij}\indc{i\in B,j\in[r]})$, and
\begin{eqnarray}
\label{eq:start}\sum_{k\geq 2}\fnorm{\Delta_{J_k*}} &\leq& \sqrt{t}\sum_{k\geq 2}\max_{j\in J_k}||\Delta_{j\cdot}|| \leq \sqrt{t}\sum_{k\geq 2}\frac{1}{t}\sum_{j\in J_{k-1}}||\Delta_{j\cdot}|| \\
\nonumber &\leq& t^{-1/2}\sum_{j\in S_u^c}||\Delta_{j\cdot}|| \leq 3t^{-1/2}\sum_{j\in S_u}||\Delta_{j\cdot}|| \\
\label{eq:end}&\leq& 3\sqrt{\frac{s_u}{t}}\sqrt{\sum_{j\in S_u}||\Delta_{j\cdot}||^2}\leq 3\sqrt{\frac{s_u}{t}}\fnorm{\Delta_{\wt{S}_u*}}.
\end{eqnarray}
In the above derivation, we have used the construction of $J_k$ and the cone condition (\ref{eq:Cone}). Hence,
$\fnorm{n^{-1/2}X\Delta}\geq \kappa \fnorm{\Delta_{\wt{S}_u*}}$
with $\kappa=\sqrt{\phi_{\min}^{\wh{\Sigma}_x}(s_u+t)}-3\sqrt{\frac{s_u}{t}}\sqrt{\phi_{\max}^{\wh{\Sigma}_x}(t)}$. In view of Lemma \ref{lem:sparsespec}, taking $t=c_1s_u$ for some sufficiently large constant $c_1$, with high probability, $\kappa$ can be lower bounded by a positive constant $\kappa_0$ only depending on $M$. Combining with (\ref{eq:Basic3}), we have
\begin{equation}
\fnorm{\Delta_{\wt{S}_u*}}\leq {C\sqrt{s_u}\rho_u}/{(2\kappa_0^2)}. \label{eq:Sum1}
\end{equation}
By (\ref{eq:start})-(\ref{eq:end}), we have
\begin{equation}
\fnorm{\Delta_{(\wt{S}_u)^c*}}\leq 
 {\sum_{k\geq 2}}
\fnorm{\Delta_{J_k*}}\leq 3\sqrt{{s_u}/{t}}\fnorm{\Delta_{\wt{S}_u*}}\leq 3c_1^{-1/2}\fnorm{\Delta_{\wt{S}_u*}}. \label{eq:Sum2}
\end{equation}
Summing (\ref{eq:Sum1}) and (\ref{eq:Sum2}), we have $\fnorm{\Delta}\leq C\sqrt{s_u}\rho$. 
By Lemma \ref{lem:ep}, we may choose $\rho_u\geq \gamma_u\sqrt{\frac{r+\log p}{n}}$ for some large $\gamma_u$ so that (\ref{eq:rhou}) holds with high probability. Hence,
$\fnorm{\Delta} \leq C\sqrt{{s_u(r+\log p)}/{n}}$
with high probability. This completes the second step.

\noindent\textbf{Step 3. } Using the same argument in Step 2 of the proof of Theorem \ref{thm:uppersep} (see supplementary material \cite{supp}), we obtain the desired bound for $L(\wh{U},U)$. The proof is complete.
\end{proof}

\bibliographystyle{plainnat}
\bibliography{CCA}


\newpage
\begin{center}
{\Large Supplement to ``Sparse CCA: Adaptive Estimation and Computational Barriers''}
	
%

\end{center}


\section{Proofs of Results in \prettyref{sec:cb}} 
\label{sec:pfcb}

In this section, we present the proofs of Theorems \ref{thm:clCCA-cont}--\ref{thm:pca-cca-reduction}. 
Here we do not consider the issue of discretization.
The main purpose is to help the readers get the intuition behind the problem without worrying about rigor at the theoretical computer science level. 
A rigorous treatment of the computational lower bounds is deferred to Section \ref{sec:discrete}
where the asymptotic equivalent discretization and the statement of rigorous results for the discretized models will be presented.

\subsection{Proof of \prettyref{thm:test-error-pca}}

The reason for introducing the two distributions \eqref{eq:f0}--\eqref{eq:f1} is to match specific mixtures of them to $\phi_0$ and $\bar\phi_{\mu}$ respectively as summarized in the following lemma.
For two probability distributions $\mathbb{P}$ and $\mathbb{Q}$, the total variation distance is defined as $\TV(\mathbb{P},\mathbb{Q})=\sup_{B}|\mathbb{P}(B)-\mathbb{Q}(B)|$.
We also write $\TV(p,q)$ if $p$ and $q$ are the densities of $\bbP$ and $\bbQ$, respectively.

\begin{lemma}\label{lem:fixedTV}
There exists an absolute constant $C>0$, such that for all integers $N\geq 12$, $k\leq N/12$ and all $|\mu|\leq 3\sqrt{\eta_N\log N}$,
\begin{align*}
\TV(h_{\mu,0},\phi_0)\leq CN^{-3}\quad \mbox{and} \quad
\TV(h_{\mu,1},\bar\phi_\mu)\leq CN^{-3},
\end{align*}
where $h_{\mu,0} = \frac{1}{2}(f_{\mu,0}+f_{\mu,1})$ and $h_{\mu,1} = \delta_N f_{\mu,1}+\left(1-\delta_N\right)\frac{1}{2}(f_{\mu,0}+f_{\mu,1})$.
\end{lemma}

The proof of the above lemma will be given in Section \ref{sec:mamamama}. 
To facilitate the proof of \prettyref{thm:test-error-pca}, we need to state and prove another two lemmas which characterize the distributions of the $W_i$'s under $H_0^G$ and $H_1^G$ respectively.
Let $\calL(\{W_i\}_{i=1}^{2n})$ denote the joint distribution of $\{W_i\}_{i=1}^{2n}$. 
In addition, denote $N_p(0,\tau\theta\theta'+I_p)$ by $\mathbb{Q}_{\theta,\tau}$. 
When $\tau=0$, $N_p(0,I_p)$ is denoted by $\mathbb{Q}_0$. 
The first lemma shows that under $H_0^G$, the joint distribution of $\{W_i\}_{i=1}^{2n}$ is close in total variation to that of a random sample of size $2n$ from $\bbQ_0$. 
\begin{lemma}
\label{lem:nullCCA}
Suppose $A\sim \mathcal{G}(N,1/2)$. There exists an absolute constant $C>0$ such that 
$$ \TV(\calL(\{W_i\}_{i=1}^{2n}), \mathbb{Q}_0^{2n}) \leq CN^{-1}.$$
\end{lemma}
\begin{proof}
Recall $\eta_N$ defined in \eqref{eq:delta-N} and $h_{\mu,0}$ 
in Lemma \ref{lem:fixedTV}.
Let $\nu$ be $N\left(0, \eta_N\right)$, 
and $\bar{\nu}$ be the distribution obtained by restricting $\nu$ on the set $[-3\sqrt{\eta_N\log N},3\sqrt{\eta_N\log N} ]$.
Then the $\mu_i$'s in \eqref{eq:red-1} are i.i.d.~r.v.'s following the distribution $\bar\nu$.
	
For each $i\in[2n]$ and each $j\in[2n]$, define i.i.d.~random variables $\wl{W}_{ij}\sim N(0,1)$. For each $i\in[2n]$ and $2n<j\leq p$, define $\wl{W}_{ij}=W_{ij}$. Let $\wl{W}_i=(\wl{W}_{i1},...,\wl{W}_{ip})'$. 
It is straightforward to verify that $\calL(\{\wl{W}_i\}_{i=1}^{2n})=\mathbb{Q}_0^{2n}$. 
By the data-processing inequality, we have
$$\TV(\calL(\{W_i\}_{i=1}^{2n}), \mathbb{Q}_0^{2n}) \leq \TV(\calL(\{W_i\}_{i=1}^{n}),\calL(\{\wl{W}_i\}_{i=1}^{n})).$$
Hence, it is sufficient to bound $\TV(\calL(\{W_i\}_{i=1}^{n}),\calL(\{\wl{W}_i\}_{i=1}^{n}))$. Conditioning on $\mu_i$, $W_{ij}$ follows $h_{\mu_i,0}$ when $A\sim \mathcal{G}(N,k)$. Therefore,
\begin{align*}
\TV(W_{ij},\wl{W}_{ij}) & =
\TV(\int h_{\mu_i,0} \diff\bar{\nu}(\mu_i),\phi_0)\\
& \leq \sup_{|\mu_i|\leq 3\sqrt{\eta_N\log N}}\TV(h_{\mu_i,0},\phi_0)\leq CN^{-3}.
\end{align*}
Here the last inequality is due to Lemma \ref{lem:fixedTV}. 
Applying Lemma 7 of \cite{ma2013computational}, we obtain
$\TV(\calL(\{W_i\}_{i=1}^{n}),\calL(\{\wl{W}_i\}_{i=1}^{n}))\leq \sum_{i=1}^n\sum_{j=1}^n\TV(W_{ij},\wl{W}_{ij})\leq CN^{-1}$.
This completes the proof.
\end{proof}

The next lemma shows that the joint distribution $\{W_i\}_{i=1}^{2n}$ is close in total variation to a mixture of the joint distribution of a random sample of size $2n$ from $\bbQ_{\theta,\tau}$. 
Here, the mixture is defined by a prior distribution $\pi$ on the $(\theta,\tau)$ pair, which is supported on a region where $\theta$ is sparse and $\tau$ is bounded away from zero.
For notational convenience, for any distribution $\bbP_\beta$ indexed by parameter $\beta\in B$ and any probability measure $\pi$ on $B$, we let 
$\int \bbP_\beta \diff \nu(\beta)$ denote the probability measure $\bbP$ defined by $\bbP(E) = \int \bbP_\beta(E) \diff \nu(\beta)$  for any event $E$.
When $\beta\sim\nu$ is a random variable and $\bbP_\beta = \calL(W|\beta)$ is the conditional distribution of $W|\beta$, we also write $\int \calL(W|\beta)\diff\nu(\beta)$ to represent the marginal distribution of $W$ after integrating out $\beta$.

\begin{lemma}\label{lem:altCCA}
Suppose $A\sim \mathcal{G}(N,1/2,k)$. 
There exists a distribution $\pi$ supported on the set 
\begin{align}
\left\{(\theta,\tau): \theta\in S^{p-1},\, |\supp(\theta)|\leq {3k}/{2},\,\tau\in [{k\eta_N}/{2}, 3{k\eta_N}/{2}] \right\}, \label{eq:noname}
\end{align}
such that for some absolute constants $C_1, C_2 > 0$,
$$ \TV(\calL(\{W_i\}_{i=1}^{2n}),\int \mathbb{Q}_{\theta,\tau}^{2n} \mathrm{d}\pi(\theta,\tau))
 \leq C_1\left(e^{-C_2k}+ \frac{1}{N}\right)+\frac{4n}{N}.$$
\end{lemma}
\begin{proof}
Recall $\eta_N$ defined in \eqref{eq:delta-N} and $h_{\mu,0}$ 
and $h_{\mu,1}$ defined 
in Lemma \ref{lem:fixedTV}.
As in the proof of \prettyref{lem:nullCCA},
let $\nu$ be $N\left(0, \eta_N\right)$, 
and $\bar{\nu}$ the distribution obtained by restricting $\nu$ on the set $[-3\sqrt{\eta_N\log N},3\sqrt{\eta_N\log N} ]$.
Then the $\mu_i$'s in \eqref{eq:red-1} are i.i.d.~r.v.'s following $\bar\nu$.
Simple calculus shows that
$\int\phi_0(x) \diff \nu(\mu) =\phi_0(x)$ is the density function of $N(0,1)$, and $\int\bar{\phi}_{\mu}(x)\diff\nu(\mu)$ gives the density function of $N\left(0,1+\eta_N\right)$.

We first focus on the case $p=2n$. 
The case of $p\geq 2n$ will be treated at the end of the proof.
Recall that $({\epsilon}_1,...,{\epsilon}_{2n})$ are the indicators of the rows of $A_0$ whether the corresponding vertices belong to the planted clique, and $({\gamma}_1,...,{\gamma}_p)$ are the corresponding indicators of the columns of $A_0$. 
Let $(\wt{\epsilon}_1,...,\wt{\epsilon}_{2n})$ and $(\wt{\gamma}_1,...,\wt{\gamma}_p)$ be i.i.d. Bernoulli random variables with mean $\delta_N=k/N$. 
Define a matrix $\wt{A}_0$, where an entry $(\wt{A}_0)_{ij}=1$ if $\wt{\epsilon}_i=\wt{\gamma}_j=1$ and is an independent instantiation of the Bernoulli$(1/2)$ distribution otherwise. 
Then, we define $\wt{W}$ with entries 
\[
\wt{W}_{ij}=(B_0)_{ij}(1-(\wt{A}_0)_{ij})+(B_1)_{ij}(\wt{A}_0)_{ij}.
\] 
Then, by Theorem 4 of \cite{diaconis1980finite} and the data-processing inequality, we have
\begin{equation}
\TV(\mathcal{L}(\wt{W}),\mathcal{L}(W))\leq \TV(\mathcal{L}(\wt{\epsilon},\wt{\gamma}),\mathcal{L}({\epsilon},{\gamma}))\leq \frac{4n}{N}.\label{eq:DF}
\end{equation}
Recall $h_{\mu,0}$ and $h_{\mu,1}$ defined in \prettyref{lem:fixedTV}.
By the definition of $\wt{W}$, conditioning on $\mu_i$ and $\wt{\gamma}_j=0$, $\wt{W}_{ij}\sim h_{\mu_i,0}$, while conditioning on $\mu_i$ and $\wt{\gamma}_j=1$, $\wt{W}_{ij}\sim h_{\mu_i,1}$. 

Further define $\wl{W}_{ij}$ by setting
$$\wl{W}_{ij}|(\wt{\gamma}_j=0,\mu_i)\sim \phi_0,\quad \wl{W}_{ij}|(\wt{\gamma}_j=1,\mu_i)\sim\bar{\phi}_{\mu_i},$$
where $\bar\phi_{\mu_i}$ is defined according to \eqref{eq:bar-phi-mu}. 
By Lemma \ref{lem:fixedTV} and Lemma 7 of \cite{ma2013computational}, uniformly over $\max_i|\mu_i|\leq 3\sqrt{\eta_N\log N}$, we have
$$\TV\Big(\mathcal{L}(\wt{W}|\wt{\gamma},\mu),\mathcal{L}(\wl{W}|\wt{\gamma},\mu)\Big)\leq\sum_{i=1}^{2n}\sum_{j=1}^p\TV\Big(\mathcal{L}(\wt{W}_{ij}|\wt{\gamma}_j,\mu_i),\mathcal{L}(\wl{W}_{ij}|\wt{\gamma}_j,\mu_i)\Big)\leq CN^{-1}$$
for some constant $C>0$. 

Next, we integrate the above bound over $\mu$. To this end, first note that
\begin{align*}
\TV(\nu,\bar\nu) = \int_{|\mu| > 3\sqrt{\eta_N\log N}} \diff\nu(\mu)  
= \int_{|x|> 3\sqrt{\log N}} \phi_0(x)\diff x \leq CN^{-4}.
\end{align*}
With slight abuse of notation, let $\int\mathcal{L}(\wt{W}|\wt{\gamma},\mu)\diff\bar{\nu}(\mu)$ (resp.~ $\int\mathcal{L}(\wt{W}|\wt{\gamma},\mu)\diff{\nu}(\mu)$) denote the conditional distribution of $\wt{W}|\wt{\gamma}$ if the coordinates of $\mu = (\mu_1,\dots, \mu_{2n})$ were i.i.d.~ following $\bar\nu$ (resp.~$\nu$), and
let $\int\mathcal{L}(\wl{W}|\wt{\gamma},\mu)\diff\bar{\nu}(\mu)$ and $\int\mathcal{L}(\wl{W}|\wt{\gamma},\mu)\diff{\nu}(\mu)$ be analogously defined.
Then, conditioning on $\wt\gamma$, we obtain
\begin{eqnarray*}
&&\TV(\int \mathcal{L}(\wt{W}|\wt{\gamma},\mu)\diff\bar{\nu}(\mu),\int\mathcal{L}(\wl{W}|\wt{\gamma},\mu) \diff\nu(\mu))  \\
&\leq& \TV(\int \mathcal{L}(\wt{W}|\wt{\gamma},\mu) \diff\bar{\nu}(\mu),\int\mathcal{L}(\wl{W}|\wt{\gamma},\mu)\diff\bar{\nu}(\mu))\\
&& ~~~ + \TV(\int\mathcal{L}(\wl{W}|\wt{\gamma},\mu)\diff\bar{\nu}(\mu),\int\mathcal{L}(\wl{W}|\wt{\gamma},\mu) \diff\nu(\mu)) \\
&\leq& \sup_{\max_i|\mu_i|\leq 3\sqrt{\eta_N\log N}}\TV(\mathcal{L}(\wt{W}|\wt{\gamma},\mu),\mathcal{L}(\wl{W}|\wt{\gamma},\mu)) + Cn\TV(\bar{\nu},\nu) 
\leq CN^{-1}.
\end{eqnarray*}
Here, the first inequality comes from the triangle inequality, the second from the definition of total variation distance. For each given $\wt{\gamma}=(\wt{\gamma}_1,...,\wt{\gamma}_n)$, define $s=\sum_{j=1}^n\wt{\gamma}_j = \sum_{j=1}^n \wt\gamma_j^2 = \|\wt\gamma_j\|^2$, $\theta=s^{-1/2}\wt{\gamma}$ and $\tau = s\eta_N$. 
Note that both $\theta$ and $\tau$ are functions of $\wt\gamma$.
Then observe that 
$\int\mathcal{L}(\wl{W}|\wt{\gamma},\mu) \diff\nu(\mu)=\mathbb{Q}^{2n}_{\theta, \tau}$,
which implies for $\mathcal{L}(\wt{W}|\wt{\gamma}) = \int \mathcal{L}(\wt{W}|\wt{\gamma},\mu)\diff\bar\nu(\mu)$,
$$\TV\left(\mathcal{L}(\wt{W}|\wt{\gamma}),\mathbb{Q}^{2n}_{\theta,\tau}\right)\leq CN^{-1}.$$
Define the event 
$
Q=\{\wt{\gamma}: |s-k|\leq k/2\}
$. 
Then, by Bernstein's inequality, $\mathbb{P}(Q^c)\leq e^{-Ck}$. Let $\wt\pi$ be the joint distribution of $(\theta,\tau)$, and ${\pi}$ be the distribution obtained from renormalizing the restriction of $\wt\pi$ on $\{(\theta(\wt\gamma),\tau(\wt\gamma)): \wt\gamma\in Q\}$ which is exactly the set in \eqref{eq:noname}. 
Then we have $\TV(\pi,\wt{\pi})\leq C\mathbb{P}(Q^c)\leq Ce^{-Ck}$.
In addition, we note that $\calL(\wt{W}|\wt\gamma) = \calL(\wt{W}|\theta,\tau)$ since there exists one-to-one identification between the pair $(\theta,\tau)$ and $\wt\gamma$. 
Therefore, we have
\begin{eqnarray*}
\TV(\mathcal{L}(\wt{W}),\int \mathbb{Q}^{2n}_{\theta,\tau}\diff{\pi}(\theta,\tau)) &\leq& \TV(\mathcal{L}(\wt{W}),\int \mathcal{L}(\wt{W}|\theta,\tau)\diff \pi(\theta,\tau)) \\
&& + \TV(\int\mathcal{L}(\wt{W}|\theta,\tau)\diff \pi(\theta,\tau),\int \mathbb{Q}^{2n}_{\theta,\tau}\diff \pi(\theta,\tau)) \\
& \leq & \TV(\wt\pi, \pi) + \sup_{\theta,\tau} \TV(\calL(\wt{W}|\theta,\tau), \bbQ_{\theta,\tau}^{2n}) \\
&\leq& C\left(e^{-Ck}+N^{-1}\right).
\end{eqnarray*}
Here, the second inequality holds since $\calL(\wt{W}) = \int \calL(\wt{W}|\theta,\tau)\diff \wt\pi(\theta,\tau)$.
Hence, by (\ref{eq:DF}),
$\TV(\mathcal{L}({W}),\int \mathbb{Q}^{2n}_{\theta,\tau}\diff \pi(\theta,\tau))\leq C\left(e^{-Ck}+N^{-1}\right)+\frac{4n}{N}$.
Note that on the support of ${\pi}$, the parameter $(\theta,\tau)$ belongs to the set (\ref{eq:noname}).
An application of data-processing inequality leads to the conclusion. When $p\geq 2n$, we may first analyze the distribution of the first $2n$ coordinates using the above arguments. The remaining $2n-p$ coordinates are exact, and the total variation bound is zero.
\end{proof}

\begin{proof}[Proof of \prettyref{thm:test-error-pca}]
Abbreviate $\frac{1}{n}\sum_{i=n+1}^{2n}W_iW_i'$ by $\wh\Sigma$. 
We can rewrite the testing function $\psi$ as 
\begin{align*}
\psi(W) = \psi(A, \mu, B_0, B_1) =\mathbf{1}\Big\{\wh{\theta}'\wh{\Sigma}\wh{\theta}\geq 1+{k\eta_N}/{4}\Big\}.	
\end{align*}
Here, $\mu = (\mu_1,\dots,\mu_{2n})$ collects the random variables in \eqref{eq:red-1}.
Thus, it is clear that $\psi$ is a randomized test for the Planted Clique detection problem \eqref{eq:pc}.
\newcommand{\Pjoint}{\mathbb{P}^{\mathrm{joint}}}
Note that for any $(\theta,\tau)$ in the support of $\pi$, we have
\begin{equation}
	\label{eq:prior-inclusion}
\mathbb{Q}_{\theta,\tau}^n  \in\mathcal{Q}\Big(n,\frac{3k}{2},p,\frac{k\eta_N}{2}\Big).	
\end{equation}

We now bound the testing errors. 
For Type-I error, Lemma \ref{lem:nullCCA} implies
$$\mathbb{P}_{H_0^G}\psi\leq \mathbb{Q}_0^n\psi+CN^{-1}.$$
Note that under $\mathbb{Q}_0^n$, $\wh{\theta}$ and $\wh{\Sigma}$ are independent. Conditioning on $\wh{\theta}$ and using Bernstein's inequality, we have
$$\wh{\theta}'\wh{\Sigma}\wh{\theta}=1+\frac{1}{n}\sum_{i=n+1}^{2n}\left(|\wh{\theta}'W_i|^2-||\wh{\theta}||^2\right)>1+\frac{k\eta_N}{4},$$
with probability at most $\exp\left(-\frac{Cnk^4}{N^2(\log N)^4}\right)$. Integrating over $\wh{\theta}$, we have
\begin{align}
\label{eq:type1-bd}
\mathbb{P}_{H_0^G}\psi\leq \exp\Big(-\frac{Cnk^4}{N^2(\log N)^4}\Big)+CN^{-1}\leq C(n^{-1}+N^{-1}),
\end{align}
where the last inequality holds under the assumptions $\frac{N(\log N)^5}{k^4}\leq c$ and $cN \leq n\leq N/12$ for some sufficiently small constant $c > 0$.

Turn to the Type-II error. Lemma \ref{lem:altCCA} implies
\begin{equation}
\mathbb{P}_{H_1^G}(1-\psi)\leq \mathbb{Q}_{{\pi}}(1-\psi)+C\Big(e^{-Ck}+N^{-1}\Big)+\frac{4n}{N},\label{eq:random}
\end{equation}
where we have used the notation $\mathbb{Q}_{{\pi}}=\int\mathbb{Q}_{\theta,\tau}^n\diff{\pi}$.  For each $\mathbb{Q}_{\theta,\tau}^n$ in the support of ${\pi}$, $W_{n+i}$ has representation
$W_{n+i}=\sqrt{\tau}g_i\theta+ \epsilon_i$,
where the $g_i$'s and the $\epsilon_i$'s are independently distributed according to $N(0,1)$ and $N_p(0,I_p)$, and are independent across $i=1,...,n$, and $\tau\geq {k\eta_N}/{2}$. Therefore,
$$
\wh{\theta}'\wh{\Sigma}\wh{\theta}=\tau|\wh{\theta}'\theta|^2\Big(\frac{1}{n}\sum_{i=1}^ng_i^2\Big)+\frac{1}{n}\sum_{i=1}^n|\wh{\theta}'\epsilon_i|^2+\frac{2\sqrt{\tau}}{n}\wh{\theta}'\theta\sum_{i=1}^ng_i\epsilon_i'\wh{\theta}.
$$
After rearrangement, we have
\begin{eqnarray*}
\Big|\wh{\theta}'\wh{\Sigma}\wh{\theta}-(1+\tau)\Big| &\leq& \Big|\frac{1}{n}\sum_{i=1}^n(g_i^2-1)\Big| + \tau\min\{|(\wh{\theta}-\theta)'\theta|^2,|(\wh{\theta}+\theta)'\theta|^2\} \\
&& + \Big|\frac{1}{n}\sum_{i=1}^n(|\wh{\theta}'\epsilon_i|^2-1)\Big| + \Big|\frac{2}{n}\sum_{i=1}^n g_i(\epsilon_i'\wh{\theta})\Big|,
\end{eqnarray*}
where $\min\{|(\wh{\theta}-\theta)'\theta|^2,|(\wh{\theta}+\theta)'\theta|^2\}$ is bounded by
$\min\{|(\wh{\theta}-\theta)'\theta|^2,|(\wh{\theta}+\theta)'\theta|^2\} 
\leq
\min\big\{||\wh{\theta}-\theta||^2,||\wh{\theta}+\theta||^2\big\}
\leq \Fnorm{P_{\wh{\theta}} - P_\theta}^2$.
Together with (\ref{eq:mazongming}), the above bound implies that for each $(\theta,\tau)$ pair in the support of $\pi$,
\begin{equation}
\mathbb{Q}^n_{\theta,\tau}\Big\{\min\{|(\wh{\theta}-\theta)'\theta|^2,|(\wh{\theta}+\theta)'\theta|^2\} > \frac{1}{3}\Big\}\leq \beta. \label{eq:haha}
\end{equation}
By Bernstein's inequality, we have
$$\mathbb{Q}^n_{\theta,\tau}\Big\{\Big|\frac{1}{n}\sum_{i=1}^n(g_i^2-1)\Big|+\Big|\frac{1}{n}\sum_{i=1}^n(|\wh{\theta}'\epsilon_i|^2-1)\Big| + \Big|\frac{2}{n}\sum_{i=1}^n g_i(\epsilon_i'\wh{\theta})\Big|>C\sqrt{\frac{\log n}{n}}\Big\}\leq n^{-C'}.$$
Combining the above analysis and using the assumptions that $\frac{N(\log N)^5}{k^4}\leq c$ and $cN\leq n\leq N/12$, we have
\begin{align}
	\label{eq:type2-bd-single}
\mathbb{Q}^n_{\theta,\tau}(1-\psi)\leq \beta+n^{-C'}.	
\end{align}
Integrating over $(\theta,\tau)$ according to the prior $\pi$ and applying
(\ref{eq:random}), we obtain
$$\mathbb{P}_{H_1^G}(1-\psi)\leq \beta+n^{-C'}+C\left(e^{-Ck}+N^{-1}\right)+\frac{4n}{N}.$$
Summing up the Type-I and Type-II errors, we have
\begin{eqnarray}
	\label{eq:errors-comb}
\mathbb{P}_{H_0^G}\psi + \mathbb{P}_{H_1^G}(1-\psi) \leq \beta+\frac{4n}{N}+C(n^{-1}+N^{-1}+e^{-C'k}).
\end{eqnarray}
Thus, the proof is complete.
\end{proof}

\subsection{Proofs of Theorems \ref{thm:clPCA-cont}, \ref{thm:clPCA-test} and \ref{thm:clCCA-cont}}

Consider a Planted Clique detection problem with $N=12n$ and $k=\floor{2s/3}$. Then the assumptions of \prettyref{thm:test-error-pca} are satisfied. Thus, $\sup_{\mathbb{Q}\in\mathcal{Q}(n,s,p,\lambda)}
\mathbb{Q}\left\{\fnorm{P_{\wh{\theta}}-P_\theta}^2> \frac{1}{3}\right\}\leq \frac{1}{4}$ implies a testing error bounded by
$\frac{1}{4}+\frac{1}{3}+C(n^{-1}+N^{-1}+e^{-C'k}),$
which is smaller than $2/3$ as $n\rightarrow\infty$. This contradicts the condition (\ref{eq:pca-cond}) that implies the condition of Hypothesis A. Hence, we must have (\ref{eq:clower}) and the proof of Theorem  \ref{thm:clPCA-cont} is complete. To prove Theorem \ref{thm:clPCA-test}, note that according to the proof of \prettyref{thm:test-error-pca}, a testing error bound $1/4$ for the sparse PCA problem implies a testing error bound $\frac{1}{4}+\frac{1}{3}+C(n^{-1}+N^{-1}+e^{-C'k})$ for the Planted Clique detection problem. The same argument we have just used leads to the desired conclusion. Finally, Theorem \ref{thm:clCCA-cont} can be derived from \prettyref{thm:pca-cca-reduction} and \prettyref{thm:test-error-pca} by applying a similar argument.

\subsection{Proof of \prettyref{thm:pca-cca-reduction}}

Let $W_i\sim N_p(0,\tau\theta\theta'+I_p)$, then $(X_i',Y_i')'\sim N_{p+m}(0,\Sigma)$ with $\Sigma$ given in (\ref{eq:scca}). We complete the proof by noting
\begin{eqnarray*}
\fnorm{P_{\wh{\theta}}-P_{\theta}}^2 &\leq& 4\min\left\{\Norm{\wh{\theta}-\theta}^2,\Norm{\wh{\theta}+\theta}^2\right\} 
\leq \frac{16\min\left\{\Norm{\wh{u}-u}^2,\Norm{\wh{u}+u}^2\right\}}{\Norm{u}^2} \\
&\leq& 16\frac{\sigma_{\max}^2(\Sigma_x)}{\sigma_{\min}^2(\Sigma_x)}L(\wh{u},u) 
\leq 16\Big(1+\frac{3}{2}\Big)^2L(\wh{u},u).
\end{eqnarray*}

\subsection{Proof of \prettyref{lem:fixedTV}}\label{sec:mamamama}
We first verify that \eqref{eq:f0}--\eqref{eq:f1} are proper density functions when $|\mu|\leq 3\sqrt{\eta_N \log N}$, which is a corollary of the following lemma.

\begin{lemma}\label{lem:1}
If $k\leq N/12$, $|\mu| \leq 3\sqrt{\eta_N\log N}$ and $|x|\leq 3\sqrt{\log N}$, then
\begin{align*}
\delta_N^{-1}\left|\bar{\phi}_{\mu}(x)-\phi_0(x)\right|\leq 
\frac{4}{5}\phi_0(x).	
\end{align*}
\end{lemma}
\begin{proof}
By definition,
$$
\delta_N^{-1}\Big|\bar{\phi}_{\mu}(x)-\phi_0(x)\Big|
= (2\delta_N)^{-1}\phi_0(x)\Big|\exp\Big(\mu x-\frac{\mu^2}{2}\Big)+\exp\Big(-\mu x-\frac{\mu^2}{2}\Big)-2\Big|.$$
Under the conditions of the lemma, we have $|\mu x|+\frac{\mu^2}{2}\leq \frac{1}{2}$, and so
$$\Big|\exp\Big(\mu x-\frac{\mu^2}{2}\Big)+\exp\Big(-\mu x-\frac{\mu^2}{2}\Big)-2\Big|\leq \mu^2+\frac{4}{3}|\mu x|^2+\frac{\mu^4}{3}\leq 8\mu^2\log N.$$
We complete the proof by combining the last two displays.
\end{proof}


The following lemma controls the rescaling constants in \eqref{eq:f0} and \eqref{eq:f1}.

\begin{lemma}\label{lem:2}
There exists an absolute constant $C > 0$ such that for any $|\mu|\leq 1$, $|M_i-1|\leq CN^{-4}$ for $i=0,1$.
\end{lemma}
\begin{proof}
Note that 
\begin{eqnarray*}
1 = \int f_{\mu,0}(x)\diff x 
& = & M_0\int\left(\phi_0(x)-\delta_N^{-1}(\bar{\phi}_{\mu}(x)-\phi_0(x))\right)\diff x \\
&&  - M_0\int_{|x|>3\sqrt{\log N}}\left(\phi_0(x)-\delta_N^{-1}(\bar{\phi}_{\mu}(x)-\phi_0(x))\right)\diff x \\
&=& M_0-M_0\int_{|x|>3\sqrt{\log N}}\left(\phi_0(x)-\delta_N^{-1}(\bar{\phi}_{\mu}(x)-\phi_0(x))\right)\diff x.
\end{eqnarray*}
The integral on the RHS is upper bounded by
\begin{eqnarray*}
\left(1+\delta_N^{-1}\right)\int_{|x|>3\sqrt{\log N}}\phi_0(x)\diff x + \delta_N^{-1}\int_{|x|>3\sqrt{\log N}}\bar{\phi}_{\mu}(x)\diff x 
\leq CN^{-4},
\end{eqnarray*}
where the last inequality comes from standard Gaussian tail bounds. This readily implies $|M_0  - 1| \leq CN^{-4}$ The desired bound on $M_1$ follows from similar arguments.
\end{proof}

\begin{proof}[Proof of Lemma \ref{lem:fixedTV}]
Define
\begin{eqnarray*}
g_i(x) = \phi_0(x)- (-1)^{i}\delta_N^{-1}(\bar{\phi}_{\mu}(x)-\phi_0(x)),\quad \mbox{for $i=0,1$}.
\end{eqnarray*}
Then we have for $i=0$ and $1$,
$$f_{\mu,i}(x)=g_i(x)- (1-M_i \indc{|x|\leq 3\sqrt{\log N}})g_i(x),$$
By \prettyref{lem:1} and \prettyref{lem:2},
\begin{align}
\int |f_{\mu,i}(x) - g_i(x)| \diff x 
& \leq  \int\left|(1-M_i \indc{|x|\leq 3\sqrt{\log N}})g_i(x)\right|\diff x 
\nonumber \\
&\leq |1-M_i|\int |g_i(x)|dx + M_i\int_{|x|>3\sqrt{\log N}}|g_i(x)|\diff x 
\leq CN^{-3}. \label{eq:lem-fixedTV-1}
\end{align}
Therefore, we have
\begin{align*}
& \TV\left(\frac{1}{2}(f_{\mu,0}+f_{\mu,1}),\phi_0\right)
= \frac{1}{2}\int\left|\phi_0(x)-\frac{1}{2}\left(f_{\mu,0}(x)+ f_{\mu,1}(x)\right)\right|\diff x \\
&\leq \frac{1}{2}\int\left|\phi_0(x)-\frac{1}{2}\left(g_0(x)+g_1(x)\right)\right|\diff x + \frac{1}{4}\sum_{i=0,1}\int |f_{\mu,i}(x)-g_i(x)|\diff x 
\leq CN^{-3},
\end{align*}
where the last inequality is due to the identity $\phi_0 = \frac{1}{2}(g_0+g_1)$ and \eqref{eq:lem-fixedTV-1}.
In addition, we have
\begin{align*}
&\TV\left(\delta_N f_{\mu,1}+\left(1-\delta_N\right)\frac{1}{2}\left(f_{\mu,0}+f_{\mu,1}\right), \bar\phi_{\mu}\right) 
\\
& = \frac{1}{2} \int\left|\delta_{N}f_{\mu,1}(x)+\frac{1-\delta_N}{2}\left(f_{\mu,0}(x)+f_{\mu,1}(x)\right)-\bar{\phi}_{\mu}(x)\right|\diff x \\
&\leq \frac{1}{2}\int\left|\delta_N g_1(x)+\frac{1-\delta_N}{2}\left(g_0(x)+g_1(x)\right)-\bar{\phi}_{\mu}(x)\right|\diff x 
\\
& ~~~~~~ + \sum_{i=0,1} \frac{1-(-1)^i\delta_N}{4}\int |f_{\mu,i}(x)-g_i(x)|dx
\\
&\leq CN^{-3}.
\end{align*}
Here, the last inequality is due to the identity $\delta_{N}g_{1}+\frac{1-\delta_N}{2}\left(g_{0}+g_{1}\right)=\bar{\phi}_{\mu}$ and \eqref{eq:lem-fixedTV-1}.
This completes the proof.
\end{proof}

\section{Discretization and Computational Lower bounds}
\label{sec:discrete}
To formally address the computational complexity issue in a continuous statistical model, we adopt the framework in \cite{ma2013computational}.
After introducing the asymptotically equivalent discretized models, we state the computational lower bounds for sparse PCA and sparse CCA under the discretized models in \prettyref{sec:cb-result}.
The necessary modifications to Algorithms \ref{algo:reduction1} and \ref{algo:reduction2} are spelled out in \prettyref{sec:reduction-discrete} to ensure that they are truly of randomized polynomial time complexity.

For any $t\in \naturals$, define the function $[\cdot]_t: \mathbb{R}\rightarrow 2^{-t}\mathbb{Z}$ by 
\begin{align}
	\label{eq:dyadic}
[x]_t=2^{-t}\floor{2^tx}.
\end{align}
For any matrix, the function is defined component-wise.
Let 
\begin{align*}
\mathcal{E}^{(p,n)}_M &= \big\{ \calL(X_1,\dots, X_n): X_i\stackrel{iid}{\sim} N_p(\mu,\Sigma), {1}/{M}\leq \sigma_{\min}(\Sigma)\leq \sigma_{\max}(\Sigma)\leq M \big\}
\end{align*}
be the class of joint distributions of $n$ i.i.d.~samples from all multivariate Gaussian distributions with spectrum contained in $[1/M,M]$, and 
\begin{align*}
\mathcal{E}^{(p,n,t)}_M = \{ \calL([X_1]_t,\dots, [X_n]_t): \calL(X_1,\dots, X_n)\in  \mathcal{E}^{(p,n)}_M \}	
\end{align*}
be its discretized counterpart.
The following lemma bounds the Le Cam distance \cite{LeCam86} between the two classes of distributions.
Its proof is given below in \prettyref{sec:pf-lecam}.
\begin{lemma} \label{lem:lecam}
When $2^{t}t^{-1/2}\geq 2(pM)^{3/2}$, the Le Cam distance between $\mathcal{E}_M^{(p,n)}$ and $\mathcal{E}_M^{(p,n,t)}$ satisfies
$\Delta(\mathcal{E}_M^{(p,n)},\mathcal{E}_M^{(p,n,t)})\leq n(pM)^{3/2}t^{1/2}2^{-t}$.
\end{lemma}

For any $t\in \naturals$, we define discretized sparse PCA parameter space as
\begin{align*}
\mathcal{Q}^t(n,s,p,\lambda)=\left\{\mathcal{L}([W]_t): \mathcal{L}(W)\in \mathcal{Q}(n,s,p,\lambda)\right\},
\end{align*}
and discretized sparse CCA probability space as
\begin{align*}
\mathcal{P}^t(n,s_u,s_v,p,m,r,\lambda;M)=\left\{\mathcal{L}([X]_t,[Y]_t): \mathcal{L}(X,Y)\in \mathcal{P}(n,s_u,s_v,p,m,r,\lambda;M)\right\}.	
\end{align*}
In view of \prettyref{thm:clCCA-cont}, we are primarily interested in $\mathcal{P}(n, s_u, s_v, p,m,1,\lambda;3)$ and its discretized counterpart.

For the sparse PCA parameter spaces, with the choice of $n,s,p$ and $\lambda$ in \prettyref{thm:clPCA-cont}, under condition \eqref{eq:pca-cond}, $\mathcal{Q}(n,s,p,\lambda) \subset \mathcal{E}_4^{(p,n)}$ and $\mathcal{Q}^t(n,s,p,\lambda) \subset \mathcal{E}_4^{(p,n,t)}$.
Thus, if we set the discretization level at $t = \ceil{4\log_2(p+n)}$, then $\mathcal{Q}(n,s,p,\lambda)$ and $\mathcal{Q}^t(n,s,p,\lambda)$ are asymptotically equivalent.
Similarly, with the choice of $n,s_u,s_v,p,m,\lambda$ in \prettyref{thm:clCCA-cont}, under condition \prettyref{eq:sample-size}, when $t = \ceil{4\log_2(p+m+n)}$, 
$\mathcal{P}(n, s_u, s_v, p,m,1,\lambda;3)\subset \mathcal{E}_{5}^{(p+m,n)}$ and $\mathcal{P}^t(n, s_u, s_v, p,m,1,\lambda;3)\subset \mathcal{E}_{5}^{(p+m,n,t)}$ are also asymptotically equivalent.
Therefore, with the foregoing discretization levels, the statistical difficulties of the original sparse PCA and CCA problems are asymptotically equivalent to those of the discretized problems.
In particular, the conditions for any procedure to be consistent are the same for the original and the discretized parameter spaces.

\subsection{Computational lower bounds for discretized models}
\label{sec:cb-result}

We now state computational lower bounds for the discretized sparse PCA and sparse CCA problems.
The meaning of ``randomized polynomial-time estimators'' is now based on the probabilistic Turing machine computation model rather than the computation model mentioned in \prettyref{rmk:real-computation}.

\begin{thm} \label{thm:clPCA}
Let $t=\ceil{4\log_2(p+n)}$.
Under the condition of \prettyref{thm:clPCA-cont},
for any randomized polynomial-time estimator $\wh{\theta}$,
\begin{equation}
\liminf_{n\to\infty}\sup_{\mathbb{Q}\in\mathcal{Q}^t(n,s,p,\lambda)}
\mathbb{Q}\Big\{\fnorm{P_{\wh{\theta}}-P_\theta}^2> \frac{1}{3}\Big\}>\frac{1}{4}.
\end{equation}
\end{thm}
\begin{thm}\label{thm:clCCA}
Let $t=\ceil{4\log_2(p+m+n)}$.
Under the condition of \prettyref{thm:clCCA-cont},
for any randomized polynomial-time estimator $\wh{u}$,
\begin{equation}
\liminf_{n\to\infty}
{\sup_{\mathbb{P}\in\mathcal{P}^t(n,s_u,s_v,p,m,1,\lambda;3)}}
\mathbb{P}\Big\{L(\wh{u},u)>\frac{1}{300}\Big\}>\frac{1}{4}.
\label{eq:ceCCA}
\end{equation}
\end{thm}

To prove these theorems, we need to modify Algorithms \ref{algo:reduction1} and \ref{algo:reduction2} which are not compatible with the Turing machine computation model.
The details are spelled out in the next subsection.
After these modifications, the proofs can be obtained by essentially following the lines of the proofs of their continuous counterparts while controlling some additional negligible terms in total variation bounds due to additional truncation. The details are omitted.

\subsection{Randomized polynomial-time reduction for discretized models}
\label{sec:reduction-discrete}
We first introduce a way to approximately sample with polynomial time complexity from a distribution obtained from discretizing a continuous distribution with density \cite[Section 4.2]{ma2013computational}.
The modifications to Algorithms \ref{algo:reduction1} and \ref{algo:reduction2} then follow.

For any $w,K\in \naturals$ and $K+w+1<b\in \naturals $, define the discrete distribution $\mathcal{A}_{w,b,K}[\calF]$ with probability mass function $\mathcal{A}_{w,b,K}[f]$ as
\begin{align}
\label{eq:dyadic-app-1}
\mathcal{A}_{w,b,K}[f]\big(-2^K + \frac{i-1}{2^{w}}\big)  =
\left[\frac{\int_{-2^K + (i-1) 2^{-w}}^{-2^K + i 2^{-w}} f(x)\diff x}{\int_{-2^K}^{2^K} f(x)\diff x}\right]_b, 
\end{align}
for $i\in [2^{K+w+1}-1]$,
and let 
\begin{align}
\label{eq:dyadic-app-2}
\mathcal{A}_{w,b,K}[f](2^K - 2^{-w}) = 1 - \sum_{i=1}^{2^{K+w+1}-1}\mathcal{A}_{w,b,K}[f](-2^K + (i-1) 2^{-w}) .
\end{align}
In \eqref{eq:dyadic-app-1}, $[\cdot]_b$ is the quantization defined previously in \eqref{eq:dyadic}, and \eqref{eq:dyadic-app-2} ensures that $\mathcal{A}_{w,b,K}[\calF]$ is a proper probability distribution.
By the definition of total variation distance, it is straightforward to verify that the approximation error in total variation distance by $\mathcal{A}_{w,b,K}[\calF]$ 
to the distribution of $[U\indc{U\in [-2^K, 2^K]}]_w$ with $U\sim \calF$ is upper bounded by $2^{K+w+1-b}$.
As discussed in Section 4.2 of \cite{ma2013computational}, regardless of the original distribution $\calF$, the computational complexity of drawing a random number from $\mathcal{A}_{w,b,K}(\calF)$ is $O(b2^{K+w})$. 
This fact is crucial in ensuring the modified reduction below is of randomized polynomial-time.

\paragraph{Randomized polynomial-time reduction}
For any $t\in \naturals$, we let
\begin{equation}
\label{eq:dyadic-const}
\begin{aligned}
w = t + \ceil{4\log_2 p},\, K = \ceil{\log_2(3\sqrt{\log (N+ p)})}, 
b  = w + K + 1 + \ceil{4\log_2 p}.
\end{aligned}
\end{equation}
The reduction in \prettyref{algo:reduction1} is modified to \prettyref{algo:reduction1-discrete} and the reduction in \prettyref{algo:reduction2} is modified to \prettyref{algo:reduction2-discrete}.
As in the continuous case, a direct reduction from Planted Clique to discretized sparse CCA can be obtained by constructing the estimator $\wh\theta$ in the third step of \prettyref{algo:reduction1-discrete} from \prettyref{algo:reduction2-discrete}.

\begin{algorithm}[hbt]
\caption{Reduction from Planted Clique to Sparse PCA (in Discretized Gaussian Single Spiked Model)}
\label{algo:reduction1-discrete}
\SetAlgoLined
\KwIn{}
1.~Graph adjacency matrix $A\in \{0,1\}^{N\times N}$;\\
2.~Estimator $\wh\theta$ for the leading eigenvector $\theta$.

\KwOut{A solution to the hypothesis testing problem \eqref{eq:pc}.}

\nl \textbf{Initialization}. 
Generate i.i.d.~random variables $\xi_1,\dots, \xi_{2n}\sim \mathcal{A}_{w,b,K}[\wt\Phi_0]$.
Set
\begin{align*}
\mu_i = [\eta_N^{1/2}]_w\, \xi_i , \quad i =1,\dots, 2n.
\end{align*}

\nl \textbf{Gaussianization}. 
Generate two matrices $B_0, B_1 \in \mathbb{R}^{2n\times 2n}$ where conditioning on the $\mu_i$'s, all the entries are mutually independent satisfying
\begin{align*}
\calL((B_0)_{ij}|\mu_i) = \mathcal{A}_{w,b,K}[\calF_{\mu_i,0}]
\quad \mbox{and} \quad
\calL((B_1)_{ij}|\mu_i) = \mathcal{A}_{w,b,K}[\calF_{\mu_i,1}].
\end{align*}
Let $A_0\in\{0,1\}^{2n\times 2n}$ be the lower--left $2n\times 2n$ submatrix of the matrix $A$.
Generate a matrix $W = [W_1',\dots,W_{2n}']'\in \mathbb{R}^{2n\times p}$ where
for each $i\in[2n]$, if $j\in [2n]$, we set
\begin{equation*}
W_{ij}=(B_0)_{ij}\left(1-(A_0)_{ij}\right)+(B_1)_{ij}(A_0)_{ij}.
\end{equation*}
If $2n < j\leq p$, we let $W_{ij}$ be an independent draw from $\mathcal{A}_{w,b,K}[N(0,1)]$.

\nl \textbf{Test Construction}. 
Let $\wh{\theta} = \wh{\theta}([W_{1}]_t,\dots, [W_{n}]_t)$ be the estimator of the leading eigenvector by treating $\{[W_{i}]_t\}_{i=1}^{n}$ as data. It is normalized to be a unit vector.
We reject $H_0^G$ if
\begin{align*}
\wh{\theta}' \Big(\big[\frac{1}{n}\sum_{i=n+1}^{2n} [W_{i}]_t [W_i]_t' \big]_w \Big) \wh{\theta} \geq 1 + [\frac{1}{4} \,k\,\eta_N]_w.
\end{align*}
\end{algorithm}

\begin{algorithm}[!b]
\caption{Reduction from Discretized Sparse PCA to Discretized Sparse CCA}
\label{algo:reduction2-discrete}
\SetAlgoLined
\KwIn{}
1.~Observations $W_1,\dots, W_n\in (2^{-t}\integers)^p$;\\
2.~Estimator $\wh{u}$ of the first leading canonical correlation coefficient $u$. 

\KwOut{An estimator $\wh\theta$ of the leading eigenvector of $\calL(W_1)$.}

\nl Generate i.i.d.~random vectors $Z_i = (Z_{i1},\dots,Z_{ip})'$ for $i\in [n]$ with $Z_{ij}\stackrel{iid}{\sim} \mathcal{A}_{w,b,K}[N(0,1)]$.
Set
\begin{align*}
X_i = \Big[\frac{1}{\sqrt{2}} \Big]_w (W_{i} + Z_i), \qquad 
Y_i = \Big[\frac{1}{\sqrt{2}} \Big]_w(W_{i} - Z_i),\qquad i=1,\dots,n.
\end{align*}

\nl Compute $\wh{u} = \wh{u}([X_1]_t,[Y_1]_t,\dots, [X_n]_t, [Y_n]_t)$. 
Set
\begin{align*}
\wh\theta = \wh\theta(W_1,\dots,W_n) = [\wh{u} / \|\wh{u}\|]_w.
\end{align*}
\end{algorithm}

By \eqref{eq:dyadic-const} and the discussion following \eqref{eq:dyadic-app-2},
the complexity for sampling any random variable in the above reduction is $O(p^8(\log p)^{3/2})$, and in total, we need to generate no more than $O(n(p+n))$ random variables. 
Hence, the total complexity for random number generation is $O(p^{10}(\log p)^{3/2})$ in view of the condition $p\geq 2n$.
On the other hand, it is straightforward to verify that all the other computations (except for the estimator $\wh{u}$ or $\wh{\theta}$) have complexity no more than $O(p^{10}(\log p)^{3/2})$.
Since the conditions of Theorems \ref{thm:clCCA} and \ref{thm:clPCA} ensure that for some constant $a>1$, $2n\leq p\leq n^a$ and $n \leq N/12$, we obtain that the additional computational complexity induced by the proposed reductions is $O(N^{10a}(\log N)^{3/2})$. Therefore, they are of randomized polynomial-time complexity.

\subsection{Proof of \prettyref{lem:lecam}}
\label{sec:pf-lecam}
We need the following lemma for the proof.
\begin{lemma}\label{lem:DMV}
For $X\sim N_p(\mu,\Sigma)$ with $M^{-1}\leq\sigma_{\min}(\Sigma)\leq\sigma_{\max}(\Sigma)\leq M$ and $U = (U_1,\dots,U_p)'$ where $U_i\stackrel{iid}{\sim}\text{Unif}[0,1]$, we have for any $t^{-1/2}2^{t}\geq 2(pM)^{3/2}$,
$$\TV(X,[X]_t+2^{-t}U)\leq (pM)^{3/2}t^{1/2}2^{-t}.$$
\end{lemma}

Since each distribution in $\mathcal{E}_M^{(p,n,t)}$ comes from discretizing a corresponding distribution in $\mathcal{E}_M^{(p,n)}$ on a grid with equal spacing $2^{-t}$, we have $\delta(\mathcal{E}_M^{(p,n)}, \mathcal{E}_M^{(p,n,t)}) = 0$.
On the other hand, \prettyref{lem:DMV} and Lemma 7 of \cite{ma2013computational} lead to 
$\delta(\mathcal{E}_M^{(p,n,t)}, \mathcal{E}_M^{(p,n)}) \leq n(pM)^{3/2}t^{1/2}2^{-t}$.
This completes the proof.	

\begin{proof}[Proof of \prettyref{lem:DMV}]
Let $f$ and $g$ denote the density functions of $X$ and $[X]_t+2^{-t}U$, respectively. 
Then $g$ is a piecewise constant function. 
For any $(x_1,...,x_p)\in B=\prod_{i=1}^p[2^{-t}i_j, 2^{-t} (i_j+1))$, where $i_j\in\mathbb{Z}$, we have
$$g(x_1,...,x_p)= \frac{1}{\nu(B)}\int_B f(x_1,...,x_p)dx_1...dx_p,$$
where $\nu$ is the Lebesgue measure. Hence,
\begin{align}
\nonumber &
\sup_{||x-\mu||_{\infty}\leq K}\left|\frac{g(x)}{f(x)}-1\right|
\leq \sup_{\substack{||x-\mu||_{\infty}\leq K\\||x-y||_{\infty}
\leq  2^{-t}}}\left|\frac{f(x)}{f(y)}-1\right| \\
\nonumber &\leq \sup_{\substack{||x-\mu||_{\infty}\leq K\\||x-y||_{\infty}
\leq  2^{-t}}}
\left|e^{{|(x-\mu)'\Sigma^{-1}(x-\mu)-(y-\mu)'\Sigma^{-1}(y-\mu)|}/{2}}-1\right| \\
\label{eq:cseasy} &\leq \sup_{\substack{||x-\mu||_{\infty}\leq K\\||x-y||_{\infty}
\leq  2^{-t}}}\left|e^{\fnorm{\Sigma^{-1}}\fnorm{(x-\mu)(x-\mu)'-(y-\mu)(y-\mu)'}/2}-1\right| \\
\label{eq:boundeasy} &\leq \exp\left(p^{3/2}MK2^{-t}\right)-1 
\\
\nonumber &\leq \frac{3}{2}p^{3/2}MK2^{-t},
\end{align}
whenever $p^{3/2}MK2^{-t}\leq \frac{1}{2}$. 
The inequality (\ref{eq:cseasy}) holds since
\begin{align*}
& |(x-\mu)'\Sigma^{-1}(x-\mu)-(y-\mu)'\Sigma^{-1}(y-\mu)|\\
&=\Tr\left(\Sigma^{-1}[(x-\mu)(x-\mu)'-(y-\mu)(y-\mu)']\right)\\
&\leq \fnorm{\Sigma^{-1}}\fnorm{(x-\mu)(x-\mu)'-(y-\mu)(y-\mu)'}
\end{align*}
by Cauchy-Schwarz inequality. 
The inequality (\ref{eq:boundeasy}) holds because $\fnorm{\Sigma^{-1}}\leq \sqrt{p}\opnorm{\Sigma^{-1}}\leq \sqrt{p}M$ and $\fnorm{(x-\mu)(x-\mu)'-(y-\mu)(y-\mu)'}\leq p||x-y||_{\infty}(||x-\mu||_{\infty}+||y-\mu||_{\infty})\leq 2pK2^{-t}$. 
Note that
$$
\int|f-g| \leq \int_{||x-\mu||_{\infty}>K} |f(x)-g(x)|dx  + \int_{||x-\mu||_{\infty}\leq K}f(x)\left|\frac{g(x)}{f(x)}-1\right|dx.
$$
According to Gaussian tail probability, the first term can be bounded by $2p\sqrt{\frac{2}{\pi}}\frac{\sqrt{M}}{K-1}e^{-\frac{(K-1)^2}{2M}}$.
The second term is bounded by $\frac{3}{2}p^{3/2}MK2^{-t}$ according to our previous analysis. Choosing $K=\sqrt{2Mt\log 2}+1$, we obtain the bound $2(pM)^{3/2}t^{1/2}2^{-t}$ for all $t^{-1/2}2^{t}\geq 2(pM)^{3/2}$. The conclusion follows the simple fact that $\TV(X,[X]_t+2^{-t}U)=\frac{1}{2}\int|f-g|$.
\end{proof}

\section{Additional Proofs}\label{sec:last-name}

\subsection{Proof of Theorem \ref{thm:uppersep}}

We first present a bound for the estimator defined by (\ref{eq:optexp}) under the joint loss.
\begin{thm} 
	\label{thm:upperjoint}
Assume
(\ref{eq:ass})
for some sufficiently small $c>0$. Then there exist constants $C,C'>0$ only depending on $c$ such that
$$\fnorm{\Sigma_x^{1/2}\left(\wh{U}^{(0)}(\wh{V}^{(0)})'-UV'\right)\Sigma_y^{1/2}}^2\leq \frac{C}{n\lambda^2}\Big(r(s_u+s_v)+s_u\log\frac{ep}{s_u}+s_v\log\frac{em}{s_v}\Big),$$
with $\mathbb{P}$-probability at least $1-\exp\left(-C'(s_u+\log(ep/s_u))\right)-\exp\left(-C'(s_v+\log(em/s_v))\right)$ uniformly over $\mathbb{P}\in \mathcal{P}(n,s_u,s_v,p,m,r,\lambda;M)$.
\end{thm}
Theorem \ref{thm:upperjoint} is similar to Theorem 1 of \cite{gao14}, except that the loss function depends on the marginal covariances so that the error bound is independent of $M$. Its proof is omitted given the similarity with that of Theorem 1 of \cite{gao14}.

From now on, we omit the superscript in $\wh{\Sigma}_x^{(1)}$, $\wh{\Sigma}_y^{(1)}$ and $\wh{\Sigma}_{xy}^{(1)}$ 
for simplicity.
Let $U^*=U\Lambda V'\Sigma_y\wh{V}^{(0)}$ and $\Delta=\wh{U}^{(1)}-U^*$. The following lemmas are needed in the proof of Theorem \ref{thm:uppersep}. Their proofs are given in Section \ref{sec:pf-name}.

\begin{lemma}\label{lem:V0}
Assume $\frac{s_u\log(ep/s_u)}{n}\leq c$ for some sufficiently small constant $c\in(0,1)$. Then, there exist some constants $C,C'>0$ only depending on $c$, such that with probability at least $1-\exp(-C's_u\log(ep/s_u))$, 
\begin{eqnarray*}
\opnorm{\Sigma_y^{1/2}\wh{V}^{(0)}} \leq 1+C\sqrt{\frac{s_u\log(ep/s_u)}{n}}, \,
\opnorm{(\wh{V}^{(0)})'\Sigma_y\wh{V}^{(0)}-I} \leq C\sqrt{\frac{s_u\log(ep/s_u)}{n}}.
\end{eqnarray*}

\end{lemma}

\begin{lemma} \label{lem:ratio2}
Assume $\frac{s_u\log(ep/s_u)}{n}\leq c$ for some sufficiently small constant $c>0$. Then, there exist some constants $C,C'>0$ only depending on $c$, such that
$$(1-\delta_C')\fnorm{\Sigma_x^{1/2}\Delta}^2\leq \fnorm{\wh{\Sigma}_x^{1/2}\Delta}^2\leq (1+\delta_C')\fnorm{\Sigma_x^{1/2}\Delta}^2,$$
with probability at least $1-\exp(-C's_u\log(ep/s_u))$, with $\delta_C'=C\sqrt{\frac{s_u\log(ep/s_u)}{n}}.$
\end{lemma}

\begin{lemma} \label{lem:l0noise1}
Assume $\frac{1}{n}\left(s_v\log(em/s_v)+s_u\log(ep/s_u)+rs_u\right)\leq c$ for some sufficiently small constant $c>0$. Then, there exist some constants $C,C'>0$ only depending on $c$, such that
$$\left|\Tr\left(\Delta'(\wh{\Sigma}_{xy}-\Sigma_{xy})\wh{V}^{(0)}\right)\right|\leq C\sqrt{\frac{rs_u+s_u\log(ep/s_u)}{n}}\fnorm{\Sigma_x^{1/2}\Delta},$$
with probability at least $1-\exp\left(-C'(s_u\log(ep/s_u)+rs_u)\right)-\exp(-C's_v\log(em/s_v))$.
\end{lemma}

\begin{lemma} \label{lem:l0noise2}
Assume $\frac{1}{n}\left(s_v\log(em/s_v)+s_u\log(ep/s_u)+rs_u\right)\leq c$ for some sufficiently small constant $c>0$. Then, there exist some constants $C,C'>0$ only depending on $c$, such that
$$\left|\Tr\left(\Delta'(\wh{\Sigma}_x-\Sigma_x)U^*\right)\right| \leq C\sqrt{\frac{rs_u+s_u\log(ep/s_u)}{n}}\fnorm{\Sigma_x^{1/2}\Delta},$$
with probability at least $1-\exp\left(-C'(s_u\log(ep/s_u)+rs_u)\right)-\exp(-C's_v\log(em/s_v))$.
\end{lemma}

\begin{proof}
The proof consists of two steps. First, we derive a bound for $\fnorm{\Sigma_x^{1/2}\Delta}$. Next, we derive the desired bound for $L(\wh{U},U)$.
\smallskip

\noindent\textbf{Step 1. } 
By the definition of the estimator, we have
$$\Tr((\wh{U}^{(1)})'\wh{\Sigma}_x\wh{U}^{(1)})-2\Tr((\wh{U}^{(1)})'\wh{\Sigma}_{xy}\wh{V}^{(0)})\leq \Tr((U^*)'\wh{\Sigma}_xU^*)-2\Tr((U^*)'\wh{\Sigma}_{xy}\wh{V}^{(0)}).$$
After rearrangement, we have
\begin{eqnarray*}
\Tr(\Delta'\wh{\Sigma}_x\Delta) 
 &\leq& 2\Tr\big(\Delta'(\wh{\Sigma}_{xy}\wh{V}^{(0)}-\wh{\Sigma}_xU^*)\big) \\
&\leq& 2\big|\Tr\big(\Delta'(\wh{\Sigma}_{xy}-\Sigma_{xy})\wh{V}^{(0)}\big)\big| + 2\big|\Tr\big(\Delta'(\wh{\Sigma}_x-\Sigma_x)U^*\big)\big|.
\end{eqnarray*}
Using Lemma \ref{lem:ratio2}, Lemma \ref{lem:l0noise1} and Lemma \ref{lem:l0noise2}, we have
$$\frac{1}{2}\fnorm{\Sigma_x^{1/2}\Delta}^2\leq 4C\sqrt{\frac{rs_u+s_u\log(ep/s_u)}{n}}\fnorm{\Sigma_x^{1/2}\Delta},$$
with high probability, which immediately implies a bound for $\fnorm{\Sigma_x^{1/2}\Delta}^2$. This completes Step 1. \smallskip

\noindent\textbf{Step 2. }  We claim that
\begin{align}
\label{eq:claim1}\sigma_{\min}^{-1}\big(\Sigma_x^{1/2}U\Lambda V'\Sigma_y\wh{V}^{(0)}\big) &\leq \frac{C}{\lambda},\\
\label{eq:claim2}\fnorm{\Sigma_x^{1/2}\wh{U}-\Sigma_x^{1/2}\wh{U}^{(1)}((\wh{U}^{(1)})'{\Sigma}_x\wh{U}^{(1)})^{-1/2}} &\leq C\sqrt{\frac{rs_u+s_u\log(ep/s_u)}{n}},
\end{align}
with high probability. The two claims (\ref{eq:claim1}) and (\ref{eq:claim2}) will be proved in the end. 
We bound $L(\wh{U},U)$ by
\begin{align*}
\sqrt{L(\wh{U},U)} &= \inf_{W\in O(r,r)}\fnorm{\Sigma_x^{1/2}(\wh{U}W-U)} \\
&\leq \fnorm{\Sigma_x^{1/2}\wh{U}-\Sigma_x^{1/2}\wh{U}^{(1)}((\wh{U}^{(1)})'{\Sigma}_x\wh{U}^{(1)})^{-1/2}}\\
&+\inf_{W\in O(r,r)}\fnorm{\Sigma_x^{1/2}\wh{U}^{(1)}((\wh{U}^{(1)})'{\Sigma}_x\wh{U}^{(1)})^{-1/2}W-\Sigma_x^{1/2}U}.
\end{align*}
With high probability, we could further bound the rightmost side by
\begin{eqnarray}
\label{eq:useproj}& & C\sqrt{\frac{rs_u+s_u\log(ep/s_u)}{n}} + \frac{1}{\sqrt{2}}\fnorm{P_{\Sigma_x^{1/2}\wh{U}}-P_{\Sigma_x^{1/2}U}} \\
\label{eq:usesin} &\leq& C\sqrt{\frac{rs_u+s_u\log(ep/s_u)}{n}} + C\sigma_{\min}^{-1}\left(\Sigma_x^{1/2}U\Lambda V'\Sigma_y\wh{V}^{(0)}\right)\fnorm{\Sigma_x^{1/2}\Delta} \\
\nonumber &\leq& C\sqrt{\frac{rs_u+s_u\log(ep/s_u)}{n\lambda^2}}.
\end{eqnarray}
The bound (\ref{eq:useproj}) is due to the claim (\ref{eq:claim2}), Lemma \ref{lem:proj} and the fact that $P_{\Sigma^{1/2}\wh{U}}=P_{\Sigma^{1/2}\wh{U}^{(1)}}$. The inequality (\ref{eq:usesin}) is derived from the sin-theta theorem \citep{wedin72}. Thus, we have obtained the desired bound for $L(\wh{U},U)$. To finish the proof, we need to prove  (\ref{eq:claim1}) and (\ref{eq:claim2}). Since $\Sigma_x^{1/2}U\in O(p,r)$, we have
$$\sigma_{\min}^{-1}\big(\Sigma_x^{1/2}U\Lambda V'\Sigma_y\wh{V}^{(0)}\big)\leq \lambda^{-1}\opnorm{(V'\Sigma_y\wh{V}^{(0)})^{-1}}.$$
Thus, it is sufficient to bound $\opnorm{(V'\Sigma_y\wh{V}^{(0)})^{-1}}$. By Theorem \ref{thm:upperjoint} and sin-theta theorem \citep{wedin72}, $\fnorm{P_{\Sigma_y^{1/2}\wh{V}^{(0)}}-P_{\Sigma_y^{1/2}V}}$ is sufficiently small. In view of Lemma \ref{lem:proj}, there exists $W\in O(r,r)$, such that
$$\fnorm{\Sigma_y^{1/2}\wh{V}^{(0)}(\wh{V}^{(0)}\Sigma_y\wh{V}^{(0)})^{-1/2}-\Sigma_y^{1/2}VW}$$
is sufficiently small. Therefore, together with Lemma \ref{lem:V0},
\begin{eqnarray*}
&& \opnorm{V'\Sigma_y\wh{V}^{(0)}-W} \\
 &\leq& \opnorm{V'\Sigma_y\wh{V}^{(0)}-V'\Sigma_yVW(\wh{V}^{(0)}\Sigma_y\wh{V}^{(0)})^{1/2}} + \opnorm{W}\opnorm{(\wh{V}^{(0)}\Sigma_y\wh{V}^{(0)})^{1/2}-I} \\
&\leq& \fnorm{\Sigma_y^{1/2}\wh{V}^{(0)}(\wh{V}^{(0)}\Sigma_y\wh{V}^{(0)})^{-1/2}-\Sigma_y^{1/2}VW}\opnorm{(\wh{V}^{(0)}\Sigma_y\wh{V}^{(0)})^{1/2}} + \opnorm{(\wh{V}^{(0)}\Sigma_y\wh{V}^{(0)})^{1/2}-I}
\end{eqnarray*}
is also sufficiently small. By Weyl's inequality \citep[p.449]{golub96}, $|\sigma_{\min}(V'\Sigma_y\wh{V}^{(0)})-1|\leq \opnorm{V'\Sigma_y\wh{V}^{(0)}-W}$ is sufficiently small. Hence, $\opnorm{(V'\Sigma_y\wh{V}^{(0)})^{-1}}\leq 2$ with high probability, which implies the desired bound in (\ref{eq:claim1}).  Finally, we need to prove (\ref{eq:claim2}). We have
\begin{eqnarray*}
&& \fnorm{\Sigma_x^{1/2}\wh{U}-\Sigma_x^{1/2}\wh{U}^{(1)}((\wh{U}^{(1)})'{\Sigma}_x\wh{U}^{(1)})^{-1/2}} \\
&\leq& \fnorm{\Sigma_x^{1/2}\wh{U}^{(1)}}\opnorm{((\wh{U}^{(1)})'\wh{\Sigma}_x^{(2)}\wh{U}^{(1)})^{-1/2}-((\wh{U}^{(1)})'{\Sigma}_x\wh{U}^{(1)})^{-1/2}} \\
&\leq& C\left(\fnorm{\Sigma_x^{1/2}U\Lambda V'\Sigma_y\wh{V}^{(0)}}+\fnorm{\Sigma_x^{1/2}\Delta}\right)\opnorm{(\wh{U}^{(1)})'(\wh{\Sigma}_x^{(2)}-\Sigma_x)\wh{U}^{(1)}}.
\end{eqnarray*}
We have already shown that $\fnorm{\Sigma_x^{1/2}\Delta}$ is sufficiently small. The term $\fnorm{\Sigma_x^{1/2}U\Lambda V'\Sigma_y\wh{V}^{(0)}}$ is bounded by $\sqrt{r}\opnorm{V'\Sigma_y\wh{V}^{(0)}}\leq \sqrt{r}(1+ \opnorm{V'\Sigma_y\wh{V}^{(0)}-W})\leq C\sqrt{r}$ by using the bound derived for $ \opnorm{V'\Sigma_y\wh{V}^{(0)}-W}$. To bound $\opnorm{(\wh{U}^{(1)})'(\wh{\Sigma}_x^{(2)}-\Sigma_x)\wh{U}^{(1)}}$, note that $\wh{\Sigma}_x^{(2)}$ only depends on $\mathcal{D}_2$ and is independent of $\wh{U}^{(1)}$.  Using union bound and an $\epsilon$-net argument (see, for example, \cite{vershynin10}) and the fact that $r\leq s_u$ (which is implied by $\Sigma_x^{1/2}U\in O(p,r)$), we have $\opnorm{(\wh{U}^{(1)})'(\wh{\Sigma}_x^{(2)}-\Sigma_x)\wh{U}^{(1)}}\leq C\sqrt{\frac{rs_u+s_u\log(ep/s_u)}{n}}$ with high probability. Hence, the proof is complete.
\end{proof}

\subsection{Proof of Theorem \ref{thm:lower}}

For any probability measures $\mathbb{P},\mathbb{Q}$, define the Kullback-Leibler(KL) divergence by $D(\mathbb{P}||\mathbb{Q})=\int\left(\log\frac{d\mathbb{P}}{d\mathbb{Q}}\right)d\mathbb{P}$. The following result is Lemma 14 in \cite{gao14}. It gives explicit formula for the KL divergence between distributions generated by a special kind of covariance matrices.
\begin{lemma}\label{lmm:KL}
For $i=1,2$, let $\Sigma_{(i)}=\begin{bmatrix}
I_p & \lambda U_{(i)}V_{(i)}' \\
\lambda V_{(i)}U_{(i)}' & I_m
\end{bmatrix}$ with $\lambda\in (0,1)$, $U_{(i)}\in O(p,r)$ and $V_{(i)}\in O(m,r)$. Let $\mathbb{P}_{(i)}$ denote the distribution of a random i.i.d. sample of size $n$ from the $N_{p+m}(0,\Sigma_{(i)})$ distribution. Then
$$D(\mathbb{P}_{(1)}||\mathbb{P}_{(2)})=\frac{n\lambda^2}{2(1-\lambda^2)}\fnorm{U_{(1)}V_{(1)}'-U_{(2)}V_{(2)}'}^2.$$
\end{lemma}

The main tool for our proof is the following Fano's lemma
\cite[Lemma 3]{yu1997assouad}.
\begin{proposition}\label{prop:fano}
Let $(\Theta,\rho)$ be a metric space and $\{\mathbb{P}_{\theta}:\theta\in\Theta\}$ a collection of probability measures. For any totally bounded $T\subset\Theta$, denote by $\mathcal{M}(T,\rho,\epsilon)$ the $\epsilon$-packing number of $T$ with respect to $\rho$, i.e., the maximal number of points in $T$ whose pairwise minimum distance in $\rho$ is at least $\epsilon$. Define the KL diameter of $T$ by
$d_{\mathrm{KL}}(T) \triangleq \sup_{\theta,\theta' \in T} D(\mathbb{P}_\theta||\mathbb{P}_{\theta'})$.
Then
\begin{equation}
\inf_{\wh{\theta}}\sup_{\theta\in\Theta}\mathbb{P}_{\theta}\left\{\rho^2\big(\wh{\theta}(X),\theta\big)\geq\frac{\epsilon^2}{4}\right\}\geq 1-\frac{d_{\mathrm{KL}}(T)+\log 2}{\log\mathcal{M}(T,\rho,\epsilon)}.
\label{eq:fano}
\end{equation}
\end{proposition}

Finally, we lower bound the prediction loss by the squared subspace distance. Its proof is given in Section \ref{sec:pf-name}.
\begin{proposition} \label{prop:loss}
Suppose the eigenvalues of $\Sigma_x$ lie in the interval $[M_1,M_2]$. Then, we have
$$\fnorm{P_{\wh{U}}-P_U}^2\leq \sqrt{2}\frac{M_2}{M_1}L(\wh{U},U).$$
A similar inequality holds for $L(\wh{V},V)$.
\end{proposition}

\begin{proof}[Proof of Theorem \ref{thm:lower}]
Let us first give an outline of the proof.
By Proposition \ref{prop:loss}, we have
$$\inf_{\wh{U}}\sup_{\mathbb{P}\in\mathcal{P}}\mathbb{P}\big\{L(\wh{U},U)\geq C\epsilon^2\big\}\geq \inf_{\wh{U}}\sup_{\mathbb{P}\in\mathcal{P}}\mathbb{P}\left(\fnorm{P_{\wh{U}}-P_U}^2\geq C_1\epsilon^2\right),$$
for any rate $\epsilon^2$. Therefore, it is sufficient to derive a lower bound for the loss $\fnorm{P_{\wh{U}}-P_U}^2$. Without loss of generality, we assume $s_u/3$ is an integer and $s_u\leq 3p/4$. The case $s_u> 3p/4$ is harder and thus it shares the same lower bound. The subset of covariance class $\mathcal{F}(p,m,s_u,s_v,r,\lambda;M)$ we consider is
\begin{align*}
T  = \bigg\{\Sigma = \begin{bmatrix} I_p & \lambda UV_0'\\ \lambda V_0 U' & I_m \end{bmatrix}: & U = \begin{bmatrix}  \wt{U} & 0 \\ 0 & u_r \end{bmatrix}, \wt{U}\in B,\\
& \hskip -2em u_r\in \mathbb{R}^{p-2s_u/3}, ||u_r||=1, |\supp(u_r)| \leq s_u/3 \bigg\},
\end{align*}
where $V_0 = \begin{bmatrix} I_r & 0' \end{bmatrix}'\in O(m,r)$ and $B$ is a subset of $O(2s_u/3,r-1)$ to be specified later. From the construction, $U$ depends on the matrix $\wt{U}$ and the vector $u_r$. As $\wt{U}$ and $u_r$ vary, we always have $U\in O(p,r)$. We use $T(u_r^*)$ to denote a subset of $T$ where $u_r=u_r^*$ is fixed, and use $T(\wt{U}^*)$ to denote a subset of $T$ where $\wt{U}=\wt{U}^*$ is fixed.

The proof has three steps. In the first step, we derive the part $\frac{rs_u}{n\lambda^2}$ using the subset $T(u_r^*)$ for some particular $u_r^*$. In the second step, we derive the other part $\frac{s_u\log (ep/s_u)}{n\lambda^2}$ using the subset $T(\wt{U}^*)$ for some fixed $\wt{U}^*$. Finally, we combine the two results in the third step.
\smallskip

\noindent \textbf{Step 1. } Let $u_r^*=(1,0,...,0)'$, and we consider the subset $T(u_r^*)$.
Let $\wt{U}_0 = \begin{bmatrix}I_{r-1} & 0' \end{bmatrix}'\in O(2s_u/3,r-1)$ and  $\epsilon_0 \in (0, \sqrt{r} ] $ to be specified later.  Define
$$
B=B(\epsilon_0) = \{\wt{U}\in O(2s_u/3,r-1):  \fnorm{\wt{U}-\wt{U}_0}\leq \epsilon_0\}.
$$
By \prettyref{lmm:KL},
\begin{equation}
\begin{aligned}
\label{eq:KL101}d_{\mathrm{KL}}\left(T(u_r^*)\right) & = \sup_{\wt{U}_{(i)} \in B(\epsilon_0)} \frac{n\lambda^2}{2(1-\lambda^2)}\fnorm{\wt{U}_{(1)} - \wt{U}_{(2)}}^2
\leq \frac{2n\lambda^2 \epsilon_0^2}{1-\lambda^2}.
\end{aligned}
\end{equation}
Here, the equality is due to the definition of $V_0$ and the inequality due to the definition of $B(\epsilon_0)$.
We now establish a lower bound for the packing number of $T(u_r^*)$.
For some $\alpha \in (0, 1)$ to be specified later,
let $\{\wt{U}_{(1)},\dots, \wt{U}_{(N)}\}\subset O(2s_u/3,r-1)$ be a maximal set such that  for any $i\neq j\in [N]$,
\begin{align}
	\label{eq:discrete-set}
\fnorm{\wt{U}_{(i)} \wt{U}_{(i)}' - \wt{U}_0 \wt{U}_0'} \leq \epsilon_0, \qquad
\fnorm{\wt{U}_{(i)} \wt{U}_{(i)}' - \wt{U}_{(j)} \wt{U}_{(j)}'} \geq \sqrt{2}\alpha\epsilon_0.
\end{align}
Then by \cite[Lemma 1]{cai13a}, for some absolute constant $C>1$,
\begin{align*}
N \geq \left(\frac{1}{C\alpha}\right)^{(r-1)(2s_u/3-r+1)}.
\end{align*}
It is easy to see that the loss function $\fnorm{P_{U_{(i)}}-P_{U_{(j)}}}^2$ on the subset $T(u_r^*)$ equals $\fnorm{\wt{U}_{(i)}\wt{U}_{(i)}'-\wt{U}_{(j)}\wt{U}_{(j)}'}^2$. Thus, for $\epsilon=\sqrt{2}\alpha\epsilon_0$ with sufficiently small $\alpha$, $\log\mathcal{M}(T(u_r^*),\rho,\epsilon)\geq (r-1)(2s_u/3-r+1)\log \frac{1}{C\alpha}\geq (r-1)(\frac{1}{6}s_u-1)\log\frac{1}{C\alpha}\geq \frac{1}{12}rs_u\log\frac{1}{C\alpha}$ when $r$ is sufficiently large and $r\leq s_u/2$. Taking $\epsilon_0^2=c_1\frac{rs_u}{n\lambda^2}$ for sufficiently small $c_1$, we have
\begin{equation}
\inf_{\wh{U}}\sup_{T(u_r^*)}\mathbb{P}\left\{\fnorm{P_{\wh{U}}-P_{U}}^2\geq \frac{\epsilon^2}{4}\right\}\geq 1-\frac{\frac{2c_1rs_u}{1-\lambda^2}+\log 2}{\frac{1}{12}rs_u\log\frac{1}{C\alpha}}.\label{eq:lower1}
\end{equation}
Since $\lambda$ is bounded away from $1$, we may choose sufficiently small $c_1$ and $\alpha$, so that the right hand side of  (\ref{eq:lower1}) can be lower bounded by $0.9$. This completes the first step.
\smallskip

\noindent\textbf{Step 2. } The part $\frac{s_u\log (ep/s_u)}{n\lambda^2}$ can be obtained from the rank-one argument spelled out in \cite{chen13}. To be rigorous, consider the  subset $T(\wt{U}^*)$ with $\wt{U}^*= \begin{bmatrix}I_{r-1} & 0' \end{bmatrix}'\in O(2s_u/3,r-1)$.
Restricting on the set $T(\wt{U}^*)$, the loss function is
$$\fnorm{P_{U_{(i)}}-P_{U_{(j)}}}^2=\fnorm{u_{r,(i)}u_{r,(i)}'-u_{r,(j)}u_{r,(j)}'}^2.$$
Let $X = [X_1\,\, X_2]$ with $X_1\in \mathbb{R}^{n\times (r-1)}$ and $X_2\in \mathbb{R}^{n\times (p-r+1)}$, and $Y = [Y_1\,\, Y_2]$ with $Y_1\in \mathbb{R}^{n\times (r-1)}$ and $Y_2\in \mathbb{R}^{n\times (m-r+1)}$.
Then it is further equivalent to estimating $u_1$ under projection loss based on the observation $(X_2, Y_2)$, because $(X_2, Y_2)$ is a sufficient statistic for $u_r$. Applying the argument in \cite[Appendix G]{chen13} and choosing the appropriate constant, we have
\begin{equation}
\inf_{\wh{U}}\sup_{T(\wt{U}^*)}\mathbb{P}\left\{\fnorm{P_{\wh{U}}-P_{U}}^2\geq C\frac{s_u\log (ep/s_u)}{n\lambda^2} \wedge c_0\right\}\geq 0.9,\label{eq:lower2}
\end{equation}
for some constant $C>0$. This completes the second step.
\smallskip

\noindent\textbf{Step 3. } For any $\mathbb{P}\in\mathcal{P}$, by union bound, we have
\begin{eqnarray*}
&& \mathbb{P}\left\{\fnorm{P_{\wh{U}}-P_{U}}^2\geq\epsilon_1^2\vee\epsilon_2^2\right\} \\
&\geq& 1-\mathbb{P}\left\{\fnorm{P_{\wh{U}}-P_{U}}^2<\epsilon_1^2\right\}-\mathbb{P}\left\{\fnorm{P_{\wh{U}}-P_{U}}^2<\epsilon_2^2\right\} \\
&=& \mathbb{P}\left\{\fnorm{P_{\wh{U}}-P_{U}}^2\geq\epsilon_1^2\right\}+\mathbb{P}\left\{\fnorm{P_{\wh{U}}-P_{U}}^2\geq\epsilon_2^2\right\}-1.
\end{eqnarray*}
Taking $\sup_{T(u_r^*)\cup T(\wt{U}^*)}$ on both sides of the inequality, and letting $\epsilon_1^2=C_1\frac{rs_u}{n\lambda^2}$ in (\ref{eq:lower1}) and $\epsilon_2^2=C_2\frac{s_u\log (ep/s_u)}{n\lambda^2}\wedge c_0$ in (\ref{eq:lower2}), we have
$$\sup_{\mathbb{P}\in\mathcal{P}}\mathbb{P}\left\{\fnorm{P_{\wh{U}}-P_{U}}^2\geq\epsilon_1^2\vee\epsilon_2^2\right\}\geq 0.9+0.9-1=0.8,$$
where we have used the identity $\sup_{\wt{U}\in T(u_r^*),u_r\in T(\wt{U}^*)}(f(u_r)+g(\wt{U}))=\sup_{u_r\in T(\wt{U}^*)}f(u_r)+\sup_{\wt{U}\in T(u_r^*)}g(\wt{U})$.
Careful readers may notice that we have assume sufficiently large $r$ in Step 1. For $r$ which is not sufficiently large, a similar rank-one argument as in Step 2 gives the desired lower bound. Thus, the proof is complete.
\end{proof}

\subsection{Proofs of technical lemmas} \label{sec:pftech}

This section gathers the proofs of all technical results used in the above sections. 
The proofs are organized according to the order of their first appearance.
To simplify  notation,
we denote $\wh{\Sigma}_x^{(j)}$, $\wh{\Sigma}_y^{(j)}$ and $\wh{\Sigma}_{xy}^{(j)}$ by $\wh{\Sigma}_x$, $\wh{\Sigma}_y$ and $\wh{\Sigma}_{xy}$ for $j\in \{0,1,2\}$ whenever there is no confusion from the context. 



\subsubsection{Proofs of lemmas in Section \ref{sec:proof}}\label{sec:pf-pf}

In order to prove Lemma \ref{lem:normalize}, we need an auxiliary result.
\begin{lemma} \label{lem:UV}
Assume $\frac{1}{n}(s_u+s_v+\log(ep/s_u)+\log(em/s_v))\leq c$ for some sufficiently small constant $c\in(0,1)$. Then there exist some constants $C,C'>0$ only depending on $c$ such that
\begin{eqnarray*}
\opnorm{U'\wh{\Sigma}_xU-I}\vee\opnorm{(U'\wh{\Sigma}_xU)^{1/2}-I} &\leq& C\sqrt{\frac{1}{n}\pth{s_u+\log\frac{ep}{s_u}}},\\
\opnorm{V'\wh{\Sigma}_yV-I}\vee\opnorm{(V'\wh{\Sigma}_yV)^{1/2}-I} &\leq& C\sqrt{\frac{1}{n}\pth{s_v+\log\frac{em}{s_v}}},
\end{eqnarray*}
with probability at least $1-\exp(-C'(s_u+\log(ep/s_u)))-\exp(C'(s_v+\log(em/s_v)))$.
\end{lemma}
\begin{proof}
Using the definition of operator norm and the sparsity of $U$, we have
\begin{eqnarray*}
&&\opnorm{U'\wh{\Sigma}_xU-I_r}=\opnorm{U'(\wh{\Sigma}_x-\Sigma_x)U}=\opnorm{(U_{S_u*})'(\wh{\Sigma}_{xS_uS_u}-{\Sigma}_{xS_uS_u})U_{S_u*}}\\
&=&\sup_{||v||=1}(U_{S_u*}v)'(\wh{\Sigma}_{xS_uS_u}-{\Sigma}_{xS_uS_u})(U_{S_u*}v)\leq \opnorm{\Sigma_{xS_uS_u}^{1/2}U_{S_u*}}^2\opnorm{\Sigma_{xS_uS_u}^{-1/2}\wh{\Sigma}_{xS_uS_u}\Sigma_{xS_uS_u}^{-1/2}-I},
\end{eqnarray*}
where $\opnorm{\Sigma_{xS_uS_u}^{1/2}U_{S_u*}}^2\leq 1$ and $\opnorm{\Sigma_{xS_uS_u}^{-1/2}\wh{\Sigma}_{xS_uS_u}\Sigma_{xS_uS_u}^{-1/2}-I}$ is bounded by the desired rate with high probability according to Lemma 16 in \cite{gao14}. Lemma 15 in \cite{gao14} implies $\opnorm{(U'\wh{\Sigma}_xU)^{1/2}-I}\leq C\opnorm{U'\wh{\Sigma}_xU-I}$, and thus $\opnorm{(U'\wh{\Sigma}_xU)^{1/2}-I}$ also shares same upper bound. The upper bound for $\opnorm{V'\wh{\Sigma}_yV-I}\vee\opnorm{(V'\wh{\Sigma}_yV)^{1/2}-I}$ can be derived by the same argument. Hence, the proof is complete.
\end{proof}

\begin{proof}[Proof of Lemma \ref{lem:normalize}]
According to the definition, we have
\begin{eqnarray*}
\opnorm{\Sigma^{1/2}_x(U-\wt{U})} &\leq& \opnorm{\Sigma_x^{1/2}U}\opnorm{(U'\wh{\Sigma}_xU)^{1/2}-I}\opnorm{(U'\wh{\Sigma}_xU)^{-1/2}}, \\
\opnorm{\Sigma_y^{1/2}(V-\wt{V})} &\leq& \opnorm{\Sigma_y^{1/2}V}\opnorm{(V'\wh{\Sigma}_yV)^{1/2}-I}\opnorm{(V'\wh{\Sigma}_yV)^{-1/2}}, \\
\opnorm{\wt{\Lambda}-\Lambda} &\leq& \opnorm{(U'\wh{\Sigma}_xU)^{1/2}-I}\opnorm{\Lambda(V'\wh{\Sigma}_yV)^{1/2}}\\
&& \hskip 8em +\opnorm{\Lambda}\opnorm{(V'\wh{\Sigma}_yV)^{1/2}-I}.
\end{eqnarray*}
Applying Lemma \ref{lem:UV}, the proof is complete.
\end{proof}

\begin{proof}[Proof of Lemma \ref{lem:feasible}]
By the definition of $\wt{U}$, we have $\wt{U}'\wh{\Sigma}_x\wt{U}=I$, and thus $\wh{\Sigma}_x^{1/2}\wt{U}\in O(p,r)$. Similarly $\wh{\Sigma}_y^{1/2}\wt{V}\in O(m,r)$. Thus,
\begin{equation}
\opnorm{\wh{\Sigma}_x^{1/2}\wt{A}\wh{\Sigma}_y^{1/2}}\leq\opnorm{\wh{\Sigma}_x^{1/2}\wt{U}}\opnorm{\wh{\Sigma}_y^{1/2}\wt{V}}\leq 1. \label{eq:Qop}
\end{equation}
Now let us use the notation $Q=\wh{\Sigma}_x^{1/2}\wt{A}\wh{\Sigma}_y^{1/2}$. Then, by the definition of $\wt{A}$, we have $Q'Q=\wh{\Sigma}_y^{1/2}V(V'\wh{\Sigma}_yV)^{-1}V'\wh{\Sigma}_y^{1/2}$, and
\begin{equation}
\Tr(Q'Q)=\Tr((V'\wh{\Sigma}_yV)^{-1}(V'\wh{\Sigma}_yV))=r. \label{eq:Qtrace}
\end{equation}
Combining (\ref{eq:Qop}) and (\ref{eq:Qtrace}), it is easy to see that all eigenvalues of $Q'Q$ are $1$. Thus, we have $\nnorm{Q}=r$ and $\opnorm{Q}=1$. The proof is complete.
\end{proof}

\begin{proof}[Proof of Lemma \ref{lem:identify}]
Denote $F=[f_1,...,f_r]$, $G=[g_1,...,g_r]$ and $c_j=f_j'Eg_j$. By $\opnorm{E}\leq 1$, we have $|c_j|\leq 1$. The left hand side of (\ref{eq:identify}) is lower bounded by
$\iprod{FKG'}{FG'-E}\geq \iprod{FDG'}{FG'-E}-\fnorm{K-D}\fnorm{FG'-E}$,
where
$\iprod{FDG'}{FG'-E} = \iprod{D}{I-F'EG} = \sum_{l=1}^rd_{l}(1-c_l)\geq d_r\sum_{l=1}^r(1-c_l)$.
The first term on the right hand side of (\ref{eq:identify}) is
\begin{align*}
 \hskip -1em \frac{d_r}{2}\fnorm{FG'-E}^2 & = \frac{d_r}{2}\Big(\fnorm{FG'}^2+\fnorm{E}^2-2\Tr(F'EG)\Big) \\
&\leq \frac{d_r}{2}\Big(\Tr(F'FG'G)+\opnorm{E}\nnorm{E}-2\sum_{j=1}^rc_j\Big)
\\
&
\leq d_r\sum_{j=1}^r(1-c_j).
\end{align*}
This completes the proof.
\end{proof}

\begin{proof}[Proof of Lemma \ref{lem:rho}]
Using triangle inequality, $||\wh{\Sigma}_{xy}-\wt{\Sigma}_{xy}||_{\infty}$ can be upper bounded by the following sum,
\begin{eqnarray*}
&& ||\wh{\Sigma}_{xy}-\Sigma_{xy}||_{\infty} + ||(\wh{\Sigma}_x-\Sigma_x)U\Lambda V'\Sigma_y||_{\infty} \\
&& + ||\Sigma_xU\Lambda V'(\wh{\Sigma}_y-\Sigma_y)||_{\infty} + ||(\wh{\Sigma}_x-\Sigma_x)U\Lambda V'(\wh{\Sigma}_y-\Sigma_y)||_{\infty}.
\end{eqnarray*}
The first term can be bounded by the desired rate by union bound and Bernstein's inequality \cite[Prop. 5.16]{vershynin10}. For the second term, it can be written as
$$\max_{j,k}\Big|\frac{1}{n}\sum_{i=1}^n(X_{ij}[X_i'U\Lambda V'\Sigma_y]_k-\mathbb{E}X_{ij}[X_i'U\Lambda V'\Sigma_y]_k)\Big|,$$
where $X_{ij}$ is the $j$-th element of $X_i$ and the notation $[\cdot]_k$ means the $k$-th element of the referred vector. Thus, it is a maximum over average of centered sub-exponential random variables. Then, by Bernstein's inequality and union bound, it is also bounded by the desired rate. Similarly, we can bound the third term. For the last term, it can be bounded by $\sum_{l=1}^r\lambda_l||(\wh{\Sigma}_x-\Sigma_x)u_lv_l'(\wh{\Sigma}_y-\Sigma_y)||_{\infty}$, where for each $l$, $||(\wh{\Sigma}_x-\Sigma_x)u_lv_l'(\wh{\Sigma}_y-\Sigma_y)||_{\infty}$ can be written as
$$\max_{j,k}\Big|\Big(\frac{1}{n}\sum_{i=1}^n(X_{ij}X_i'u_l-\mathbb{E}X_{ij}X_i'u_l)\Big)\Big(\frac{1}{n}\sum_{i=1}^n(Y_{ik}Y_i'v_l-\mathbb{E}Y_{ik}Y_i'v_l)\Big)\Big|.$$
It can be bounded by the rate $\frac{\log (p+m)}{n}$ with the desired probability using union bound and Bernstein's inequality. Hence, the last term can be bounded by $\frac{\lambda_1r\log (p+m)}{n}$. Under the assumption that $r\sqrt{\frac{\log (p+m)}{n}}$ is bounded by a constant, it can further be bounded by the rate $\sqrt{\frac{\log (p+m)}{n}}$ with high probability. Combining the bounds of the four terms, the proof is complete.
\end{proof}

\begin{proof}[Proof of Lemma \ref{lem:proj}]
By the property of least squares, we have
\begin{eqnarray*}
\inf_W\fnorm{F-GW}^2 &=& \fnorm{F-G(G'G)^-G'F}^2 
= \fnorm{F-P_GF}^2 
= r-\Tr(P_FP_G).
\end{eqnarray*}
Since $\fnorm{P_F-P_G}^2=2r-2\Tr(P_FP_G)$, the proof is complete.
\end{proof}

\begin{proof}[Proof of Lemma \ref{lem:ep}]
By the definition of $U^*$,  we have $\Sigma_{xy}\wh{V}^{(0)}=\Sigma_x U^*$. Thus,
$$\max_{1\leq j\leq p}||[\wh{\Sigma}_{xy}\wh{V}^{(0)}-\wh{\Sigma}_xU^*]_{j\cdot}||\leq \max_{1\leq j\leq p}||[(\wh{\Sigma}_{xy}-\Sigma_{xy})\wh{V}^{(0)}]_{j\cdot}||+\max_{1\leq j\leq p}||[(\wh{\Sigma}_x-\Sigma_x)U^*]_{j\cdot}||.$$
Let us first bound $\max_{1\leq j\leq p}||[(\wh{\Sigma}_x-\Sigma_x)U^*]_{j\cdot}||$. Note that the sample covariance can be written as
$$\wh{\Sigma}_x=\Sigma_x^{1/2}\Big(\frac{1}{n}\sum_{i=1}^n Z_iZ_i'\Big)\Sigma_x^{1/2},$$
where $\{Z_i\}_{i=1}^n$ are i.i.d. Gaussian vectors distributed as $N(0,I_p)$. Let $T_j'$ be the $j$-th row of $\Sigma_x^{1/2}$, and then we have
$$[(\wh{\Sigma}_x-\Sigma_x)U^*]_{j\cdot}=\frac{1}{n}\sum_{i=1}^n (T_j'Z_i Z_i'\Sigma_x^{1/2}U^*-T_j'\Sigma_x^{1/2}U^*).$$
For each $i$ and $j$, define vector
$$W_i^{(j)}=\begin{bmatrix}
T_j'Z_i \\
(U^*)'\Sigma_x^{1/2}Z_i
\end{bmatrix}.$$
Since $T_j'Z_i Z_i'\Sigma_x^{1/2}U^*$ is a submatrix of $W_i^{(j)}(W_i^{(j)})'$, we have
$||[(\wh{\Sigma}_x-\Sigma_x)U^*]_{j\cdot}||\leq \opnorm{\frac{1}{n}\sum_{i=1}^n (W_i^{(j)}(W_i^{(j)})'-\mathbb{E}W_i^{(j)}(W_i^{(j)})')}$.
Hence, for any $t>0$, we have
\begin{eqnarray}
\nonumber&& \mathbb{P}\Big\{\max_{1\leq j\leq p}||[(\wh{\Sigma}_x-\Sigma_x)U^*]_{j\cdot}||>t\Big\} \\
\nonumber&\leq& \sum_{j=1}^p \mathbb{P}\Big\{\opnorm{\frac{1}{n}\sum_{i=1}^n (W_i^{(j)}(W_i^{(j)})'-\mathbb{E}W_i^{(j)}(W_i^{(j)})')}>t\Big\} \\
\label{eq:unionbound}&\leq& \sum_{j=1}^p\exp\Big(C_1r-C_2n\min\Big\{\frac{t}{\opnorm{\mathcal{W}^{(j)}}},\frac{t^2}{\opnorm{\mathcal{W}^{(j)}}^2}\Big\}\Big),
\end{eqnarray}
where $\mathcal{W}^{(j)}=\mathbb{E}W_i^{(j)}(W_i^{(j)})'$, and we have used concentration inequality for sample covariance \cite[Thm. 5.39]{vershynin10}. Since $\opnorm{\mathcal{W}^{(j)}}\leq C_3$ for some constant $C_3$ only depending on $M$, (\ref{eq:unionbound}) can be bounded by
$$\exp\Big(C_1'(r+\log p)-C_2'n (t\wedge t^2)\Big).$$
Take $t^2=C_4\frac{r+\log p}{n}$ for some sufficiently large $C_4$, and under the assumption $n^{-1}(r+\log p)\leq C_1$, $\max_{1\leq j\leq p}||[(\wh{\Sigma}_x-\Sigma_x)U^*]_{j\cdot}||\leq C\sqrt{\frac{r+\log p}{n}}$ with probability at least $1-\exp(-C'(r+\log p))$. Similar arguments lead to the bound of $\max_{1\leq j\leq p}||[(\wh{\Sigma}_{xy}-\Sigma_{xy})\wh{V}^{(0)}]_{j\cdot}||$. Let us sketch the proof. Note  that we may write
$$[(\wh{\Sigma}_{xy}-\Sigma_{xy})\wh{V}^{(0)}]_j=\frac{1}{n}\sum_{i=1}^n\big(T_j'Z_iY_i'\wh{V}^{(0)}-\mathbb{E}(T_j'Z_iY_i'\wh{V}^{(0)})\big).$$
Then, define
$H_i^{(j)}=\begin{bmatrix}
T_j'Z_i \\
(\wh{V}^{(0)})'Y_i
\end{bmatrix}$,
and we have
$$\max_{1\leq j\leq p}||[(\wh{\Sigma}_{xy}-\Sigma_{xy})\wh{V}^{(0)}]_j||\leq \max_{1\leq j\leq p}\opnorm{\frac{1}{n}\sum_{i=1}^n (H_i^{(j)}(H_i^{(j)})'-\mathbb{E}H_i^{(j)}(H_i^{(j)})')}.$$
Using the same argument, we can bound this term by $C\sqrt{\frac{r+\log p}{n}}$ with probability at least $1-\exp(-C'(r+\log p))$. Thus, the proof is complete.
\end{proof}

\subsubsection{Proofs of lemmas in Section  \ref{sec:last-name}}\label{sec:pf-name}

\begin{proof}[Proof of Lemma \ref{lem:V0}]
Let $T_v=\wh{S}_v\cup S_v$, where $\wh{S}_v=\supp(\wh{V}^{(0)})$.
First, let us bound $\opnorm{\Sigma_{yT_vT_v}^{1/2}\wh{V}_{T_v*}^{(0)}}$. Since $(\wh{V}^{(0)})'\wh{\Sigma}_y \wh{V}^{(0)}=I_r$, we have
\begin{align*}
\opnorm{\Sigma_{yT_vT_v}^{1/2}\wh{V}_{T_v*}^{(0)}} &\leq \opnorm{\Sigma_{yT_vT_v}^{1/2}\wh{\Sigma}_{yT_vT_v}^{-1/2}}\opnorm{\wh{\Sigma}_{yT_vT_v}^{1/2}\wh{V}^{(0)}}\leq \opnorm{\Sigma_{yT_vT_v}^{1/2}\wh{\Sigma}_{yT_vT_v}^{-1/2}} \\
&= \opnorm{\Sigma_{yT_vT_v}^{1/2}\wh{\Sigma}_{yT_vT_v}^{-1}\Sigma_{yT_vT_v}^{1/2}}^{1/2} = \sigma_{\min}(\Sigma_{yT_vT_v}^{-1/2}\wh{\Sigma}_{yT_vT_v}\Sigma_{yT_vT_v}^{-1/2})^{-1/2} \\
&\leq \left(1-\opnorm{\Sigma_{yT_vT_v}^{-1/2}\wh{\Sigma}_{yT_vT_v}\Sigma_{yT_vT_v}^{-1/2}-I}\right)^{-1/2} 
\leq 1+ C\sqrt{\frac{s_u\log(ep/s_u)}{n}},
\end{align*}
with probability at least $1-\exp(-C's_u\log(ep/s_u))$, where the last inequality is by Lemma 12 of \cite{gao14}.
Hence,
\begin{align*}
& \opnorm{(\wh{V}^{(0)})'\Sigma_y\wh{V}^{(0)}-I} = \opnorm{(\wh{V}^{(0)})'(\Sigma_y-\wh{\Sigma}_y)\wh{V}^{(0)}} 
= \opnorm{(\wh{V}^{(0)}_{T_v*})'(\Sigma_{yT_vT_v}-\wh{\Sigma}_{yT_vT_v})\wh{V}^{(0)}_{T_v*}} \\
&~~~ \leq \opnorm{\Sigma_{yT_vT_v}^{1/2}\wh{V}_{T_v*}^{(0)}}^2\opnorm{\Sigma_{yT_vT_v}^{-1/2}\wh{\Sigma}_{yT_vT_v}\Sigma_{yT_vT_v}^{-1/2}-I} 
\leq 4C\sqrt{\frac{s_u\log(ep/s_u)}{n}},
\end{align*}
with probability at least $1-\exp(-C's_u\log(ep/s_u))$. The proof is completed by realizing $\opnorm{\Sigma_{yT_vT_v}^{1/2}\wh{V}_{T_v*}^{(0)}} =\opnorm{\Sigma_y^{1/2}\wh{V}^{(0)}}$.
\end{proof}

\begin{proof}[Proof of Lemma \ref{lem:ratio2}]
Let $T_u=\wh{S}_u\cup S_u$, where $\wh{S}_u=\supp(\wh{U})$.
Using the definition of Frobenius norm, we have
\begin{align*}
& \big|\fnorm{\Sigma_x^{1/2}\Delta}^2-\fnorm{\wh{\Sigma}_x^{1/2}\Delta}^2\big| = \big|\Tr(\Delta'(\wh{\Sigma}_x-\Sigma_x)\Delta)\big| 
= \big|\Tr((\Delta_{T_u*})'(\wh{\Sigma}_{xT_uT_u}-\Sigma_{xT_uT_u})\Delta_{T_u*})\big| \\
&~~~ \leq \fnorm{\Sigma_{xT_uT_u}\Delta_{T_u*}}^2\opnorm{\Sigma_{xT_uT_u}^{-1/2}\wh{\Sigma}_{xT_uT_u}\Sigma_{xT_uT_u}^{-1/2}-I} 
\leq C\sqrt{\frac{s_u\log(ep/s_u)}{n}}\fnorm{\Sigma_x^{1/2}\Delta}^2,
\end{align*}
with high probability, where we have used  $\fnorm{\Sigma_{xT_uT_u}\Delta_{T_u*}}^2=\fnorm{\Sigma_x^{1/2}\Delta}^2$ and Lemma 12 in \cite{gao14} in the last inequality. After rearrangement, the proof is complete.
\end{proof}

\begin{proof}[Proof of Lemma \ref{lem:l0noise1}]
In this proof, $\wh{\Sigma}_x$ is constructed from $\mathcal{D}_0$ and $\wh{\Sigma}_y$ is constructed from $\mathcal{D}_1$. We use the notation $T_u=S_u\cup \wh{S}_u$ and $T_v=S_v\cup \wh{S}_v$, where $\wh{S}_u=\supp(\wh{U})$ and $\wh{S}_v=\supp(\wh{V}^{(0)})$. Note that $T_u$ depends on $\mathcal{D}_1$ and $T_v$ depends on $\mathcal{D}_0$.
We first condition on $\mathcal{D}_0$, and then we have
\begin{eqnarray*}
&& \big|\Tr\big(\Delta'(\wh{\Sigma}_{xy}-\Sigma_{xy})\wh{V}^{(0)}\big)\big|
= \big|\iprod{\wh{\Sigma}_{xyT_uT_v}-\Sigma_{xyT_uT_v}}{\Delta_{T_u*}(\wh{V}^{(0)}_{T_v*})'}\big| \\
&\leq& \opnorm{\Sigma_{yT_vT_v}^{1/2}\wh{V}_{T_v*}^{(0)}}\fnorm{\Sigma_{xT_uT_u}^{1/2}\Delta_{T_u*}}\big|\iprod{\Sigma_{xT_uT_u}^{-1/2}(\wh{\Sigma}_{xyT_uT_v}-\Sigma_{xyT_uT_v})\Sigma_{yT_vT_v}^{-1/2}}{K_{T_u}}\big| \\
&\leq& \opnorm{\Sigma_{yT_vT_v}^{1/2}\wh{V}_{T_v*}^{(0)}}\fnorm{\Sigma_{xT_uT_u}^{1/2}\Delta_{T_u*}}\sup_{T}\big|\iprod{\Sigma_{xTT}^{-1/2}(\wh{\Sigma}_{xyTT_v}-\Sigma_{xyTT_v})\Sigma_{yT_vT_v}^{-1/2}}{K_{T}}\big|
\end{eqnarray*}
where $T$ ranges over all subsets with cardinality bounded by $2s_u$, and for each such $T$, $K_{T}=\fnorm{\Sigma_{xTT}^{1/2}\Delta_{T*}(\wh{V}^{(0)}_{T_v*})'\Sigma_{yT_vT_v}^{1/2}}^{-1}\Sigma_{xTT}^{1/2}\Delta_{T*}(\wh{V}^{(0)}_{T_v*})'\Sigma_{yT_vT_v}^{1/2}$ satisfying $\fnorm{K_{T}}=1$. We do not put $T_v$ in the subscript of $K$ because conditioning on $\mathcal{D}_0$, $T_v$ is fixed. For each $T$, we can use Lemma 7 in \cite{gao14} to bound $\left|\iprod{\Sigma_{xTT}^{-1/2}(\wh{\Sigma}_{xyTT_v}-\Sigma_{xyTT_v})\Sigma_{yT_vT_v}^{-1/2}}{K_{T}}\right|$. A direct union bound argument leads to
$$\sup_{T}\left|\iprod{\Sigma_{xTT}^{-1/2}(\wh{\Sigma}_{xyTT_v}-\Sigma_{xyTT_v})\Sigma_{yT_vT_v}^{-1/2}}{K_{T}}\right|\leq C\sqrt{\frac{rs_u+s_u\log(ep/s_u)}{n}},$$
with probability at least $1-\exp\left(-C'(s_u\log(ep/s_u)+rs_u)\right)$. By Lemma \ref{lem:V0}, we have $\opnorm{\Sigma_{yT_vT_v}^{1/2}\wh{V}_{T_v*}^{(0)}}=\opnorm{\Sigma_y^{1/2}\wh{V}^{(0)}}\leq 2$ with high probability. Finally, observing that $\fnorm{\Sigma_{xT_uT_u}^{1/2}\Delta_{T_u*}}=\fnorm{\Sigma_{x}^{1/2}\Delta}$, we have completed the proof.
\end{proof}

\begin{proof}[Proof of Lemma \ref{lem:l0noise2}]
It is omitted due to similarity to that of Lemma \ref{lem:l0noise1}.
\end{proof}

\begin{proof}[Proof of Proposition \ref{prop:loss}]
Let the singular value decomposition of $U$ be $U=\Theta RH'$. Then we have $HR\Theta'\Sigma_x\Theta RH'=U'\Sigma_xU=I$, from which we derive $\Theta'\Sigma_x\Theta=R^{-2}$. Using Lemma \ref{lem:proj}, we have
\begin{align*}
& \fnorm{P_{\wh{U}}-P_U} = \sqrt{2}\inf_W\fnorm{\wh{U}W-\Theta} \\
~~~&\leq \sqrt{2}\inf_W\fnorm{\wh{U}WHR^{-1}-\Theta R H'HR^{-1}} 
\leq \sqrt{2}\inf_W\fnorm{\wh{U}W-U}\opnorm{R^{-1}} \\
~~~&\leq \sqrt{2}M_1^{-1/2}\inf_W\fnorm{\Sigma_x^{1/2}(\wh{U}W-U)}\opnorm{\Theta'\Sigma_x\Theta}^{1/2} \\
~~~& \leq \sqrt{2}(M_2/M_1)^{1/2}\inf_W\fnorm{\Sigma_x^{1/2}(\wh{U}W-U)}\\
~~~&\leq \sqrt{2}(M_2/M_1)^{1/2}\inf_{W\in O(r,r)}\fnorm{\Sigma_x^{1/2}(\wh{U}W-U)}.
\end{align*}
Finally, by $\fnorm{\Sigma_x^{1/2}(\wh{U}W-U)}^2=\Tr((\wh{U}W-U)'\Sigma_x(\wh{U}W-U))$, the proof is complete.
\end{proof}


\section{Implementation of \eqref{eq:optpoly}}

\label{sec:admm}

To implement the convex programming \eqref{eq:optpoly}, we turn to the Alternating Direction Method of Multipliers (ADMM) \citep{douglas1956, Boyd11}. In the rest of this section, we write $\wh\Sigma_x$ and $\wh\Sigma_y$ for $\wh\Sigma_x^{(0)}$ and $\wh\Sigma_y^{(0)}$ for notational convenience.

First, note that \eqref{eq:optpoly} can be rewritten as
\begin{equation}
\label{eq:admm}
\begin{aligned}
& \mbox{minimize} &&~ f(F) + g(G),\\
& \mbox{subject to} &&~ \wh\Sigma_x^{\hf} F \wh\Sigma_y^{\hf} - G = 0,
\end{aligned}
\end{equation}
where
\begin{align}
	\label{eq:admm-f}
f(F) & = - \iprod{\wh\Sigma_{xy}}{F} + \rho \norm{F}_1,\\
\label{eq:admm-g}
g(G) & = \infty \indc{\norm{G}_* > r} + \infty \indc{\opnorm{G} > 1}.
\end{align}
Thus, the augmented Lagrangian form of the problem is
\begin{align}
\label{eq:admm-lag}
\calL_\eta(F, G, H) = f(F) + g(G) + 
\iprod{H}{\wh\Sigma_x^{\hf} F \wh\Sigma_y^{\hf} - G}
+ \frac{\eta}{2}\Fnorm{\wh\Sigma_x^{\hf} F \wh\Sigma_y^{\hf} - G}^2. 
\end{align}

Following the generic algorithm spelled out in Section 3 of \cite{Boyd11}, suppose after the $k$-th iteration, the matrices are $(F^k, G^k, H^k)$, then we update the matrices in the $(k+1)$-th iteration as follows:
\begin{align}
F^{k+1} & = \argmin_F \calL_\eta(F, G^k, H^k), 
\label{eq:admm-F-update}\\
G^{k+1} & = \argmin_G \calL_\eta(F^{k+1}, G, H^k), 
\label{eq:admm-G-update}\\
H^{k+1} & = H^k + \eta (\wh\Sigma_x^{\hf} F^{k+1} \wh\Sigma_y^{\hf} - G^{k+1}).
\label{eq:admm-H-update}
\end{align}
The algorithm iterates over \eqref{eq:admm-F-update} -- \eqref{eq:admm-H-update} till some convergence criterion is met.
It is clear that the update \eqref{eq:admm-H-update} for the dual variable $H$ is easy to calculate.
Moreover the updates \eqref{eq:admm-F-update} and \eqref{eq:admm-G-update} can be solved easily and have explicit meaning in giving solution to sparse CCA. We are going to show that  \eqref{eq:admm-F-update} can be viewed as a Lasso problem. Thus, this step targets at the sparsity of the matrix $UV'$. The update \eqref{eq:admm-G-update} turns out to be equivalent to a singular value capped soft thresholding problem, and it targets at the low-rankness of the matrix $\Sigma_x^{1/2}UV'\Sigma_y^{1/2}$.
In what follows, we study in more details the updates for $F$ and $G$.

First, we note that \eqref{eq:admm-F-update} is equivalent to
\begin{align}
F^{k+1} 
& = \argmin_F f(F) + \iprod{H^k}{\wh\Sigma_x^{\hf} F \wh\Sigma_y^{\hf}} 
+ \frac{\eta}{2}\Fnorm{\wh\Sigma_x^{\hf} F \wh\Sigma_y^{\hf} - G^k}^2
\nonumber \\
& = \argmin_F \frac{\eta}{2}\Fnorm{\wh\Sigma_x^{\hf} F \wh\Sigma_y^{\hf} - (G^k - \frac{1}{\eta}H^k + \frac{1}{\eta} \wh\Sigma_{x}^{-1/2} \wh\Sigma_{xy} \wh\Sigma_{y}^{-1/2})}^2 + \rho\norm{F}_1.
\label{eq:admm-F-update1}
\end{align}
Thus, it is clear that the update of $F$ in \eqref{eq:admm-F-update} can be viewed as a Lasso problem as summarized in the following proposition.
Here and after, for any positive semi-definite matrix $A$, $A^{-1/2}$ denotes the principal square root of its pseudo-inverse.
\begin{proposition}
\label{prop:F-update}
Let $\mathrm{vec}$ be the vectorization operation of any matrix and $\otimes$ the Kronecker product. 
Then $\mathrm{vec}(F^{k+1})$ is the solution to the following standard Lasso problem
\begin{align*}
\min_x~~ \Fnorm{ \Gamma x - b }^2 + \frac{2\rho}{\eta} \norm{x}_1
\end{align*}
where $\Gamma = \wh\Sigma_y^{1/2}\otimes \wh\Sigma_x^{1/2}$ and $b = \mathrm{vec}(G^k - \frac{1}{\eta}H^k + \frac{1}{\eta} \wh\Sigma_{x}^{-1/2} \wh\Sigma_{xy} \wh\Sigma_{y}^{-1/2})$.
\end{proposition}

\begin{remark}
It is worth mentioning that the vectorized formulation in \prettyref{prop:F-update} is for illustration only. 
In practice, we solve the problem in \eqref{eq:admm-F-update1} directly, since the vectorized version, especially the Kronecker product, would great increase the computation cost.
The solver to \eqref{eq:admm-F-update1} can be easily implemented in standard software packages for convex programming, such as TFOCS \cite{becker11}.
\end{remark}

Since each update of $F$ is the solution of some Lasso problem, it should be sparse in the sense that its vector $\ell_1$ norm is well controlled. 

Turning to the update for $G$, we note that \eqref{eq:admm-G-update} is equivalent to 
\begin{align}
G^{k+1} 
& = \argmin_G g(G) - \iprod{H^k}{G} + \frac{\eta}{2}\Fnorm{\wh\Sigma_x^{\hf} F^{k+1} \wh\Sigma_y^{\hf} - G}^2
\nonumber \\
& = \argmin_G \frac{\eta}{2}\Fnorm{G - (\frac{1}{\eta} H^k + \wh\Sigma_x^{\hf} F^{k+1} \wh\Sigma_y^{\hf} )}^2 \nonumber \\
& \hskip 8em + \infty \indc{\norm{G}_* > r} + \infty \indc{\opnorm{G} > 1}
\nonumber \\
& = \argmin_G \Fnorm{G - (\frac{1}{\eta} H^k + \wh\Sigma_x^{\hf} F^{k+1} \wh\Sigma_y^{\hf} )}^2\nonumber \\
& \hskip 8em + \infty \indc{\norm{G}_* > r} + \infty \indc{\opnorm{G} > 1}.
\label{eq:admm-G-update1}
\end{align}
The solution to the last display has a closed form according to the following result.

\begin{proposition}
\label{prop:capped-soft-th}
Let $G^*$ be the solution to the optimization problem:
\begin{equation*}
\begin{aligned}
\mathrm{minimize}~~ & \Fnorm{G - W}\\
\mathrm{subject~to}~~ & \norm{G}_* \leq r,\quad \opnorm{G} \leq 1.
\end{aligned}
\end{equation*}
Let the SVD of $W$ be $W = \sum_{i=1}^m \omega_i a_i b_i'$ with $\omega_1\geq \cdots\geq \omega_m \geq 0$ the ordered singular values.
Then $G^* = \sum_{i=1}^m g_i a_i b_i'$ where for any $i$, $g_i = 1 \wedge (\omega_i - \gamma^*)_+$ for some  $\gamma$ which is the solution to
\begin{equation*}
\begin{aligned}
\mathrm{minimize} ~~ & \gamma, \qquad
\mathrm{subject~to}~~\gamma > 0 ,~~ \sum_{i=1}^m 1 \wedge (\omega_i - \gamma)_+ \leq r.
\end{aligned}
\end{equation*}
\end{proposition}
\begin{proof}
The proof essentially follows that of Lemma 4.1 in \cite{Vu13}. In addition to the fact that the current problem deals with asymmetric matrix, the only difference that we now have an inequality constraint $\sum_i g_i \leq r$ rather than an equality constraint as in \cite{Vu13}. 
The asymmetry of the current problem does not matter since it is orthogonally invariant.
\end{proof}

Here and after, we call the operation in Proposition \ref{prop:capped-soft-th} singular value capped soft thresholding (SVCST) and write $G^* = \mathrm{SVCST}(W)$. 
Thus, any update for $G$ results from the SVCST operation of some matrix, and so it has well controlled singular values.


In summary, the convex program (\ref{eq:optpoly}) is implemented as Algorithm \ref{algo:admm}.

\begin{algorithm}[tbh]
\KwIn{}
1. Sample covariance matrices $\wh\Sigma_x$, $\wh\Sigma_y$ and $\wh\Sigma_{xy}$, \\
2. Penalty parameter $\rho$, \\
3. Rank~$r$, \\
4. ADMM parameter $\eta$ and tolerance level $\epsilon$.
\smallskip

\KwOut{Estimated sparse canonical correlation signal $\wh{A}$.}
\smallskip

\nl Initialize: $k = 0$, $F^0 = \mathrm{SVCST}(\wh\Sigma_{xy}), G^0 = 0, H^0 = 0$.

\Repeat{$\max\{\fnorm{F^{k+1} - F^k}, \rho \fnorm{G^{k+1} - G^k} \}\leq \epsilon$}{
\nl Update $F^{k+1}$ as in \eqref{eq:admm-F-update} ~~~ (Lasso) \; 

\nl Update $G^{k+1} \leftarrow \mathrm{SVCST}(\eta^{-1} H^k + \wh\Sigma_x^{\hf} F^{k+1} \wh\Sigma_y^{\hf})$ ~~~ (SVCST) \; 

\nl Update $H^{k+1} \leftarrow H^k + \eta (\wh\Sigma_x^{\hf} F^{k+1} \wh\Sigma_y^{\hf} - G^{k+1})$ \;

\nl $k \leftarrow k+1$ \;
}

\nl Return $\wh{A} = F^k$.
\caption{An ADMM algorithm for SCCA\label{algo:admm}}
\end{algorithm}

\section{Numerical Studies}\label{sec:num}

This section presents numerical results demonstrating the competitive finite sample performance of the proposed adaptive estimation procedure CoLaR on simulated datasets.

\paragraph{Simulation settings}
We consider three simulation settings.
In all these settings, we set $p = m$, $\Sigma_x = \Sigma_y = \Sigma$, and
$r = 2$ with $\lambda_1 = 0.9$ and $\lambda_2 = 0.8$. 
Moreover, the nonzero rows of both $U$ and $V$ are set at $\{1, 6, 11, 16, 21\}$.
The values at the nonzero coordinates are obtained from normalizing (with respect to $\Sigma$) random numbers drawn from the uniform distribution on the finite set $\{-2,1,0,1,2\}$.
The choices of $\Sigma$ in the three settings are as follows:
\begin{enumerate}
\item \texttt{Identity}: $\Sigma = I_p$.

\item \texttt{Toeplitz}: $\Sigma = (\sigma_{ij})$ where $\sigma_{ij} = 0.3^{|i-j|}$ for all $i,j\in [p]$.
In other words, $\Sigma_x$ and $\Sigma_y$ are Toeplitz matrices.

\item \texttt{SparseInv}: $\Sigma = ({\sigma^0_{ij}}/{\sqrt{\sigma^0_{ii} \sigma^0_{jj}}})$. We set $\Sigma^0 = (\sigma^0_{ij}) = \Omega^{-1}$ where $\Omega = (\omega_{ij})$ with
\begin{align*}
\omega_{ij} = \indc{i=j} + 0.5 \times \indc{|i-j| = 1} + 0.4 \times \indc{|i-j| = 2}, \quad i,j\in [p].
\end{align*}
In other words, $\Sigma_x$ and $\Sigma_y$ have sparse inverse matrices.
\end{enumerate}
In all three settings, we normalize the variance of each coordinate to be one.

%
%
%

\paragraph{Implementation details}

The proposed CoLaR estimator in \prettyref{sec:adapt-est} has two stages.
The convex program \eqref{eq:optpoly} in the first stage can be solved via an ADMM algorithm \citep{Boyd11}. 
The details of the ADMM approach are presented in Section \ref{sec:admm}. 
The optimization problem \eqref{eq:refinepoly} in the second stage can be solved by a standard group-Lasso algorithm \citep{yuan06}. 

In all numerical results reported in this section, we used the same penalty level $\rho = 0.55\sqrt{\log(p+m)/n}$ in \eqref{eq:optpoly} and we used $\eta = 2$ in \eqref{eq:admm-H-update}.
In \eqref{eq:refinepoly}, we used five-fold cross validation to select a common penalty parameter $\rho_u = \rho_v = b\sqrt{(r+\log p)/n}$.
In particular,
for $l = 1,\dots, 5$, we use one fold of the data as the test set $(X_{(l)}^{\mathrm{test}}, Y_{(l)}^{\mathrm{test}})$ and the other four folds as the training set $(X_{(l)}^{\mathrm{train}}, Y_{(l)}^{\mathrm{train}})$. 
For any choice of $b$, we solved \eqref{eq:refinepoly} on $(X_{(l)}^{\mathrm{train}}, Y_{(l)}^{\mathrm{train}})$ to obtain estimates $(\wh{U}_{(l)}, \wh{V}_{(l)})$. 
Then we computed the sum of canonical correlations between $X_{(l)}^{\mathrm{test}} \wh{U}_{(l)} \in \reals^{n\times r}$ and  $Y_{(l)}^{\mathrm{test}} \wh{V}_{(l)} \in \reals^{n\times r}$ to obtain $\mbox{CV}_l(b)$.
Finally, $\mbox{CV}(b) = \sum_{l=1}^5 \mbox{CV}_l(b)$.
Among all the candidate penalty parameters, we select the $b$ value such that $\mbox{CV}(b)$ is maximized.
The candidate $b$ values used in the simulation below are $\sth{0.5,1,1.5,2}$.
Throughout the simulation, we used all the sample $\{(X_i, Y_i)\}_{i=1}^n$ to form the sample covariance matrices used in \eqref{eq:optpoly} -- \eqref{eq:mmxpoly}.

In addition to the performance of CoLaR,
we also report that of the method proposed in \cite{witten09} (denoted by PMA here and on). The PMA seeks the solution to the optimization problem
$$\max_{u,v} u'\wh{\Sigma}_{xy}v,\quad \text{subject to }||u||\leq 1, ||v||\leq 1, ||u||_1\leq c_1, ||v||_1\leq c_2.$$
The solution is used to estimate the first canonical pair $(\wh{u}_1,\wh{v}_1)$. 
Then the same procedure is repeated after $\wh{\Sigma}_{xy}$ is replaced by $\wh{\Sigma}_{xy}-(\wh{u}_1'\wh{\Sigma}_{xy}\wh{v}_1)\wh{u}_1\wh{v}_1'$, and the solution gives the estimator of the second canonical pair $(\wh{u}_2,\wh{v}_2)$. This process is repeated until $\wh{u}_r,\wh{v}_r$ is obtained. 
Note that the normalization constraint $\norm{u}\leq 1$ and $\norm{v}\leq 1$
implicitly assumes that the marginal covariance matrices ${\Sigma}_x$ and $\Sigma_y$ are identity matrices.
We used the R implementation of the method (function \texttt{CCA} in the \texttt{PMA} package in R) by the authors of \cite{witten09}. 
To remove undesired amplification of error caused by normalization, we renormalized each individual $\wh{u}_j$ with respect to $\wh\Sigma_x$ and each individual $\wh{v}_j$ with respect to $\wh{\Sigma}_y$ before calculating the error under the loss \eqref{eq:recerror}.
For each simulated dataset, we set the sparsity penalty parameters \texttt{penaltyx} and \texttt{penaltyz} of the function \texttt{CCA} at each of the eleven different values $\{0.6^l: l=0,1,\dots, 10\}$ and only the smallest estimation error out of all eleven trials was used to compute the error reported in the tables below.

\paragraph{Results} 
Tables \ref{tab:res-id} -- \ref{tab:res-sp} report, in each of the three settings, the medians of the prediction errors of CoLaR and PMA out of $100$ repetitions for four different configurations of $(p, m, n)$ values. 

In each table, the columns {$U$-PMA} and $V$-PMA report the medians of the smallest estimation errors out of the eleven trials on each simulated dataset.
The columns $U$-init and $V$-init report the median estimation errors of the renormalized $r$ left singular vectors and right singular vectors of the solutions to the initialization step \eqref{eq:optpoly}, where the renormalization is the same as in \eqref{eq:mmxpoly} and in both \eqref{eq:optpoly} and renormalization we used all the $n$ pairs of observations.
Last but not least, the columns $U$-CoLaR and $V$-ColaR report the median estimation errors of the CoLaR estimators where both stages were carried out.

In all simulation settings, both the renormalized initial estimators and the CoLaR estimators consistently outperform PMA.
Comparing the last four columns within each table, we also find that the CoLaR estimators with both stages carried out significantly improve over the renormalized initial estimators,
which is in accordance with our theoretical results in Section \ref{sec:adaptive}.

In summary, the proposed method delivers consistent and competitive performance in all three covariance settings across all dimension and sample size configurations, and its behavior agrees well with the theory.

\begin{table}[!h]
\centering
	\smallskip
\begin{small}
\begin{tabular}{c||c|c|c|c|c|c}
\hline
$(p,m,n)$ & $U$-PMA & $V$-PMA & 
$U$-init & $V$-init & $U$-CoLaR & $V$-CoLaR \\
\hline
\hline
(300, 300, 200) & 
2.1316 & 2.1297 & 
0.2653 & 0.1712 & 0.0498 & 0.0646\\
\hline
(600, 600, 200) &
3.4154 & 3.3584 & 
0.3167 & 0.2087 & 0.0671 & 0.0776 \\
\hline
(300, 300, 500) &
0.2683 & 0.2701 & 
0.1207 & 0.0665 & 0.0135 & 0.0159 \\ 
\hline
(600, 600, 500) & 
2.0335 & 2.0368 & 
0.1448 & 0.0817 & 0.0166 & 0.0203 \\
\hline
\end{tabular}
\end{small}
\caption{Prediction errors (\texttt{Identity}): Median in 100 repetitions.
\label{tab:res-id}}
\end{table}

\begin{table}[!h]
\centering
	\smallskip
\begin{small}
	\begin{tabular}{c||c|c|c|c|c|c}
	\hline
	$(p,m,n)$ & $U$-PMA & $V$-PMA & 
	$U$-init & $V$-init & $U$-CoLaR & $V$-CoLaR \\
\hline
\hline
(300, 300, 200) & 
2.1853 & 2.1840 & 
0.2885 & 0.1706 & 0.0511 & 0.0601 \\
\hline
(600, 600, 200) &
3.4247 & 3.4852 & 
0.3236 & 0.2004 & 0.0638 & 0.0764 \\
\hline
(300, 300, 500) &
0.2358 & 0.2191 &  
0.1202 & 0.0664 & 0.0135 & 0.0166 \\
\hline
(600, 600, 500) & 
2.1214 & 2.0889 & 
0.1408 & 0.0811 & 0.0176 & 0.0209 \\
\hline
\end{tabular}
\end{small}
\caption{Prediction errors (\texttt{Toeplitz}): Median in 100 repetitions.
\label{tab:res-to}}
\end{table}

\begin{table}[!h]
\centering
	\smallskip
\begin{small}
	\begin{tabular}{c||c|c|c|c|c|c}
	\hline
	$(p,m,n)$ & $U$-PMA & $V$-PMA & 
	$U$-init & $V$-init & $U$-CoLaR & $V$-CoLaR \\
\hline
\hline
(300, 300, 200) & 
2.9697 & 2.9619 & 
0.5552 & 0.5718 & 0.1568 & 0.1194 \\
\hline
(600, 600, 200) &
4.6908 & 4.3339 & 
0.5596 & 0.6133 & 0.2123 & 0.1572 \\
\hline
(300, 300, 500) &
2.3967 & 2.0620 & 
0.2695 & 0.1917 & 0.0242 & 0.0219 \\
\hline
(600, 600, 500) & 
2.8707 & 2.8609 & 
0.3068 & 0.2368 & 0.0338 & 0.0271 \\
\hline
\end{tabular}
\end{small}
\caption{Prediction errors (\texttt{SparseInv}): Median in 100 repetitions.
\label{tab:res-sp}}
\end{table}

\paragraph{{Model misspecification}}
{We now examine the performance of our estimator when the model is misspecified.
To this end, we consider the case where there are three pairs of non-trivial canonical correlations present in the data but we set $r=2$ in our algorithm.
As before, we consider three different types of marginal covariance matrices: \texttt{Identity}, \texttt{Toeplitz} and \texttt{SparseInv}. 
In addition, we generate the first two pairs of canonical correlation vectors in the same way as before.
For the generation of the third pair of canonical directions, we consider two different scenarios. 
In the first scenario, the support of the third pair of canonical direction vectors are set at $\{1,6,11,16,21\}$ and so they are the same as those of the first two pairs.
In the second scenario, we put no constraint on the support of these vectors.
For both scenarios, we set $\lambda_3 = 0.3$.
\prettyref{tab:res-mis} reports the prediction errors of the first two pairs of canonical correlations in both scenarios when $(p,m,n) = (300, 300, 500)$. 
The implementation details are exactly the same as before.
The first two columns contain results in the first scenario, and the third and the fourth columns the second scenario.
Comparing these results with their counterparts in correctly specified models (the last two cells in the second last rows of Tables \ref{tab:res-id}--\ref{tab:res-sp}), we have found that the performance of our estimator was robust to model misspecification in both scenarios.
} 

\begin{table}[bth]
\centering
	\smallskip
\begin{small}
	\begin{tabular}{c||c|c|c|c}
	\hline
\multirow{2}{*}{Covariance} & \multicolumn{2}{c|}{Scenario 1} &
\multicolumn{2}{c}{Scenario 2} \\
\cline{2-5}
	 & $U$-CoLaR & $V$-CoLaR & 
	$U$-CoLaR & $V$-CoLaR \\
\hline
\hline
\texttt{Identity} & 
0.0190 & 0.0195 & 
0.0144 & 0.0160  \\
\hline
\texttt{Toeplitz} &
0.0197 & 0.0197 & 
0.0138 & 0.0147  \\
\hline
\texttt{SparseInv} &
0.0348 & 0.0263 & 
0.0221 & 0.0328  \\
\hline
\end{tabular}
\end{small}
\caption{Prediction errors with misspecified models: Median in 100 repetitions.
$(p,m,n) = (300, 300, 500).$
\label{tab:res-mis}}
\end{table}

\end{document}